\documentclass{aa} 

\usepackage[varg]{txfonts} 

\bibpunct{(}{)}{;}{a}{}{,} 
\usepackage{natbib}

\usepackage{amsmath} 
\usepackage{graphicx} 
\usepackage{xspace} 
\usepackage{xcolor} 
\usepackage{upgreek} 
\usepackage[caption = false]{subfig} 

\definecolor{linkcolor}{rgb}{0.6,0,0} 
\definecolor{citecolor}{rgb}{0,0,0.75}
\definecolor{urlcolor}{rgb}{0.12,0.46,0.7}
\usepackage[breaklinks, colorlinks, urlcolor=urlcolor, linkcolor=linkcolor,citecolor=citecolor,pdfencoding=auto]{hyperref}
\hypersetup{linktocpage}

\newcommand{\average}[1]{\langle#1\rangle}

\newcommand{\sform}[2]{{#1}\times 10^{#2}}
\renewcommand{\pi}{\uppi}
\newcommand{\kb}{k_\mathrm{B}}
\newcommand{\order}[1]{\mathcal{O}(#1)}
\newcommand{\Dirac}[1]{\delta_\mathrm{D}(#1)}

\newcommand{\Mpc}{\,h^{-1}\mathrm{Mpc}}

\newcommand{\iMpc}{\,h\mathrm{Mpc}^{-1}}

\newcommand{\Msun}{\,h^{-1}\mathrm{M_\odot}}
\newcommand{\Kelvin}{\,\mathrm{K}}

\newcommand{\Om}{\Omega_\mathrm{m}}

\newcommand{\Ol}{\Omega_\Lambda}
\newcommand{\Ob}{\Omega_\mathrm{b}}

\newcommand{\Oc}{\Omega_\mathrm{c}}

\newcommand{\dc}{\delta_\mathrm{c}}
\newcommand{\Dv}{\Delta_\mathrm{v}}

\newcommand{\eg}{e.g.,\xspace}
\newcommand{\ie}{i.e.\xspace}

\newcommand{\nbody}{$N$-body\xspace}
\newcommand{\LCDM}{$\Lambda$CDM\xspace}
\newcommand{\halofit}{\textsc{halofit}\xspace}

\newcommand{\hmcode}{\textsc{hmcode}\xspace}
\newcommand{\emu}{\textsc{cosmic emu}\xspace}

\newcommand{\planck}{\textit{Planck}\xspace}

\newcommand{\wmap}{\textit{WMAP}\xspace}
\newcommand{\cfhtlens}{CFHTLenS\xspace}
\newcommand{\kids}{KiDS\xspace}

\newcommand{\tsz}{tSZ\xspace}
\newcommand{\bahamas}{\textsc{bahamas}\xspace}

\newcommand{\sgenic}{\textsc{sgenic}\xspace}
\newcommand{\owls}{\textsc{owls}\xspace}
\newcommand{\cosmoowls}{\textsc{cosmo-owls}\xspace}

\newcommand{\agn}{\textsc{agn}\xspace}
\newcommand{\hi}{\textsc{agn-hi}\xspace}
\newcommand{\lo}{\textsc{agn-lo}\xspace}
\newcommand{\dmonly}{\textsc{dmonly}\xspace}

\newcommand{\meadaddress}{https://github.com/alexander-mead/collapse}

\newcommand{\ngenicaddress}{https://www.h-its.org/2014/11/05/ngenic-code/}

\newcommand{\bahamasaddress}{http://www.astro.ljmu.ac.uk/~igm/BAHAMAS/}
\newcommand{\bahamaslink}{\href{\bahamasaddress}{\bahamasaddress}}
\newcommand{\sgenicaddress}{https://github.com/sbird/S-GenIC}
\newcommand{\sgeniclink}{\href{\sgenicaddress}{\sgenicaddress}\xspace}

\def\apjl{ApJ Let.}

\def\aap{A\&A}

\def\apjs{ApJS}

\begin{document}
\title{A hydrodynamical halo model for weak-lensing cross correlations}
\author{A. J. Mead\inst{\ref{ins:Barcelona}, \ref{ins:Vancouver}\thanks{alexander.j.mead@googlemail.com}}
\and T. Tr\"oster\inst{\ref{ins:Edinburgh}}
\and C. Heymans\inst{\ref{ins:Edinburgh}, \ref{ins:Germany}}
\and L. Van Waerbeke\inst{\ref{ins:Vancouver}}
\and I. G. McCarthy\inst{\ref{ins:Liverpool}}
}
\institute{Institut de Ci\`encies del Cosmos, Universitat de Barcelona, Mart\'i Franqu\`es 1, E08028 Barcelona, Spain\label{ins:Barcelona}
\and Department of Physics and Astronomy, University of British Columbia, 6224 Agricultural Road, Vancouver, BC V6T 1Z1, Canada\label{ins:Vancouver}
\and Institute for Astronomy, University of Edinburgh, Royal Observatory, Blackford Hill, Edinburgh EH9 3HJ, UK\label{ins:Edinburgh}
\and Ruhr-University Bochum, Astronomical Institute, German Centre for Cosmological Lensing, Universit\"{a}tsstr. 150, 44801 Bochum, Germany\label{ins:Germany}
\and Astrophysics Research Institute, Liverpool John Moores University, 146 Brownlow Hill, Liverpool L3 5RF, UK\label{ins:Liverpool}
}
\date{Accepted June 30, 2020} 

\label{firstpage}

\abstract{ 
On the scale of galactic haloes, the distribution of matter in the cosmos is affected by energetic, non-gravitational processes; so-called baryonic feedback. A lack of knowledge about the details of how feedback processes redistribute matter is a source of uncertainty for weak-lensing surveys, which accurately probe the clustering of matter in the Universe over a wide range of scales. We develop a cosmology-dependent model for the matter distribution that simultaneously accounts for the clustering of dark matter, gas and stars. We inform our model by comparing it to power spectra measured from the \bahamas suite of hydrodynamical simulations. As well as considering matter power spectra, we also consider spectra involving the electron-pressure field, which directly relates to the thermal Sunyaev-Zel'dovich (tSZ) effect. We fit parameters in our model so that it can simultaneously model both matter and pressure data and such that the distribution of gas as inferred from tSZ has influence on the matter spectrum predicted by our model. We present two variants; one that matches the feedback-induced suppression seen in the matter--matter power spectrum at the per-cent level and a second that matches the matter--matter data slightly less well ($\simeq 2$ per cent), but that is able to simultaneously model the matter--electron pressure spectrum at the $\simeq 15$ per-cent level. We envisage our models being used to simultaneously learn about cosmological parameters and the strength of baryonic feedback using a combination of tSZ and lensing auto- and cross-correlation data.
} 


\keywords{
cosmology: theory
-- 
large-scale structure of Universe
}

\maketitle 

\section{Introduction}
\label{sec:introduction}

Weak gravitational lensing (reviewed by \citealt{Kilbinger2015}) is the name given to small, correlated distortions in the observed shapes of galaxies that are caused by coherent gravitational light deflections between the galaxies and the observer. These shape distortions are of small amplitude compared to the intrinsic shapes of galaxies and therefore one must have large samples of galaxies in order to tease out signal from the background `shape noise'. The correlation of shape distortions can be used to make maps of (a galaxy-redshift-weighted-version of) the matter distribution, and these can be used to understand the growth and evolution of matter perturbations in the cosmos. Many forthcoming surveys are being designed to measure this effect very precisely, with the ultimate goal of using structure to learn about the accelerated expansion of the Universe.

Weak lensing on its own is a promising tool to infer the distribution of structure, but in order to extract accurate cosmological-parameter constraints from data sets it is necessary to have theoretical models that are significantly more accurate than the survey error bars. On very large scales, density fluctuations are small and linear perturbation theory describes the distribution and evolution or the matter accurately enough for all conceivable future data sets \citep{Lesgourgues2011b}. However, the weak-lensing signal is predominantly determined by high amplitude, `non-linear', fluctuations on small scales \citep{Jain1997} and these have not successfully been described by any ab-initio calculation. The standard scheme for modelling the weak-lensing signal is therefore to run \nbody simulations over the range of cosmological scenarios under consideration and then to simply measure the fluctuation spectrum as seen in the simulations. One can then either devise fitting functions for the spectra \citep[\eg][]{Smith2003, Takahashi2012, Mead2015b} or build so-called `emulation' schemes \citep[\eg][]{Lawrence2010, Agarwal2012, Lawrence2017, Knabenhans2019,Angulo2020} to interpolate between simulation outputs. The lensing signal can then be calculated by integrating over these power spectra along the line-of-sight with weightings to account for the projection and lensing.

Most cosmological simulations consider only the gravitational interaction\footnote{Often simulations that consider only the gravitational interaction are called `dark-matter only'. We prefer `gravity only' because these simulations \emph{do} contain baryons, and the initial conditions are set up to account for baryons, but they consider only the gravitational interaction between particles when they are subsequently run.}, partly because it is simpler and partly because non-gravitational effects are less well understood. Early analytic work by \cite{White2004b} demonstrated that baryons cooling in a halo may contract and core the inner halo profile, and it is now known that non-gravitational processes can impact the power spectrum of the matter distribution in a way that will significantly bias cosmological parameter constraints if not accounted for \citep[\eg][]{Semboloni2011, Mohammed2014b, Schneider2016, Copeland2018, Huang2019}. The \owls suite of hydrodynamical simulations of \cite{Schaye2010} were used by \cite{vanDaalen2011} to demonstrate that the matter field can be particularly affected by active-galactic nuclei (AGN) feedback and showed that the main effect that reshapes the matter field is the redistribution of gas that is heated as a result of accretion energy generated near the central black hole. This heating is extreme enough that gas can even be expelled from the host haloes. Dismayingly, there is a huge range in the possible amplitude of the effects of feedback \citep[][]{LeBrun2014, McCarthy2017, Chisari2018} and if this remains unaddressed then the ability of the next generation of weak-lensing surveys to learn about the accelerated expansion will be severely compromised. The effect of baryonic feedback on matter clustering and on the structure of haloes has also been investigated using other hydrodynamic simulations: \cite{Martizzi2012, Martizzi2013} look at the effect of feedback on halo profiles. \cite{Velliscig2015} does the same with the \cosmoowls simulations and \cite{Mummery2017} using the more recent `BAryons and HAloes of MAssive Systems' (\bahamas\footnote{\bahamaslink}) simulations. All authors agree that the statistics of the matter field on scales relevant to forthcoming weak lensing surveys can be affected at the tens of per cent level.

Several techniques have been put forward to mitigate the impact that the unknown feedback strength may have on cosmological-parameter constraints from weak-lensing surveys: \cite{Eifler2015} advocate determining a set of `principal components' from libraries of hydrodynamical simulations and then marginalising over these in a data analysis. \cite{Mead2015b} use a modified version of the halo model with parameters that determine the effects of feedback on the halo profiles via a change in concentration and an overall bloating of the halo. The relative merits of these two approaches have been explored by \cite{Huang2019}. Several authors \citep[][]{Rudd2008, Semboloni2011, Fedeli2012, Semboloni2013, Fedeli2014a, Fedeli2014b, Debackere2019} have developed halo models that specifically include a gas component that can be used to model the effect of feedback on the matter--matter spectrum. Recently, \cite{vanDaalen2020} have shown that the differences in the effect of feedback on the power spectrum can be mainly attributed to different gas fractions in haloes of $\sim 10^{14}\Msun$, and showed that the feedback amplitude is close to `universal' when expressed in this variable. Another promising technique is the `baryonification' method \citep{Schneider2015, Schneider2018, Arico2019} where haloes in gravity-only simulations are manually deformed in post-processing in such a way that they have profiles appropriate for their gas content. The deformation can be informed either from hydrodynamic simulations or from observational data. A variant of this method is investigated by \cite*{Dai2018} who use the actual equations of motion governing gas physics to perform the halo deformation.

A more optimistic position to take is to regard the impact of feedback on observables as a positive \citep[\eg][]{Harnois-Deraps2015a, Foreman2016, MacCrann2017}; we may be able to learn about energetic processes within galaxies by analysing the weak-lensing data. An analysis of \cfhtlens data by \cite{MacCrann2015} included the effect of feedback via a single parameter and a functional form extracted from hydrodynamical simulations. \cite{Joudaki2017a} performed a similar analysis and used the model of \cite{Mead2015b} to investigate feedback. Both authors found a modest preference in the data for the presence of feedback. Recent Kilo Degree Survey (\kids) analyses \citep{Hildebrandt2017, Joudaki2017b, Hildebrandt2019} use the model of \cite{Mead2015b} and have similar results for the feedback amplitude. Recent results from the Deep Lens Survey \citep[][]{Yoon2019} using the same model show a stronger preference for strong feedback. However, the matter--matter spectrum may not be the ideal tool if one is interested in energetic processes in galaxies since it is not probing the gas directly.

To measure the gas distribution more directly one can use the thermal Sunyaev-Zel'dovich (tSZ) effect: the inverse Compton scattering of the relatively cold cosmic-microwave background (CMB) photons by relatively energetic thermal electrons that predominantly exist in galaxy clusters. This scattering results in an increase in energy of the CMB photons, with a characteristic frequency-dependence that distorts the standard black-body CMB spectrum in the direction of the galaxy cluster on the sky. The amplitude of the effect depends on the product of electron number density and temperature along the line of sight, a quantity with the same units as pressure and that is therefore known as `electron pressure'. As a scattering process, the tSZ amplitude does not decay with distance and therefore traces the large-scale structure out to high redshift. The characteristic frequency dependence of the induced deviation of the CMB spectrum from blackbody can be exploited to make high fidelity maps of the electron pressure distribution in the Universe \citep[\eg][]{Planck2015XXII}. Due to the fact that tSZ emanates from hot gas in dense clusters, hydrodynamic simulations have been an essential tool for modelling and understanding the effect. Early simulations include those by \cite{Suto1998} and \cite{Bond2005}. \cite{Trac2011} developed a technique to paint electron pressure onto a gravity-only simulation and therefore to quickly make mock tSZ maps and power spectrum templates. It was later realised that AGN feedback could have a large impact on the tSZ signal \citep{Dolag2009, Arnaud2010, Schaye2010, McCarthy2010, Battaglia2010, McCarthy2011}, and therefore that these violent feedback episodes needed to be carefully included in simulations, which in turn lead to work to investigate the magnitude of this effect \citep[\eg][]{Martizzi2012, Battaglia2012a, Battaglia2012b, Puchwein2013a, Velliscig2014, LeBrun2014, McCarthy2014, LeBrun2015, Sijacki2015, Dolag2016}.

The lensing--tSZ cross correlation has been measured by \cite{VanWaerbeke2014}, \cite{Hill2014} and \cite{Hojjati2017}. Observational work is often conducted using a stacking approach, where the tSZ map is stacked around the locations of suspected tSZ emitters. \cite{Makiya2018} does the same but with galaxies rather than haloes. \cite{Hill2018} and \cite{Tanimura2019b} look at tSZ around SDSS galaxies, see a strong signal and both find evidence for a `two-halo term' -- evidence of correlation that extends to scales much beyond the halo boundary.\cite{Tanimura2019a} and \cite{deGraaff2019} looked at residual tSZ emission \emph{between} close pairs of galaxies and found evidence of the filamentary structure that would be expected to exist given the cosmic web. \cite{Tanimura2018} mask cluster galaxies and find evidence for a residual `intra-cluster' tSZ signal. All evidence that supports tSZ emanating from low-density regions of the Universe as well as from dense clusters. \cite{Hojjati2015} used the \cosmoowls simulations to investigate the lensing--tSZ cross correlation measured by \cite{VanWaerbeke2014}, similar work was carried out by \cite{Battaglia2015}. An advantage of using the cross correlation between lensing and tSZ is that we can ignore many systematic effects that may plague the individual data sets. However, in some cases systematic effects may be correlated between data sets, which may be the case for thermal dust emission from dusty galaxies that will correlate with the lensing signal and may contaminate tSZ maps during the map-making process. \cite{Yan2018} showed that this may affect the cross correlation between CMB lensing and tSZ, more than between lower-redshift galaxy lensing and tSZ, due to the higher redshift contribution to CMB lensing. It may be possible to address issues of contamination from thermal dust using the methods of \cite{Addison2012} and \cite{Addison2013}.  More recently, the cross correlation has been measured and investigated by \cite{Baxter2019} and \cite{Omori2019} in the context of a cross correlation of CMB and galaxy lensing with the Dark Energy Survey, and by \cite{Osato2018} and \cite{Osato2020} with the Hyper Suprime-Cam survey.

To understand the tSZ signal \cite{Refregier2002} construct a halo model and compare this model to a simulation in order to investigate the tSZ auto-spectrum, demonstrating good agreement with early simulations. \cite{Komatsu2002} construct a similar model and demonstrate how the tSZ spectrum scales with cosmological parameters, particularly demonstrating an extreme sensitivity to $\sigma_8$, the amplitude of the linear power spectrum. These early works suggested that the tSZ auto-spectrum could contain a wealth of information about cosmological parameters. \cite*{Holder2007} examined the relative impacts of feedback (preheating) and changes in cosmology on various tSZ statistics, including the power spectrum. More recent works attempting to utilise tSZ for cosmology are by \cite{Shaw2010}, \cite{Bolliet2018} and \cite{Hill2018}. In this paper we do not utilise the tSZ auto-spectrum as we are concerned that it may contain poorly understood internal systematics (\citealt{Planck2015XXIV}; although see \citealt{Horowitz2017}). \cite{Hill2014} measured and investigated CMB lensing--tSZ correlation while \cite{Ma2015} investigated the galaxy lensing--tSZ detection from \cite{VanWaerbeke2014}. In both cases the cross correlation was modelled using the language of the halo model, with NFW profiles being taken for the dark matter and `universal pressure profiles' (UPP; \citealt{Arnaud2010}) for the electron pressure.

The purpose of this paper is to develop a halo model that may be used for a joint analysis of lensing and tSZ. We demand that the model be able to reproduce the lensing--lensing auto correlation function and the lensing--tSZ cross correlation. We also demand that the physics underpinning the model be the same so that the two signals depend on each other in a physical way. This is different from previous work in which one profile is used for the matter and a completely separate profile is used for the electron pressure. In this way we hope to be able to learn about cosmological parameters and feedback using both data sets simultaneously. Unless otherwise stated, all figures in this paper are made with cosmological parameters inspired by the \wmap9 \citep{Hinshaw2013} data analysis: $\Om=0.2793$, $\Ob=0.0463$, $\Ol=1-\Om$, $\sigma_8=0.821$, $n_\mathrm{s}=0.972$, $h=0.700$. These are the same cosmological parameters that are used in the \bahamas hydrodynamical simulations.

This paper is set out as follows: In Section~\ref{sec:halomodel} we show how the halo model can be used to make predictions for the cross spectrum of any field pair, and how we can improve the model by specifically describing a halo in terms of a cold-dark matter (CDM), gas and stellar component. In Section~\ref{sec:ingredients} we list our ingredient choices for our halo-model calculations. In Section~\ref{sec:simulations} we detail the \bahamas simulations, against which we compare our model. In Section~\ref{sec:improved} we tune our halo model by fitting its parameters to simulation data, which is the main results of this paper. Our conclusions are in Section~\ref{sec:summary}. We also provide a number of appendices: Appendix~\ref{app:two_halo} examines a common pitfall regarding computation of the two-halo term. Appendix~\ref{app:power_measurement} discusses how we measure power spectra and shot noise from the \bahamas simulations where we have particles with different masses and different properties. In Appendix~\ref{app:variations} we present a series of plots that show how changing our halo-model parameters affects our power spectra. Finally, Appendix~\ref{app:projections} details how we compute projected power spectra of tracers of our underlying three-dimensional fields and we show how angular scales of the projected power spectra receive contributions from the underlying three-dimensional scales.

\section{Halo model}
\label{sec:halomodel}

In this section we review the basic halo-model computation of the cross spectrum between two three-dimensional cosmological fields \citep{Seljak2000,Peacock2000,Scoccimarro2001,Cooray2002}. We start with a general calculation that can be used to calculate the cross spectrum for \emph{any} cosmological field pair as long as the halo profiles of the fields are known. Some examples of fields for which this calculation could be applied are `matter overdensity', `gas overdensity', `electron pressure', or `galaxy overdensity'. We then specialise to the details of the calculation appropriate for the lensing--lensing and \tsz--lensing power spectra that are the main focus of this paper, where lensing is generated by the matter overdensity field and \tsz by the electron-pressure field\footnote{Note that we use matter overdensity but straight electron pressure, not the pressure overdensity, because these are the exact quantities probed by lensing and tSZ respectively.}. The modelling presented here draws heavily on work by \cite{Refregier2002}, \cite{Shaw2010}, \cite{Fedeli2014a,Fedeli2014b}, \cite{Schneider2015} and \cite{Debackere2019}.

\subsection{Cross correlations}
\label{sec:cross_correlations}

In the following we will omit time dependence from function arguments to make the notation less cluttered. Consider two 3D cosmological fields, $u(\mathbf{r})$ and $v(\mathbf{r})$. The halo model cross spectrum between the two fields at a fixed redshift can be written as the sum of a two- and a one-halo term, respectively
\begin{equation}
P_{2\mathrm{H},uv}(k)=P_{\mathrm{lin}}(k)
\prod_{i=u,v}\left[\int_0^\infty b(M)W_i(M,k)n(M)\;\mathrm{d}M\right]\ ,
\label{eq:two_halo}
\end{equation}
\begin{equation}
P_{1\mathrm{H},uv}(k)=\int_0^\infty W_u(M,k)W_v(M,k)n(M)\;\mathrm{d}M\ ,
\label{eq:one_halo}
\end{equation}
where $P_{\mathrm{lin}}(k)$ is the linear power spectrum of matter fluctuations, $M$ is the halo mass and $b(M)$ is the linear halo bias with respect to the linear matter density (a dimensionless quantity):
\begin{equation}
\delta_\mathrm{h}(M, k\to0) = b(M)\delta_\mathrm{m}(k\to0)\ .
\label{eq:halo_bias}
\end{equation}
$n(M)$ is the halo mass function: the distribution function for the comoving number density of haloes in a mass range (sometimes denoted $\mathrm{d}n/\mathrm{d}M$ in the literature; units always per mass interval per volume). Equations~(\ref{eq:two_halo}) and (\ref{eq:one_halo}) contain (spherical) Fourier transforms of the halo profiles of the fields $u(\mathbf{r})$ and $v(\mathbf{r})$ that are being cross correlated:
\begin{equation}
W_u(M,k)=\int_0^\infty4\pi r^2\frac{\sin(kr)}{kr}\theta_u(M,r)\;\mathrm{d}r\ ,
\label{eq:window_function}
\end{equation}
where $\theta_u(M,r)$ is the averaged radial profile for the field $u(\mathbf{r})$ in a host halo of mass $M$. For example, if one were interested in the electron-pressure field then $\theta_{u}$ would be the pressure profile of free electrons, if one were interested in the matter power spectrum, then $\theta_{u}$ would be the halo matter overdensity profile. To use these equations to compute the matter--matter power spectrum then one would set $\theta_{u}=\theta_{v}=\rho_\mathrm{m}(M,r)/\bar\rho$ where $\rho_\mathrm{m}$ is the halo matter density profile and $\bar\rho$ is the mean matter density. The division by $\bar\rho$ ensures that the end result is the overdensity spectrum and we note that $W_\mathrm{m}(M,k\to0)=M/\bar\rho$. We work using comoving $k$, so $r$ in the previous equations is also comoving, as is $\bar\rho$, which is therefore constant. The units of the halo profile $\theta_u$ are the units of the field $u$. The units of the halo-Fourier-transform functions $W_u$ (equation~\ref{eq:window_function}) are those of the field $u(\mathbf{r})$ multiplied by volume. We will interchange between $\Delta^2(k)$ and $P(k)$ power spectrum definitions:
\begin{equation}
\Delta^2_{uv}(k)=4\pi\left(\frac{k}{2\pi}\right)^3P_{uv}(k)\ ,
\label{eq:power_definitions}
\end{equation}
where $\Delta^2_{uv}(k)$ has units of the product of fields $u$ and $v$, while $P_{uv}(k)$ has this and an additional unit of volume. 

The dimensionless mass function, $g(\nu)$, normalised such that the integral over all $\nu$ gives unity, is related to $n(M)$ via
\begin{equation}
g(\nu)\;\mathrm{d}\nu=\frac{M}{\bar\rho}n(M)\;\mathrm{d}M\ ,
\label{eq:mass_function}
\end{equation}
and is written in terms of the mass variable $\nu=\dc/\sigma(M)$: $\dc$ is the critical linear density threshold for halo collapse: $\dc\simeq 1.686$, which has a weak cosmology dependence. $\sigma(M)$ is the variance in the linear matter field when filtered on a Lagrangian scale $R$ corresponding to mass $M=4\pi R^3/3$:
\begin{equation}
\sigma^2(R)=\int_0^{\infty}\Delta_\mathrm{lin}^2(k)\left[\frac{3}{(kR)^3}(\sin{kR}-kR\cos{kR})\right]^2\;\mathrm{d}\ln{k}\ ,
\label{eq:variance_density}
\end{equation}
where the expression in square brackets is the normalised Fourier transform of a real-space top-hat filter.

Note that the adopted halo mass function and bias \emph{must} satisfy the following properties for matter power spectra to have the correct large-scale limit\footnote{Achieving these limits is difficult numerically because of the large amount of mass contained in low-mass haloes according to most popular mass functions. Special care must be taken with the two-halo integral in the case of power spectra that involve the matter field. See Appendix \ref{app:two_halo}.}:
\begin{equation}
\int_0^\infty Mn(M)\;\mathrm{d}M=\bar\rho\ ,
\label{eq:mf_normalisation}
\end{equation}
\begin{equation}
\int_0^\infty b(M) M n(M)\;\mathrm{d}M=\bar\rho\ .
\label{eq:bias_normalisation}
\end{equation}
In words: these equations enforce that all matter be in a halo of some mass and that, on average, matter is unbiased with respect to itself. 

It is worth examining the approximations that lead to the halo-model equations~(\ref{eq:two_halo}) and (\ref{eq:one_halo}): It has been assumed that haloes trace the underlying linear matter distribution with a linear halo bias, that halo profiles are perfectly spherical with no substructure and that there is no scatter in profile properties at fixed host halo mass. There is also nothing in the modelling to prevent haloes from overlapping. These approximations will break down, and the errors in the eventual power spectrum that they contribute will vary with the fields that are being considered. Unfortunately remedies for these assumptions, where they exist, make the halo model calculation increasingly complex \citep[\eg][]{Smith2007,Giocoli2010,Valageas2011,Smith2011b,vandenBosch2013} and are beyond the scope of the work presented here. 

\subsection{Halo composition and profiles}
\label{sec:halo_composition}

In this paper we consider each halo of total mass $M$ to be made of separate CDM, gas and stellar components with (possibly time-dependent) mass fraction $f_i(M)$ in component $i$, such that these sum to unity
\begin{equation}
f_\mathrm{c}(M)+f_\mathrm{g}(M)+f_\mathrm{*}(M) = f_\mathrm{m}(M) = 1\ .
\label{eq:halo_composition}
\end{equation}
The gas is further separated into that bound to the halo and that ejected from the halo by some feedback process, 
\begin{equation}
f_\mathrm{g}(M) = f_\mathrm{bnd}(M)+f_\mathrm{ejc}(M) \ .
\label{eq:gas_composition}
\end{equation}
We use the terms `bound' and `ejected' to differentiate gas that is within the halo virial radius from that outside it. In our formalism it is not necessary that the ejected gas component to be gravitationally bound to the host halo. Indeed, it could be that the ejected gas is far outside the virial radius. All that is required is that the gas component was `associated' with the halo, in that it was part of the initial overdensity from which the halo formed. This means that $M$ should not literally be interpreted as the halo mass, because some mass that was initially associated with the halo may be located outside the virial radius. $M$ is therefore a label, and we take this label to coincide with the measured halo mass in gravity-only simulations, thus justifying our use of a mass function and halo bias calibrated on such simulations. When haloes that are identified in hydrodynamic simulations are compared to non-hydrodynamic simulations that are run with the same initial conditions the halo masses differ object-to-object. However, there is almost always an object-to-object correspondence \citep[\eg][]{vanDaalen2014}. Differing halo masses in hydrodynamical simulations arise predominantly because matter has been moved across the (somewhat-arbitrary) halo boundary. 

We also separate the stellar mass into that in central and satellite galaxies,
\begin{equation}
f_\mathrm{*}(M) = f_\mathrm{cen}(M)+f_\mathrm{sat}(M) \ .
\label{eq:star_composition}
\end{equation}
where any black hole mass is included in the stellar component.

Each component of the halo is given a spherical density profile, $\rho_i(M, r)$, that describes how the component is distributed. These are normalised such that
\begin{equation}
f_i(M) M = \int_0^{r_\mathrm{v}} 4\pi r^2 \rho_i(M,r)\;\mathrm{d}r\ .
\label{eq:profile_normalisation}
\end{equation}
Note that we can write $\rho_\mathrm{m}=\rho_\mathrm{c}+\rho_\mathrm{g}+\rho_\mathrm{*}$ and a similar equation for $W_\mathrm{m}$.

In this paper we also compute halo-model spectra under the assumption that all matter is CDM, mainly for comparisons with gravity-only simulations. In this case we ignore the gas and stellar components of the halo and set the CDM fraction to unity.

\subsection{Large-scale limit for matter fields}
\label{sec:k_to_zero}

It is instructive to consider the $k\to0$ limit of both the two- and the one-halo terms in equations~(\ref{eq:two_halo}) and (\ref{eq:one_halo}) for the specific case of matter fields. Let us illustrate this for the example of the cross spectrum of field $u(\mathbf{r})$ with field $v(\mathbf{r})$, in this case $\theta_u(M,r)=\rho_u(M,r)/\bar\rho$ and equations~(\ref{eq:window_function}) and (\ref{eq:profile_normalisation}) give $W_u(M,k\to0)=f_u(M)M/\bar\rho$, therefore:
\begin{equation}
\frac{P_{2\mathrm{H},uv}(k\to0)}{P_\mathrm{lin}(k\to0)}=
\prod_{i=u,v}\left[\frac{1}{\bar\rho}\int_0^\infty b(M)f_i(M)Mn(M)\;\mathrm{d}M\right]\ ,
\label{eq:two_halo_kzero}
\end{equation}
\begin{equation}
P_{1\mathrm{H},uv}(k\to0)=\frac{1}{\bar\rho^2}\int_0^\infty f_u(M) f_v(M) M^2 n(M) \;\mathrm{d}M\ .
\label{eq:one_halo_kzero}
\end{equation}
If we adopt the notation
\begin{equation}
\average{x}=\frac{1}{\bar\rho}\int_0^\infty x(M) M n(M)\;\mathrm{d}M\ ,
\label{eq:halo_average}
\end{equation}
for an average of property $x$ over halo mass then equation~(\ref{eq:two_halo_kzero}) becomes
\begin{equation}
\frac{P_{2\mathrm{H},uv}(k\to0)}{P_\mathrm{lin}(k\to0)}=\average{b f_u}\average{b f_v}\, .
\label{eq:two_halo_kzero_eval}
\end{equation}
We see that the amplitude of the two-halo term is governed by the product of the mean-halo-bias-weighted abundances for fields $u(\mathbf{r})$ and $v(\mathbf{r})$. If we consider the auto-spectrum of matter overdensity, then $u=v=\mathrm{m}$, $f_\mathrm{m}(M)=1$ and equation~(\ref{eq:two_halo_kzero_eval}) tells us that the two-halo term is the linear power spectrum because $\average{b}=1$ (equation~\ref{eq:bias_normalisation}). If $f_u(M)$ does not depend on halo mass then the average would reduce to $\Omega_u/\Om$, where $\Omega_u$ is the cosmological density parameter for species $u$. More generally $f_u(M)$ will depend on halo mass, and the two-halo term on large scales is then the linear power spectrum weighted by the field abundance multiplied by the halo bias of the haloes in which the field is to be found.

If instead we focus on the one-halo term, equation~(\ref{eq:one_halo_kzero}) becomes
\begin{equation}
P_{1\mathrm{H},uv}(k\to0)=\frac{\average{M f_u f_v}}{\bar\rho}\ ,
\label{eq:one_halo_kzero_eval}
\end{equation}
and we see that the amplitude of the one-halo term is governed by the halo-mass-weighted product of the abundances. For the matter--matter case, $u=v=\mathrm{m}$ then $f_\mathrm{m}=1$ and the amplitude of the one-halo term is governed by $\average{M}$, which is the mean halo mass that a unit of matter is to be found in (\emph{not} the mean halo mass). For our fiducial cosmology at $z=0$ this mass is $\sim 10^{13.5}\Msun$, which gives some indication of the typical halo mass responsible for power in the one-halo term of the matter--matter spectrum.


\section{Specific ingredients}
\label{sec:ingredients}

The method presented in Section~\ref{sec:halomodel} has been general, but here we list the specific ingredients used in this paper. We are interested in both the lensing auto correlation function and the cross correlation between \tsz and lensing. This means that we will compute the 3D matter--matter and matter--electron pressure power spectra. We are also interested in testing how our model for the different components of a halo compares to those components measured in hydrodynamical simulations, so we shall also compute auto- and cross-spectra of CDM, gas and stars. In this section we take default values for hydrodynamical parameters, but we later free and fit these in Section~\ref{sec:improved}.

\subsection{Halo demographics}

In our calculations we adopt the mass function ($g(\nu)$ from equation~\ref{eq:mass_function}) of \cite{Sheth1999}
\begin{equation}
g(\nu)\;\mathrm{d}\nu=A\left[1+\frac{1}{(q\nu^2)^p}\right]\mathrm{e}^{-q\nu^2/2}\;\mathrm{d}\nu\ ,
\label{eq:st_mf}
\end{equation}
with $p=0.3$, $q=0.707$ and $A\simeq0.216$. Note that this function has no explicit redshift dependence. We use the halo bias derived from equation~(\ref{eq:st_mf}) using the peak-background split formalism \citep{Mo1996,Sheth2001}
\begin{equation}
b(\nu)=1-\frac{1}{\dc}\left[1+\nu\frac{\mathrm{d}}{\mathrm{d}\nu}\ln g(\nu)\right]\ ,
\label{eq:peak_background_split}
\end{equation}
which then automatically fulfils the mean-bias condition in equation~(\ref{eq:bias_normalisation}). Explicitly for the mass function in equation~(\ref{eq:st_mf}) the peak-background split gives
\begin{equation}
b(\nu)=1+\frac{1}{\dc}\left[q\nu^2-1+\frac{2p}{1+(q\nu^2)^p}\right]\ .
\label{eq:st_bias}
\end{equation}

The \cite{Sheth1999} relation was fitted to haloes that were identified in \nbody simulations with a cosmology-dependent virial overdensity criterion taken from the spherical-collapse model: spherical-overdense regions that are a factor of $\Dv$ times denser that the background \emph{matter} density. We take $\Dv(z)$ from the \LCDM fitting function of \cite{Bryan1998}
\begin{equation}
\Dv(z)=\frac{1}{\Om(z)}\left\{18\pi^2-82[1-\Om(z)]-39[1-\Om(z)]^2\right\}\ .
\label{eq:Deltav_Bryan}
\end{equation}
$\Dv$ is known as the virial-collapse density\footnote{The non-standard factor of $\Om(z)$ in the denominator of equation~(\ref{eq:Deltav_Bryan}) arises because we work with respect to the matter density, rather than critical density.}. The cosmology dependence of the linear-collapse density, $\dc$, calculated according to the spherical-collapse model, was also taken into account in the mass function of \cite{Sheth1999}. Even though this is a small numerical change it has a larger impact upon the mass function than one might expect due to the fact that it is exponentiated \citep{Courtin2011,Mead2017}. We use the \LCDM fitting formula from \cite{Nakamura1997}:
\begin{equation}
\dc(z)=\frac{3}{20}(12\pi)^{2/3}\left\{1+0.0123\log_{10}\Om(z)\right\}\ .
\label{eq:deltac_Nakamura}
\end{equation}
Note that it is important to use halo structural parameters that are consistent with the way haloes are defined in the mass function. We take haloes defined with the virial definition in equation~(\ref{eq:Deltav_Bryan}) and any conversions between halo mass $M$ and $\nu$ use the cosmology-dependent $\dc$ from equation~(\ref{eq:deltac_Nakamura}).

\subsection{Halo composition}

We consider haloes to be made from CDM, gas, and stars. We set the halo CDM fraction to be the constant universal fraction
\begin{equation}
f_\mathrm{c}(M)=\frac{\Omega_\mathrm{c}}{\Omega_\mathrm{m}}\ ,
\label{eq:CDM_fraction}
\end{equation}
which we consider appropriate because, while feedback may indirectly redistribute CDM within a halo, it will not be able to eject CDM.

Gas is split into bound and ejected gas, with the fraction of bound gas in haloes taken from \cite{Schneider2015}:
\begin{equation}
f_\mathrm{bnd}(M)=\frac{\Omega_\mathrm{b}}{\Omega_\mathrm{m}}\frac{(M/M_0)^\beta}{1+(M/M_0)^\beta}\ .
\label{eq:bound_fraction}
\end{equation}
This function ensures that the bound gas fraction transitions from the universal baryon fraction in high-mass haloes ($M\gg M_0$) to zero in lower mass systems, with $\beta$ governing the rate of transition; haloes of mass $M_0$ having lost half of their initial baryon content. By default we take $M_0=10^{14}\Msun$ and $\beta=0.6$ as in \cite{Schneider2015} as this in in rough accordance with observational results \citep[\eg][]{Sun2009,Vikhlinin2009,Gonzalez2013}. The ejected gas fraction is then fixed such that it accounts for all remaining baryons that were originally associated with the halo assuming that the perturbation was adiabatic
\begin{equation}
f_\mathrm{ejc}(M)=\frac{\Omega_\mathrm{b}}{\Omega_\mathrm{m}}-f_\mathrm{bnd}(M)-f_\mathrm{*}(M)\ .
\label{eq:ejected_fraction}
\end{equation}
With these definitions, the requirement that the total halo mass is accounted for (equation~\ref{eq:halo_composition}) is automatically satisfied\footnote{In the unusual situation that the halo mass fraction in bound gas plus that in stars would be greater than the universal baryon fraction we subtract the excess mass from the stars. This only happens for very high halo masses or unusual sets of parameters.}.

We follow \cite{Fedeli2014a} and set the stellar fraction to be
\begin{equation}
f_\mathrm{*}(M)=A_*\exp\left[-\frac{\log^2_{10}(M/M_*)}{2\sigma_*^2}\right]\ ,
\label{eq:star_fraction}
\end{equation}
a form that expresses the fact that star formation efficiency peaks in haloes of mass $M_*$ with halo stellar mass fraction $A_*$, while being suppressed for higher and lower halo masses with a logarithmic width of $\sigma_*$. By default we take $A_*=0.03$, $\sigma_*=1.2$ and $M_*=10^{12.5}\Msun$, values motivated by \cite*{Moster2013} and \cite{Kravtsov2018}. We also impose a limit for the high-mass end of $f_*$: $f_*(M>M_*)\geq A_*/3$, suggested by observational data on the high-halo-mass saturation of the relation between stellar mass and halo mass \citep[\eg][]{Leauthaud2012}. The stellar halo mass fraction is further split into central and satellite stars, with the stellar content of low-mass haloes assumed to be dominated by a single central galaxy while the stellar content of high-mass haloes is dominated by satellites.  For $M<M_*$ we take
\begin{equation}
f_\mathrm{cen}(M) = f_*(M)\ ,
\quad
f_\mathrm{sat}(M) = 0\ ,
\label{eq:star_fractions_low_mass}
\end{equation}
while for $M>M_*$ haloes we have
\begin{equation}
f_\mathrm{cen}(M) = f_*(M)\left(\frac{M}{M_*}\right)^\eta\ ,
\label{eq:central_star_fraction_high_mass}
\end{equation}
\begin{equation}
f_\mathrm{sat}(M) = f_*(M)\left[1-\left(\frac{M}{M_*}\right)^\eta\right]\ ,
\label{eq:satellite_star_fraction_high_mass}
\end{equation}
with a default of $\eta = -0.3$, taken from \cite{Moster2013}, such that the stellar mass content of high-mass haloes is dominated by satellites while that of low-mass haloes is dominated by centrals.

\begin{figure}
\begin{center}
\includegraphics[height=9cm,angle=270]{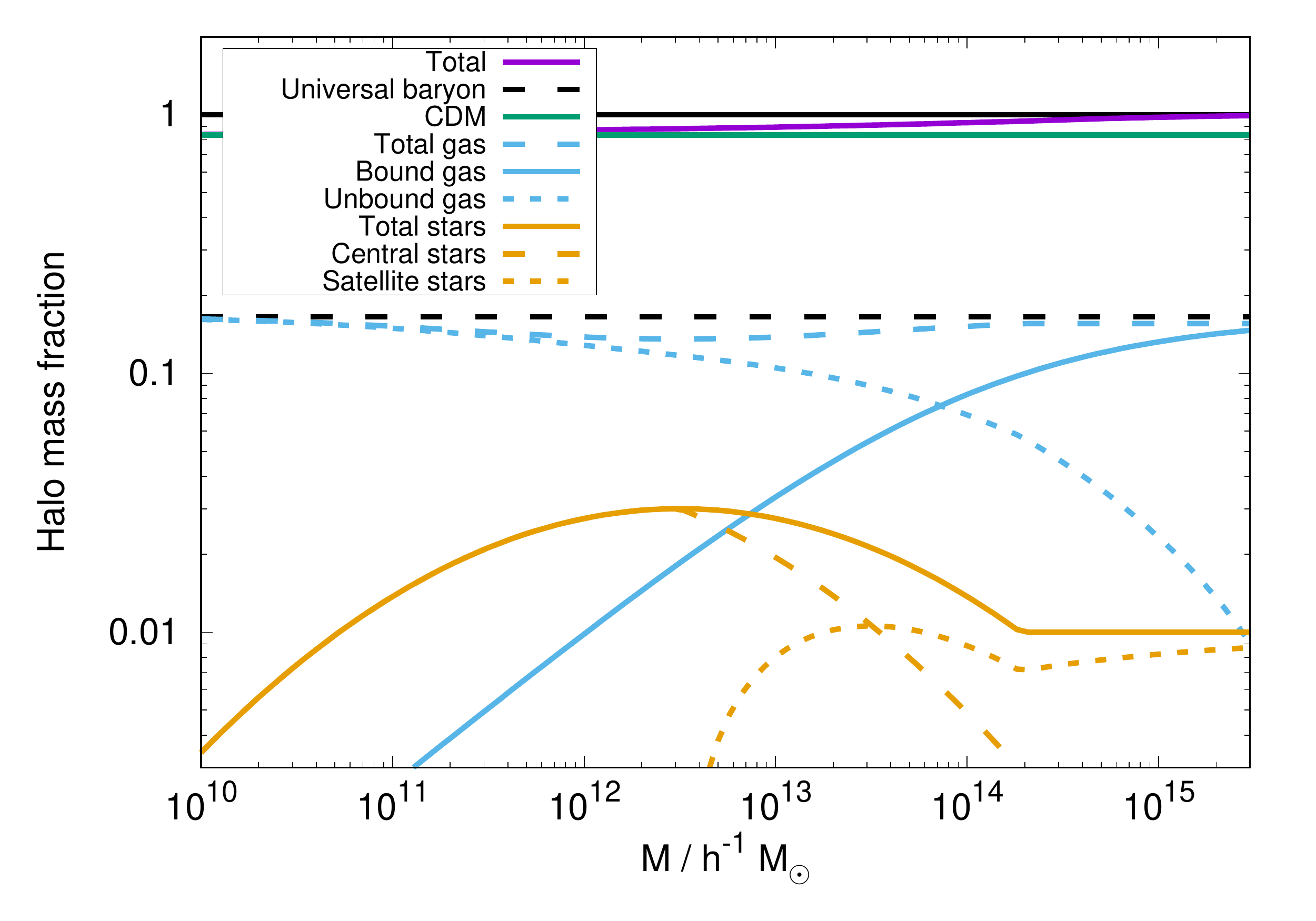}
\end{center}
\caption{The halo mass fractions $f_u(M)$ as a function of halo mass for the different components in our model at $z=0$: The halo CDM fraction is constant at $\Oc/\Om$ for all halo masses. Star formation efficiency peaks around $M_*=10^{12.5}\Msun$ with a mass fraction $A_*=0.03$, it saturates at high mass at a value of $f_* =A_*/3$. Stars are split into central and satellite, with the halo stellar content being dominated by centrals at low masses and by satellites at high mass. The bound gas fraction is near to the universal baryon fraction for high halo masses, but decreases below half the universal fraction for halo masses lower than $M_0=10^{14}\Msun$, the remaining baryons that are neither bound gas or stars are considered to be in unbound gas that is not in the halo. The total mass fraction drops from unity at lower masses due to the ejected gas.}
\label{fig:halo_mass_fractions}
\end{figure}

In Fig.~\ref{fig:halo_mass_fractions} we show the halo mass fractions of the different components as a function of halo mass: The CDM fraction is constant across halo mass. The bound gas fraction is close to the universal baryon fraction for high mass haloes but depletes below half the universal content for $M_0 < 10^{14}\Msun$. Low-mass haloes have more baryon content in stars than in gas. The stellar fraction peaks at $M_* = 10^{12.5}\Msun$ with a stellar mass fraction $A_* = 0.03$. Stellar mass in satellite galaxies dominates higher-mass haloes, while stellar mass in central galaxies dominates the lower-mass haloes.

\subsection{Halo profiles}

In gravity-only simulations it has been known since \citeauthor*{Navarro1997} (NFW; \citeyear{Navarro1997}) that gravitational collapse causes dark matter to arrange itself into approximately spherical structures following the `NFW' profile
\begin{equation}
\rho_\mathrm{c}(M,r)\propto\frac{1}{r/r_\mathrm{s}(1+r/r_\mathrm{s})^2}\ .
\label{eq:CDM_profile}
\end{equation}
For our halo-model calculation, the profile is truncated at the virial radius, defined such that this encloses an average density of $\Dv$ times the mean background density:
\begin{equation}
 M=\frac{4\pi}{3} r_\mathrm{v}^3\Dv\bar\rho\ ,
 \end{equation}
where $\Dv$, the virial-collapse density, is taken from equation~(\ref{eq:Deltav_Bryan}). As we always work in comoving units we take $\bar\rho$ to be the (constant) comoving matter density and $r_\mathrm{v}$ is therefore also comoving. The remaining halo scale radius parameter, $r_\mathrm{s}$ is then usually specified via the concentration $c=r_\mathrm{v}/r_\mathrm{s}$ and we use the concentration-mass relation of \cite{Duffy2008} appropriate for the virial definition of halo radius
\begin{equation}
c(M)=7.85\left(\frac{M}{\sform{2}{12}\Msun}\right)^{-0.081}(1+z)^{-0.71}\ .
\label{eq:Duffy_concentration}
\end{equation}
It has been shown \citep[\eg][]{Rudd2008,Velliscig2014,Mummery2017} that feedback physics changes the dark-matter halo profiles seen in hydrodynamic simulations when compared to profiles in gravity-only simulations. However, the profiles are still well fit by the NFW form, but have slightly altered concentration parameters. In a similar way to \cite{Mead2015b}, we introduce a change in concentration parameter caused by the ejection of gas in a way that modifies the standard concentration multiplicatively, 
\begin{equation}
c(M)\to c(M)\left[1+\epsilon_1 +(\epsilon_2-\epsilon_1)\frac{f_\mathrm{bnd}(M)} {\Omega_\mathrm{b}/\Omega_\mathrm{m}}\right]\ .
\label{eq:concentration_modification}
\end{equation}
The concentrations of (low-mass) haloes that have lost all of their gas are multiplied by $(1+\epsilon_1)$, while those of (high-mass) haloes that have not lost any gas are multiplied by $(1+\epsilon_2)$. We take $\epsilon_1=\epsilon_2=0$ as default, which corresponds to no modification. It may be possible to implement an analytical recipe for how gas changes the dark matter profile, such as that presented in \cite{White2004b} or \cite{Schneider2015} but we find that this is not necessary for our study.


Gas that is gravitationally bound to a halo is described by the \citeauthor{Komatsu2001} (KS; \citeyear{Komatsu2001}) density profile, which we take as
\begin{equation}
\rho_\mathrm{bnd}(M,r)\propto \left[\frac{\ln(1+r/r_\mathrm{s})}{r/r_\mathrm{s}}\right]^{1/(\Gamma-1)}\ ,
\label{eq:bound_profile}
\end{equation}
with $r_\mathrm{s}$ being identical to that in equation~(\ref{eq:CDM_profile}). $\Gamma$ is the polytropic index for the gas, which we take to be $\Gamma=1.17$ by default. Decreasing $\Gamma$ makes the gas profile more centrally concentrated. Equation~(\ref{eq:bound_profile}) is taken from \cite{Martizzi2013}, and is a slightly simplified version of the original KS profile (see \citealt{Komatsu2002}), which we use because it is more convenient to numerically Fourier transform. \cite{Rabold2017} have shown that the KS model provides an extremely good description of haloes that have not undergone violent feedback and \cite{Yan2020} have shown that it also provides a good model for \bahamas haloes if $\Gamma$ is allowed to vary.
 
We treat the ejected gas in a manner similar to that in \cite{Fedeli2014a} and \cite{Debackere2019}, which in turn is similar to the way that warm dark matter is treated in \cite{Smith2011b}: we make the approximation that the ejected gas does not contribute to the one-halo term, but only to the two-halo term\footnote{Numerically this is achieved by setting the ejected gas profile to zero in the one-halo term, and to a delta function in the two-halo term.}. This gives an additive contribution to the two-halo term which is exactly the linear power spectrum multiplied by $\average{bf_\mathrm{ejc}}$ per gas field, where
\begin{equation}
\average{bf_\mathrm{ejc}}=\frac{1}{\bar\rho}\int_0^\infty b(M) f_\mathrm{ejc}(M) M n(M)\;\mathrm{d}M\ .
\label{eq:smooth_gas_correction}
\end{equation}
This is equivalent to taking the ejected gas to be a biased tracer of the linear matter density field. It remains to be determined if this is a reasonable assumption, but it does have the virtue of ensuring that we have accounted for all of the mass and thus preserving the physical relations in equations~(\ref{eq:mf_normalisation}) and (\ref{eq:bias_normalisation}).

We take the stellar density profile for the central galaxies to be a delta function located at the halo centre
\begin{equation}
\rho_\mathrm{cen}(M,r)\propto \delta_\mathrm{D}(r)\ ,
\label{eq:star_profile}
\end{equation}
while we assume that satellite galaxies $\rho_\mathrm{sat}(M,r)$ follow the NFW profile, given in equation~(\ref{eq:CDM_profile}).



To determine the constant of proportionality for the density profile of each component (equations~\ref{eq:CDM_profile}, \ref{eq:bound_profile} and \ref{eq:star_profile}) we use equation~(\ref{eq:profile_normalisation}): assuring that the component makes up the correct mass fraction of a given halo.

To compute any spectra involving the electron pressure we also need to know the temperature of ionised electrons in haloes. We assume all gas to be ionised, and for the bound gas we can use the KS profile to determine the gas temperature:
\begin{equation}
T_\mathrm{g}(M,r)= T_\mathrm{v}(M)\frac{\ln(1+r/r_\mathrm{s})}{r/r_\mathrm{s}}\ ,
\label{eq:bound_temperature}
\end{equation}
which amounts to assuming hydrostatic equilibrium. We compute the central temperature $T_\mathrm{v}(M)$ as the halo virial temperature for the gas:
\begin{equation}
\frac{3}{2}\kb T_\mathrm{v}(M)=\alpha \frac{GMm_\mathrm{p}\mu_\mathrm{p}}{a r_\mathrm{v}}\ ,
\label{eq:virial_temperature}
\end{equation}
where $m_\mathrm{p}$ is the proton mass, $\mu_\mathrm{p}$ is the mean gas particle mass divided by the proton mass, and the factor of $a$ converts the comoving $r_\mathrm{v}$ to a physical radius. The total halo electron pressure $P_\mathrm{e}$ is then given by the ideal gas law
\begin{equation}
P_\mathrm{e}(M,r)=
\frac{\rho_\mathrm{bnd}(M,r)}{m_\mathrm{p}\mu_\mathrm{e}}
k_\mathrm{B}T_\mathrm{g}(M,r)\ ,
\label{eq:electron_pressure}
\end{equation}
where $\rho_\mathrm{bnd}$ is the halo bound gas density, $m_\mathrm{p}$ is the proton mass and $\mu_\mathrm{e}$ is mean gas particle mass per electron divided by the proton mass\footnote{If the gas were comprised only of ionised hydrogen and helium, with a hydrogen mass fraction $f_\mathrm{H}$, we would have $\mu_\mathrm{p}=4/(3+5f_\mathrm{H})$ and $\mu_\mathrm{e}=2/(1+f_\mathrm{H})$. For the \bahamas with $f_\mathrm{H}=0.752$, $\mu_\mathrm{e}\simeq 0.59$ and $\mu_\mathrm{e}\simeq 1.14$. However, we adopt the values $\mu_\mathrm{p}=0.61$ and $\mu_\mathrm{e}=1.17$ to be compatible with gas metallicity in the simulations and the way the electron pressure field was measured in post processesing.}. The first term in this equation is exactly the electron number density. $\alpha$ in equation~(\ref{eq:virial_temperature}) is a parameter that encapsulates deviations from a simple virial relation, \eg unvirialized or unthermalised gas or turbulence, we take the standard $\alpha=1$ by default. In our calculations we have neglected relativistic corrections to the effective pressure that contributes to the tSZ spectrum \citep{Itoh1998,Nozawa2000}, these corrections may be important to consider in the future \citep{Remazeilles2018}.

The ejected gas is treated as if it is distributed with the background linear perturbations, so that it contributes only to the two-halo term. It is given a constant temperature, $T_\mathrm{w}$, which is supposed to be the temperature of the warm-hot intergalactic medium (WHIM; \citealt{Cen1999}), but we caution the reader against taking this name too literally. We take $T_\mathrm{w}=10^{6.5}\Kelvin$ as the default temperature of the WHIM \citep{VanWaerbeke2014}.

\subsection{Large-scale limit for electron pressure}
\label{sec:k_to_zero_pressure}

It is informative to calculate the amplitude scaling of the two- and one-halo terms for the case of the electron pressure, to compare to matter. If we take the $k\to0$ limit for the halo Fourier transforms (equation~\ref{eq:window_function}) of matter we get $W_\mathrm{m}(M,k\to0)\propto M$, but for the electron pressure that originates from bound halo gas this changes to
\begin{equation}
W_\mathrm{p}(M,k\to0) \propto M^{5/3}\ .
\end{equation}
The extra $M^{2/3}$ for electron pressure comes from the fact that pressure is the product of gas density, which is $\propto M$, and gas temperature (equation~\ref{eq:virial_temperature}), which is $\propto M^{2/3}$. Using the notation from Section~\ref{sec:k_to_zero} we can write the large-scale limit of the two-halo term as
\begin{equation}
\frac{P_\mathrm{2H,mp}(k\to0)}{P_\mathrm{lin}(k\to0)}=
A\average{b f_\mathrm{bnd} M^{2/3}} +
BT_\mathrm{w}\average{b f_\mathrm{ejc}}\ ,
\label{eq:two_halo_kzero_mp}
\end{equation}
where the second term originates from the pressure of non-halo WHIM gas ($A$ and $B$ are constants). For the one-halo term we have
\begin{equation}
P_\mathrm{1H,mp}(k\to0)\propto\average{f_\mathrm{bnd} M^{5/3}}\ .
\label{eq:one_halo_kzero_mp}
\end{equation}
The extra $M^{2/3}$ in these equations compared to those for the matter--matter power spectrum cause the matter--electron pressure cross spectrum to be dominated by relatively more massive haloes. A corollary of this is that the transition between the two- and one-halo term occurs at larger scales for the matter--electron pressure spectrum compared to the matter--matter spectrum.

\subsection{Example power spectra}

\begin{figure}
\begin{center}
\includegraphics[height=8.5cm, angle=270]{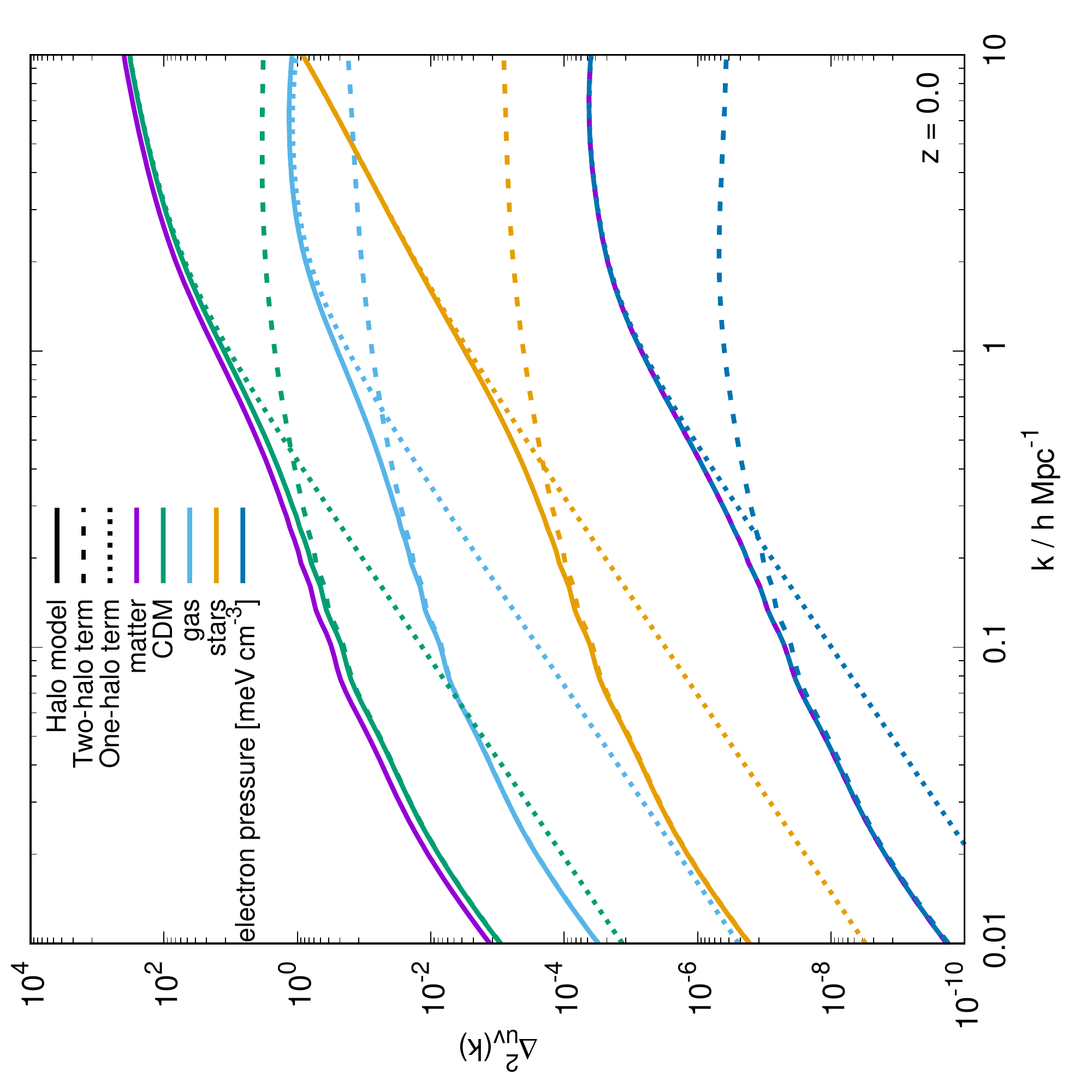}
\end{center}
\caption{The $z=0$ halo-model predictions for the matter--matter (top), CDM--CDM, gas--gas, stars--stars and the matter--electron pressure (bottom) power spectra. We also show the two-halo (long-dashed) and the one-halo (short-dashed) terms for each spectrum apart from matter--matter. At large scales, the shapes of the two-halo terms are all identical, and are that of the linear power spectrum, at small scales the one-halo terms dominates and have significantly different shapes for each spectrum. Note that the transition scale between the two- and one-halo term is at much larger scales for the matter--electron pressure cross spectrum, which is a consequence of this spectrum being dominated by contributions from more massive haloes compared to those that contribute to the matter spectra. The down-turn at small scales in the gas--gas and matter--electron pressure spectra is partly due to the underrepresentation of low-mass haloes and partly due to their relatively smooth halo profiles.}
\label{fig:power_components}
\end{figure}

In Fig.~\ref{fig:power_components} we show the halo-model predictions the spectra of matter--matter, CDM--CDM, gas--gas, stars--stars and matter--electron pressure. We see that at large scales the shape of all curves are identical, a consequence of all components following the linear perturbation distribution (equation~\ref{eq:two_halo_kzero}). For the CDM--CDM spectrum, the ratio at large scales compared to the matter spectrum is exactly $(\Oc/\Om)^2$. For gas--gas and stars--stars it is roughly $(\Omega_\mathrm{g}/\Om)^2$ and $(\Omega_\mathrm{*}/\Om)^2$, but this is not exact because the bias of the haloes in which each field lives also contributes to the two-halo term amplitude (see equation~\ref{eq:two_halo_kzero_eval}). At smaller scales we see that the shapes of the one-halo terms are all quite different, which arises from the different halo profiles of each component. Note that for $k\simeq10\iMpc$ the stellar density distribution eventually has more power than the gas, despite the lower abundance: a consequence of the gas being smoothly distributed on small scales while stars are tightly clustered in halo centres. The CDM--CDM spectrum is very similar to the total matter spectrum, which makes sense since CDM dominates the matter budget; however, there are shape differences at smaller scales where the different scale-dependence of contributions of gas--gas and stars--stars are important. The shape of the matter--electron pressure spectrum looks superficially like that of the gas--gas spectrum, but with the two- to one-halo transition taking place at a different scale, due to the different scalings of the two- and one-halo terms. Note that this spectrum cannot be directly compared with the others since it has units of pressure whereas the others are dimensionless.

\begin{figure*}
\begin{center}
\includegraphics[height=18cm, angle=270]{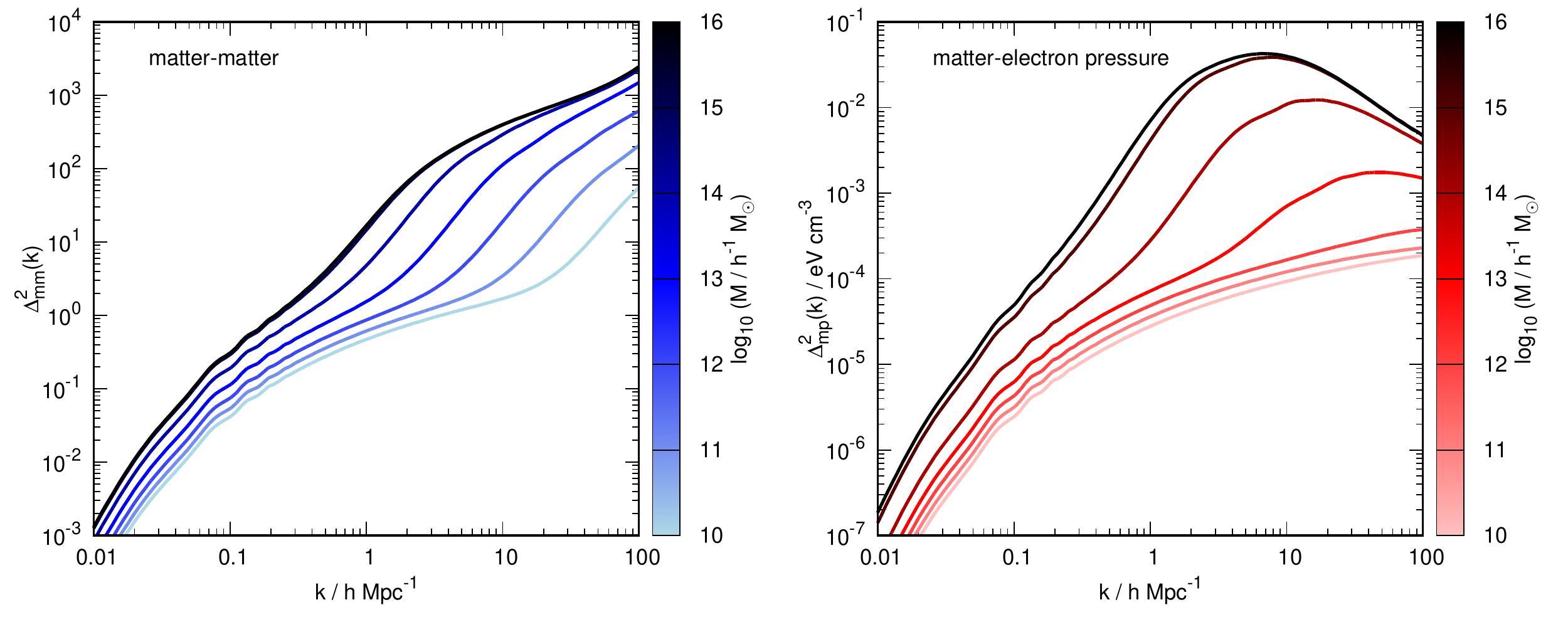}
\end{center}
\caption{Cumulative halo model power spectra as a function of the upper-limit of halo mass at $z=0$. The left-hand panel shows the matter--matter power spectrum as the upper limit is raised from $10^{10}$ to $10^{16}\Msun$ (\ie the lightest coloured curve shows the contribution from only haloes below $10^{10}\Msun$), the right-hand panel shows the same but for the matter--electron pressure spectrum. In both cases the converged spectrum from all halo masses is shown in thick black and this has been reached with an upper limit of $10^{16}\Msun$. We see that the matter--matter spectrum builds up fairly gradually with halo mass and that all scales receive contributions from a wide range of halo masses. In contrast, the matter--electron pressure spectrum receives very little contribution from haloes less massive than $10^{13}\Msun$, with the vast majority of the power coming from haloes between $10^{14}$ and $10^{15}\Msun$.}
\label{fig:power_mass_contribution}
\end{figure*}

\begin{figure*}
\begin{center}
\includegraphics[height=18cm, angle=270]{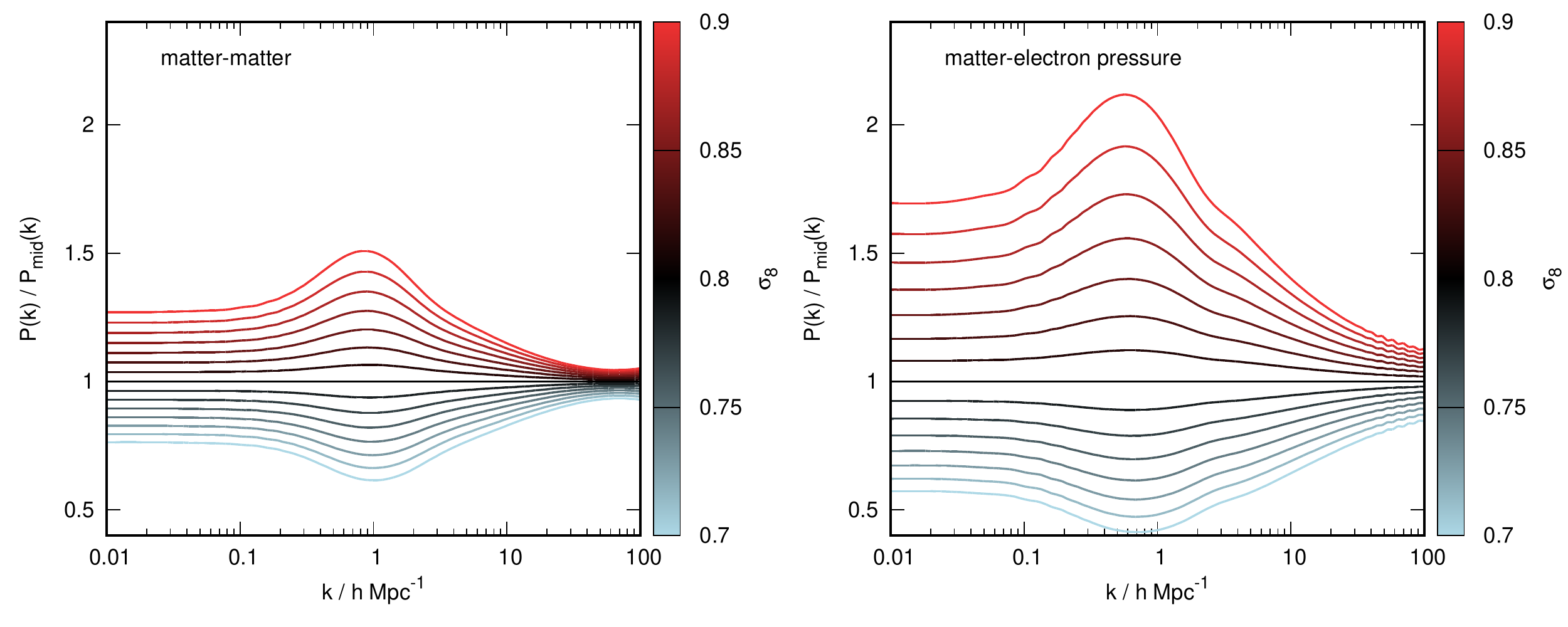}
\end{center}
\caption{The effect of $\sigma_8$ variations on the matter--matter power spectrum (left) and on the matter--electron pressure power spectrum (right) at $z=0$. We show halo model power spectra for $\sigma_8 = 0.7$ (blue) to $\sigma_8 = 0.9$ (red) divided by a central $\sigma_8 = 0.8$ model. For linear theory the plot would simply be horizontal lines. We see that boosting the amplitude of fluctuations increases the amplitude of the matter--electron power spectrum far more than it does for the matter-matter. At a wavenumber of $\simeq 1\iMpc$ the matter-matter power spectrum scales like $\sigma_8^{3.5}$ while matter--electron pressure scales like $\sigma_8^{5.8}$. This is because the electron pressure field is dominated by high-mass haloes from the tail of the mass function that are very sensitive to the linear power spectrum amplitude.}
\label{fig:power_cosmology}
\end{figure*}

In Fig.~\ref{fig:power_mass_contribution} we show how the $z=0$ matter--matter spectrum and the matter--electron pressure spectrum are built up as a function of halo mass. We show cumulative power spectra computed as we vary the upper limit of halo mass in the integrals in equations~(\ref{eq:two_halo}) and (\ref{eq:one_halo}) from $10^{10}$ to $10^{16}\Msun$. In both cases the integration has converged with an upper limit of $10^{16}\Msun$. The matter--matter spectrum builds up more gradually as a function of halo mass compared to the matter--electron pressure spectrum, the latter receiving the majority of the power from haloes between $10^{14}$ and $10^{15}\Msun$ and very little power from haloes below $10^{13}\Msun$. This, combined with the relatively smooth pressure profiles, contributes to the existence of a peak in the model cross spectrum at $k\sim 7\iMpc$, although we caution the reader that this peak may not be physical since we do not model halo substructure. By extension, this implies that the matter--electron pressure cross spectrum is sensitive to higher halo masses than the standard matter--matter spectrum \cite[\eg][]{Makiya2018}. The reason for this is two fold: First, equation~(\ref{eq:bound_fraction}) tells us that low-mass haloes are deficient in gas and therefore contribute less free-electron number density to the signal. Second, equations~(\ref{eq:two_halo_kzero_mp}) and (\ref{eq:one_halo_kzero_mp}) tell us that the overall amplitude of both the two- and the one-halo terms is more sensitive to high-mass haloes. This also ensures that the non-linear scaling of these spectra with $\sigma_8$ is more extreme, as can be seen in Fig.~\ref{fig:power_cosmology}, since increasing the power spectrum amplitude a small amount has a comparatively large impact on the high-mass tail of the halo-mass function. We find that roughly $P_\mathrm{mm}\propto \sigma_8^{3.5}$, $P_\mathrm{mp}\propto \sigma_8^{5.8}$ and $P_\mathrm{pp}\propto \sigma_8^8$ with the exact exponent depending on wavenumber. This fact has been noticed in the past \cite[\eg][]{Komatsu2002,Hill2013b,Hill2014} and motivates the use of the tSZ auto-spectrum \citep{Komatsu2002,McCarthy2014}, or the cross correlation of tSZ with lensing \citep{Hill2014,Ma2015,Hojjati2017}, as a sensitive probe of the power-spectrum amplitude. This statement is equivalent to noting that the high-mass tail of the halo mass function is a sensitive probe of the power spectrum amplitude.

\section{Hydrodynamic simulations}
\label{sec:simulations}

In this paper we compare auto- and cross-spectra from our halo model to those measured from the \bahamas simulations \citep{McCarthy2017}. These are a set of smooth-particle hydrodynamical simulations with separate dark matter and gas particles. The dark matter interacts only gravitationally but the gas also experiences a hydrodynamic pressure force. In addition, processes like star formation, gas heating and cooling, supernovae explosions and AGN formation and evolution are also modelled using `sub-grid' recipes. A full discussion of the sub-grid implementation can be found in \cite{Schaye2010}, \cite{LeBrun2014}, and \cite{McCarthy2017}. Hydrodynamic gas particles have temperature and density as properties, as well as the standard position and velocity, and these can be used to calculate an electron pressure per particle (Appendix \ref{app:power_measurement}). Star (macro) particles are also created from the gas as the simulation evolves, and these keep track of star formation and death. Black holes are also considered, which can swallow gas particles and grow, although they contribute little to the overall mass.

\begin{figure*}
\begin{center}
\includegraphics[width=18.5cm]{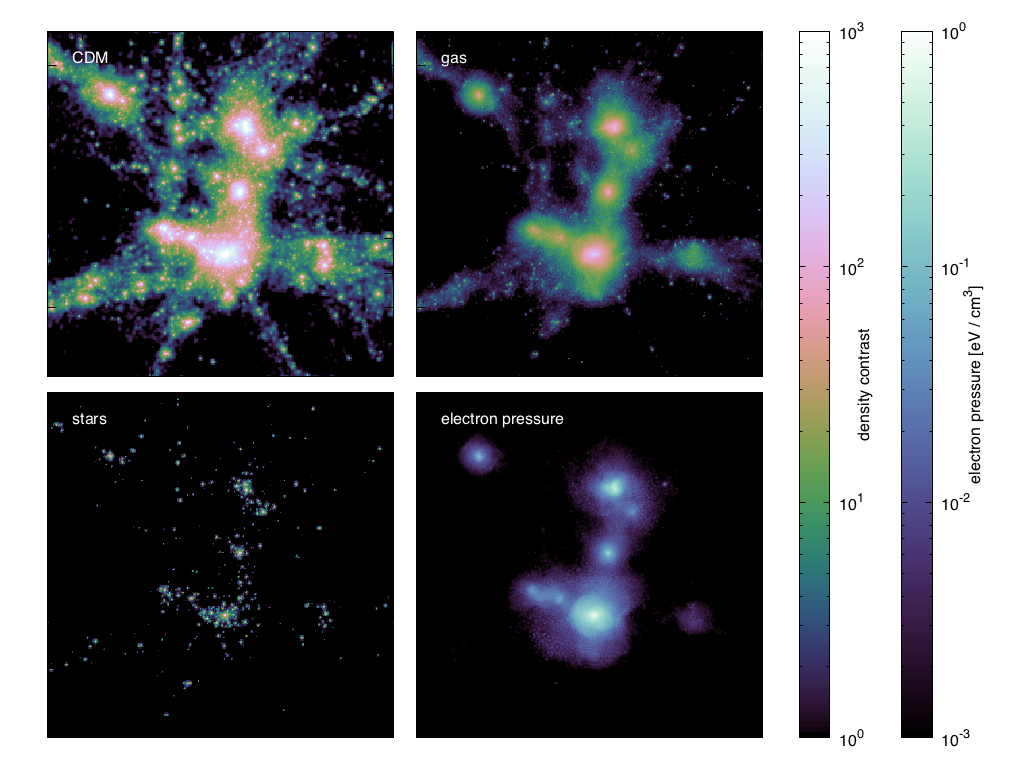}
\end{center}
\caption{The different fields measured from the \bahamas \agn simulation at $z=0$ around a $M\sim 10^{15}\Msun$ halo averaged through a square slab of side $20\Mpc$ and thickness $5\Mpc$. We show CDM (top left), gas (top right) and star (bottom left) density contrast as well as electron pressure (bottom right). Colour intensity increases with field value logarithmically and we show a dynamic range of $10^{3}$ for both density contrast and electron pressure. We see that the gas is less clumped than the CDM and that low-mass sub haloes show a large gas depletion compared to the host. Stars are more tightly clustered in the halo centre than the CDM, but are absent in many of the substructures. The total matter density-contrast field would be a sum of the CDM, gas and star fields. The electron pressure broadly follows the gas but can only be seen emanating from the highest gas-density peaks over the dynamic range shown.}
\label{fig:bahamas_fields}
\end{figure*}

The \bahamas simulation boxes are $400\Mpc$ cubes with periodic boundary conditions. Each simulation is of $1024^3$ dark matter and $1024^3$ gas particles with cosmological parameters inspired by the \wmap9 \citep{Hinshaw2013} data analysis. Initial conditions are created at $z=127$ using second-order Lagrangian perturbation theory via the code \sgenic\footnote{\sgenic: \sgeniclink} \citep{Bird2020} with appropriate, different transfer functions used for dark matter and baryons. We utilise three different hydrodynamic simulations that differ only in the `strength' of their AGN feedback. This strength is determined by a single parameter, the AGN sub-grid heating temperature, which is the temperature increase given to gas particles that are targeted for feedback. The default \agn model, for which we have 3 realisations, was calibrated to reproduce the amplitude of the observed gas--halo mass relation of groups and clusters, as inferred from X-ray observations \citep{McCarthy2018}. The \lo and \hi models lowered and raised the sub-grid heating temperature so as to approximately bracket the scatter in the observed relation. For comparison purposes there is also a non-hydrodynamic `gravity-only' simulation (\dmonly). This simulation still uses two sets of particles for gas and dark matter, and these different particles start with different initial conditions. The only difference compared to the full hydrodynamic simulations is in the subsequent evolution, where only gravity is considered in \dmonly.

In Fig.~\ref{fig:bahamas_fields} we show different fields around a massive halo for the \agn simulation at $z=0$. We see that the CDM forms a skeleton around which the other fields are clustered, with the gas density being less tightly clustered than CDM and stellar density being more tightly clustered. Both gas and stars are missing from the lower-mass haloes. The electron pressure follows the gas density, but is restricted to emanate from only the highest gas-density peaks. This is because high-mass haloes also have a higher temperature compared to low-mass haloes, which boosts their electron-pressure signal relative to the density. Low-mass haloes are therefore severely under represented in the electron-pressure distribution.

\begin{figure*}
\begin{center}
\includegraphics[height=18cm, angle=270]{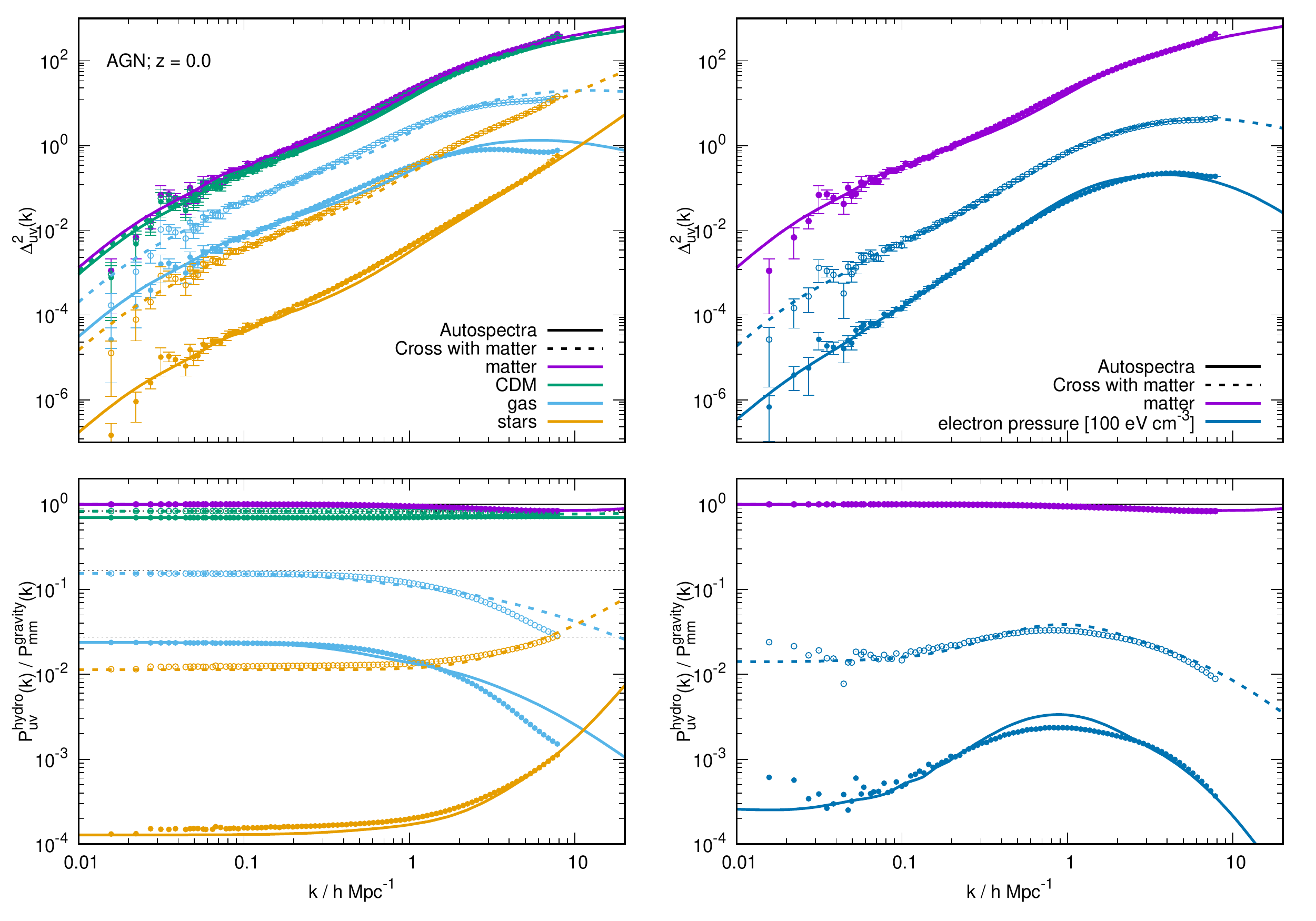}
\end{center}
\caption{A comparison of the default halo-model prediction with power spectra from the \bahamas \agn simulation at $z=0$. Points with errors show the measured power spectra with an error-on-the-mean calculated from the finite number of modes that contribute to each $k$. The upper two panels show spectra for `all matter' (purple), CDM (green), gas (light blue), stars (orange) and electron pressure (dark blue). The electron pressure field has units of $100\mathrm{eV}\,\mathrm{cm}^{-3}$. The effect of power aliasing can be seen as an upturn in power at the highest $k$ ($\sim 7\iMpc$) shown for each spectrum (see Appendix~\ref{app:power_measurement}). Lines show the halo-model predictions using the default model discussed in Section~\ref{sec:ingredients}: solid lines and points show auto-spectra while broken lines and open points show cross-spectra with the total matter field. We see that the trends in the halo model agree reasonably well with the simulations but that there are disagreements in the details. In the lower two panels we show the `response function' calculated with respect to the matter--matter power in the gravity-only model: $P^\mathrm{hydro}_{uv}(k) / P^\mathrm{gravity}_\mathrm{mm}(k)$. The simulations spectra have been divided by those from the \dmonly simulation and the halo model has been divided by a standard halo-model prediction that assumes all matter to be in NFW haloes. The horizontal-dashed-black lines in the lower left panel show the expected large scale $\Oc/\Om$, $(\Oc/\Om)^2$, $\Ob/\Om$, and $(\Ob/\Om)^2$ asymptotes for the CDM and gas cross--matter and auto-spectra. We see that the response functions are generally smoother than the raw power spectra: a consequence of cosmic variance cancellations at large scales and cancellation of aliasing effects at small scales.}
\label{fig:hydro}
\end{figure*}

In this paper, we are interested in two-point statistics that pertain to weak gravitational lensing and to the tSZ effect. We therefore measure the power spectrum of density fluctuations and the electron-pressure field from the simulations. We work with the full 3D data, rather than with 2D projections, as these are more directly tied to the modelling. Details of how we measure these power spectra from the particles and how we consider the effects of shot noise can be found in Appendix~\ref{app:power_measurement}. We compute the auto- and cross-spectra of all combinations of the fields: total matter overdensity $\delta_\mathrm{m}$, CDM overdensity, $\delta_\mathrm{c}$, gas overdensity, $\delta_\mathrm{g}$, stellar overdensity, $\delta_\mathrm{*}$ and electron pressure $P_\mathrm{e}$. In our work, all overdensities are defined relative to the total matter density (\ie $1+\delta_u=\rho_u/\bar\rho$), which ensures
\begin{equation}
1+\delta_\mathrm{m}=(1+\delta_\mathrm{c})+(1+\delta_\mathrm{g})+(1+\delta_\mathrm{*})\ .
\label{eq:fluctuation_conservation}
\end{equation}

The measured power spectra from the \bahamas \agn simulation are shown in Fig.~\ref{fig:hydro} together with the halo-model prediction from the model presented in Section~\ref{sec:ingredients} with the default parameter values. We note that the halo model provides a reasonable model of the data for all spectra shown, but that it is not perfect (the log-log scale can hide some serious defects). At large scales we note that the amplitudes are generally in good agreement, which tells us that we have the overall abundances and halo occupation of each species approximately correct and that we have the mean background electron pressure reasonably well modelled. At smaller scales we note differences that must be due to an incorrect mass function, choice of halo profiles or else due to physical effects that are missing from our simple halo model. For the matter spectra we also note that the standard problem of a power deficit in the transition region between the two- and one-halo terms \citep[\eg][]{Tinker2005, Valageas2011, Mead2015b} is present for all spectra shown. This defect seems to be less of a problem in the spectra involving the electron pressure. With reference to Fig.~\ref{fig:power_components} we can conjecture that this may be because these spectra are more dominated by the one-halo term and because the transition between the two- and one-halo terms takes place at comparatively larger scales. The existence of a peak in both the matter--electron pressure cross spectrum and electron pressure auto spectrum is hinted at, but by no means confirmed, by the simulation measurements.

The lower panels of Fig.~\ref{fig:hydro} show the power spectrum response functions \citep[\eg][]{Mead2017, Cataneo2019}, which we define as the ratio of any power spectrum of any field combination to the matter--matter power spectrum measured in a `gravity-only' model. In the case of the \bahamas simulations we calculate this response with respect to the \dmonly simulation. The response has the virtue that it cancels the Gaussian cosmic variance at large scales, which leads to a smooth large-scale response. The measured response functions for the \bahamas spectra that involve the electron pressure are less smooth at large scales, which is probably because of the comparative dominance of the one-halo term at large scales and therefore that the large-scale noise is likely to be dominated by Poisson fluctuations in massive halo numbers, rather than the Gaussian variance present in the initial conditions. For the halo model, we calculate this response ratio with respect to a halo model calculation assuming all mass to be in NFW haloes ($f_\mathrm{c}=1$). That the response functions are constant at large scales is indicative that the power on these scales is simply the linear matter spectrum multiplied by some weighting (equations~\ref{eq:two_halo_kzero_eval} and \ref{eq:two_halo_kzero_mp}). For the halo model, the response has the virtue of cancelling some of the standard problems such as the lack of power in the transition region \citep[\eg][]{Tinker2005, Mead2015b}, as can be seen in  Fig.~\ref{fig:hydro}, and helps to a lesser extent with the general underestimation of power at smaller scales \citep{Giocoli2010}. Working at the level of the response may also alleviate problems that may arise from not using the most up-to-date ingredients for our halo model. For example, we use the mass function of \cite{Sheth1999} whereas there exist more recent mass functions such as \cite{Tinker2008, Tinker2010} or \cite{Despali2016}. We prefer \cite{Sheth1999} because it was calibrated to a wider range of cosmological parameter space than more modern mass functions.

\section{Fitting halo model parameters}
\label{sec:improved}

The level of agreement between the simulations and our model shown in Fig.~\ref{fig:hydro} in encouraging, but demonstrates that the model is not sufficiently accurate to use to draw robust conclusions from forthcoming survey data. Note that the signal-to-noise is roughly 3:1 in the lensing--tSZ $C(\ell)$ measurements of \citealt{Hojjati2015} and is expected to be 5:1 for forthcoming KiDS measurements (Tr{\"o}ster et al. in prep). It is possible that some of this modelling inaccuracy arises from incorrect choices for ingredients, but we think that a substantial amount must also arise from missing features in our basic halo-model calculation. Therefore, following the logic in \cite{Mead2015b} we improve our model by fitting parameters that pertain to gas physics to the \bahamas simulations at the level of the power spectra. The aim is to find parameters that govern a population of `effective' haloes that describe the power spectra well for the different feedback strengths. We do this at the level of the halo-model `response', discussed in Section~\ref{sec:simulations}. Once an accurate model for the response has been developed, we must then multiply this response by an accurate prescription for the matter--matter power spectrum in the gravity-only scenario. In our case we take the accurate prediction to be from \hmcode \citep{Mead2015b, Mead2016}, but we could have used \halofit \citep{Smith2003, Takahashi2012} or an emulator prediction \citep[\eg \emu;][]{Lawrence2010, Lawrence2017}. Note that our model does not generally require us to fit parameters in this way, its utility is more general, fitting just seems to us to be the most obvious way for us to proceed to construct an accurate model.

We present several different models, the utility of each depending on the eventual use case:
\begin{enumerate}
\item stars
\item matter
\item matter \& electron pressure
\item matter, CDM, gas \& stars
\end{enumerate}
Because the stars are a subdominant component of the total matter budget, we find it necessary to fit parameters that govern the stars separately (1), and then to hold these parameters fixed while we fit models to the matter (2) and to the matter and electron pressure (3). In case (4), we fit a model to all the matter fields simultaneously, refitting parameters that govern the star distribution. The free parameters that we have fitted are listed in Table~\ref{tab:parameter_descriptions} together with descriptions of their physical meaning and their default values. These parameters were chosen by trial-and-error to form minimal set that were able to model the data well, without opening the parameter space too widely.

To fit we use the \cite{Nelder1965} simplex algorithm with a large number of initial starting locations so as to avoid local minima. To determine the redshift dependence and relevant parameters, we initially fitted our model to different redshifts separately and we then used this information to parameterise sensible redshift dependencies for some of the parameters. The parameters are fitted to $z<1$ \bahamas data across redshifts simultaneously with a linear weighting in $z$ and a log weighting in $k$ between $0.015$ and $7\iMpc$. Our model currently takes $\sim 0.5$s per $z$ to evaluate all 15 cross spectra we are interested in. The numerical bottle-neck is the computation of the Fourier transforms of the gas density and electron pressure profiles, which must be done numerically.

\begin{table*}
\begin{center}
\caption{The default hydrodynamical halo model parameters. In the default case these are all independent of halo mass and redshift, but we allow for some redshift and mass dependence in our fitted halo models.}
\begin{tabular}{c c c c}
\hline
Parameter & Default value & Equation & Physical meaning \\
\hline
$\epsilon_1$ & $0$ & \ref{eq:concentration_modification} & Halo concentration modification for gas-poor haloes \\
$\epsilon_2$ & $0$ & \ref{eq:concentration_modification} & Halo concentration modification for gas-rich haloes \\
\hline
$M_0$ & $10^{14}\Msun$ & \ref{eq:bound_fraction} & Halo mass below which haloes have lost more than half of their initial gas content \\
$\beta$ & $0.6$ & \ref{eq:bound_fraction} & Low-mass power-law slope of halo bound gas fraction \\
$\Gamma$ & $1.17$ & \ref{eq:bound_profile} & Polytropic index for the equation of state of gas that is bound in haloes \\
\hline
$A_*$ & $0.03$ & \ref{eq:star_fraction} & Peak fraction of halo mass that is in stars \\
$M_*$ & $10^{12.5}\Msun$ & \ref{eq:star_fraction} & Halo mass of peak star-formation efficiency \\
$\sigma_*$ & $1.2$ & \ref{eq:star_fraction} & Logarithmic width of star-formation efficiency distribution \\
$\eta$ & $-0.3$ & \ref{eq:central_star_fraction_high_mass} & Power-law index for central--satellite galaxy split \\
\hline
$\alpha$ & $1$ & \ref{eq:virial_temperature} & Ratio of halo temperature to that of virial equilibrium \\
$T_\mathrm{w}$ & $10^{6.5}\Kelvin$ & below \ref{eq:electron_pressure} & Temperature of the warm-hot intergalactic medium \\
\hline
\end{tabular}
\label{tab:parameter_descriptions}
\end{center}
\end{table*}

Before presenting our results we caution the reader against taking the \emph{details} of our modelling as a serious physical description of the underlying haloes. We are fitting parameters in a halo model to power spectra taken from simulations; we are not fitting the simulations at the level of individual halo profiles. Therefore, the parameters of our hydrodynamical halo model should be thought of as pertaining to `effective' haloes that, through the apparatus of the halo model, provide an accurate model of the spectra we are interested in. They should not be confused with the exact parameters that govern the actual internal structure of actual haloes. Although they may relate to these, it would be necessary to check explicitly that this was so. We remind the reader of the shortcomings of the halo model approach: There may be features in the power spectra of the fields we are interested in that will never be accurately described by a linearly biased population of spherical, virialized haloes. Some of our parameter fitting may account for some of these deficiencies, although working with the response also helps to ameliorate some of these problems. Recall Fig.~\ref{fig:power_mass_contribution}, where we saw that haloes of different masses are more-or-less important to different power spectra. This implies that we may drastically alter the profiles of haloes that contribute negligibly to a particular power spectrum and leave the spectrum almost unchanged. It would obviously be incorrect to over interpret the physical meaning of fitted parameters in this instance. Also remember that computing a power spectrum via the halo model represents a huge compression of the information that is potentially available from the halo profiles individually, which is smeared out across $k$ by integrations. It is perfectly possible for inaccuracies in the modelling of the halo population to add or cancel in various unintuitive ways during these numerical compressions. \cite{Huffenberger2003} demonstrated that halo-structural parameters derived directly from haloes extracted from a cosmological simulation were not the same as those derived by fitting the structural parameters in a halo-model power spectrum calculation to the measured power spectrum from the same simulation. \cite{Mead2015b} demonstrated that relatively drastic non-physical additions were required in order to improve halo model predictions for the matter--matter power spectrum from the $30$ per cent level to the $5$ per cent level, and that these were particularly important in the transition region. \cite{Tinker2005} and \cite{vandenBosch2013} both demonstrated that problems exist when trying to model galaxy clustering and galaxy--galaxy lensing using the halo model. To our knowledge, it has never been demonstrated that power spectra from simplistic halo-model calculations agree with those measured from simulations, even if the ingredients for the halo model calculation are taken directly from the simulation of interest.

\begin{table*}
\begin{center}
\caption{The best-fitting effective halo model parameters for different combinations of power spectra from the \bahamas simulations. In each case, the halo model is fitted to the three different AGN heating temperatures separately ($10^{7.6}$, $10^{7.8}$ and $10^{8.0}\Kelvin$). We interpolate between parameters as a function of $T_\mathrm{AGN}$ to get a model for intermediate temperatures and that we think is robust to modest extrapolation.}
\begin{tabular}{c c c c c c c}
\hline
Model & Parameter & Equation & Fitted parameter & $10^{7.6}\Kelvin$ & $10^{7.8}\Kelvin$ & $10^{8.0}\Kelvin$ \\
\hline
(1) stars 
& $A_*=A_{*,0}+A_{*,1}z$ & \ref{eq:star_fraction} & $A_{*,0}$ 
& 0.0348 & 0.0330 & 0.0309 \\
& & & $A_{*,1}$ 
& -0.0093 & -0.0088 & -0.0082 \\
& $M_*=M_{*,0}\exp{(M_{*,1}z)}$ & \ref{eq:star_fraction} & $\log_{10}(M_{*,0}/\Msun)$ 
& 12.4620 & 12.4479 & 12.3923 \\
& & & $M_{*,1}$
& -0.3664 & -0.3521 & -0.3073 \\
& $\eta$ & \ref{eq:central_star_fraction_high_mass} & $\eta$ 
& -0.3428 & -0.3556 & -0.3505 \\
\hline
(2) matter 
& $\epsilon_1=\epsilon_{1,0}+\epsilon_{1,1}z$ & \ref{eq:concentration_modification} & $\epsilon_{1,0}$
& 0.2841 & 0.2038 & 0.0526 \\
& & & $\epsilon_{1,1}$ 
& -0.0046 & -0.0047 & 0.0365 \\
& $\Gamma$ & \ref{eq:bound_profile} & $\Gamma$ 
& 1.2363 & 1.3376 & 1.6237 \\
& $M_0$ & \ref{eq:bound_fraction} & $\log_{10}(M_0/\Msun)$ 
& 13.0020 & 13.3658 & 14.0226 \\
\hline
(3) matter, electron pressure 
& $\epsilon_1=\epsilon_{1,0}+\epsilon_{1,1}z$ & \ref{eq:concentration_modification} & $\epsilon_{1,0}$
& -0.1002 & -0.1065 & -0.1253 \\
& & & $\epsilon_{1,1}$
& -0.0456 & -0.1073 & -0.0111 \\
& $\Gamma$ & \ref{eq:bound_profile} & $\Gamma$ 
& 1.1647 & 1.1770 & 1.1966 \\
& $M_0$ & \ref{eq:bound_fraction} & $\log_{10}(M_0/\Msun)$ 
& 13.1949 & 13.5937 & 14.2480 \\
& $\alpha$ & \ref{eq:virial_temperature} & $\alpha$ 
& 0.7642 & 0.8471 & 1.0314 \\
& $T_\mathrm{w} = T_\mathrm{w,0}\exp{(T_\mathrm{w,1}z)}$ & below \ref{eq:electron_pressure} &$\log_{10}(T_\mathrm{w,0}/\Kelvin)$ 
& 6.6762 & 6.6545 & 6.6615 \\
& & & $T_\mathrm{w,1}$ 
& -0.5566 & -0.3652 & -0.0617 \\
\hline
(4) matter, CDM, gas, stars
& $A_*=A_{*,0}+A_{*,1}z$ & \ref{eq:star_fraction} & $A_{*,0}$ 
& 0.0346 & 0.0342 & 0.0321 \\
& & & $A_{*,1}$ 
& -0.0092 & -0.0105 & -0.0094 \\
& $M_*=M_{*,0}\exp{(M_{*,1}z)}$ & \ref{eq:star_fraction} & $\log_{10}(M_{*,0}/\Msun)$ 
& 12.5506 & 12.3715 & 12.3032 \\
& & & $M_{*,1}$ 
& -0.4615 & 0.0149 & -0.0817 \\
& $\eta$ & \ref{eq:central_star_fraction_high_mass} & $\eta$ 
& -0.4970 & -0.4052 & -0.3443 \\
& $\epsilon_1=\epsilon_{1,0}+\epsilon_{1,1}z$ & \ref{eq:concentration_modification} & $\epsilon_{1,0}$
& 0.4021 & 0.1236 & -0.1158 \\
& & & $\epsilon_{1,1}$
& 0.0435 & -0.0187 & 0.1408 \\
& $\Gamma=\Gamma_0+z\Gamma_1$ & \ref{eq:bound_profile} & $\Gamma_0$ 
& 1.2763 & 1.2956 & 1.2861 \\
& & & $\Gamma_1$
& -0.0554 & -0.0937 & -0.1382 \\
& $M_0$ & \ref{eq:bound_fraction} & $\log_{10}(M_0/\Msun)$ 
& 13.0978 & 13.4854 & 14.1254 \\
\hline
\end{tabular}
\label{tab:fitted_parameters}
\end{center}
\end{table*}

\begin{figure*}
\begin{center}
\includegraphics[width=18cm]{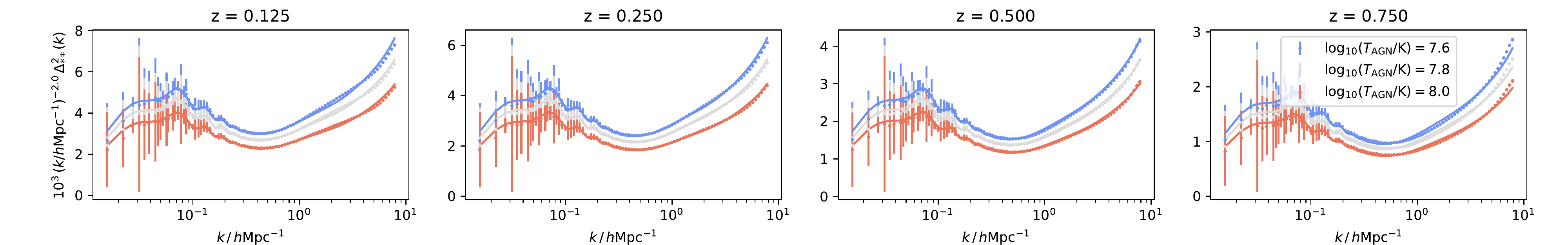}
\end{center}
\caption{Halo model (1) for the star--star power spectrum is shown as sold lines while simulation measurements are shown as points with error bars that come from the scatter in power between three different realisations of the \agn \bahamas simulation. Note that the power spectra shown here is the measured response multiplied by \hmcode, so cosmic variance is removed at small scales, but we still show the error bar for comparison. Different colours denote the three different AGN feedback temperatures: $10^{7.6}$ (blue), $10^{7.8}$ (grey) and $10^{8.0}\Kelvin$ (red). Parameters for the model can be found in Table~\ref{tab:fitted_parameters}. We see that the spectrum is reduced in amplitude as feedback temperature increases, a consequence of AGN feedback suppressing star formation. This is reflected in the modelling as the value of $A_*$ and $M_*$ decreases as AGN strength increases.}
\label{fig:model_stars}
\end{figure*}

\subsection{Stars}

The first model (1) we present is for the star--star power spectrum in Fig.~\ref{fig:model_stars}. The free parameters are $A_*$, $M_*$ and $\eta$ with their best-fitting values listed in Table~\ref{tab:fitted_parameters}. We find that we are able to fit a reasonable model to the different simulations with different AGN feedback strengths (temperatures), and that our model follows the data with an accuracy of $\simeq5$ per cent for the response, nicely demarcating the different feedback scenarios. We note a preference for an inverse correlation between both $A_*$ and $M_*$ and the feedback temperature. This makes physical sense, since increased AGN activity is suppressing star formation. This could result in both an overall suppression in the number of stars in a halo of a given mass and also in pushing the peak star-formation efficiency to lower halo masses. In contrast, the best-fitting $\eta$, which governs the split between central and satellite stellar mass, is very similar for each of the different feedback temperatures. Note that there is a scatter in the efficiency of this suppression of star formation by AGN between simulations from different groups. Indeed a correct stellar mass--halo mass relation is one of the targets that hydro simulations strive for, and in practice this can be quite difficult to achieve. We also looked at additionally fitting the parameter $\sigma_*$, but found that this did not produce a significant improvement in the quality of the fit.

\begin{figure*}
\begin{center}
\includegraphics[width=18cm]{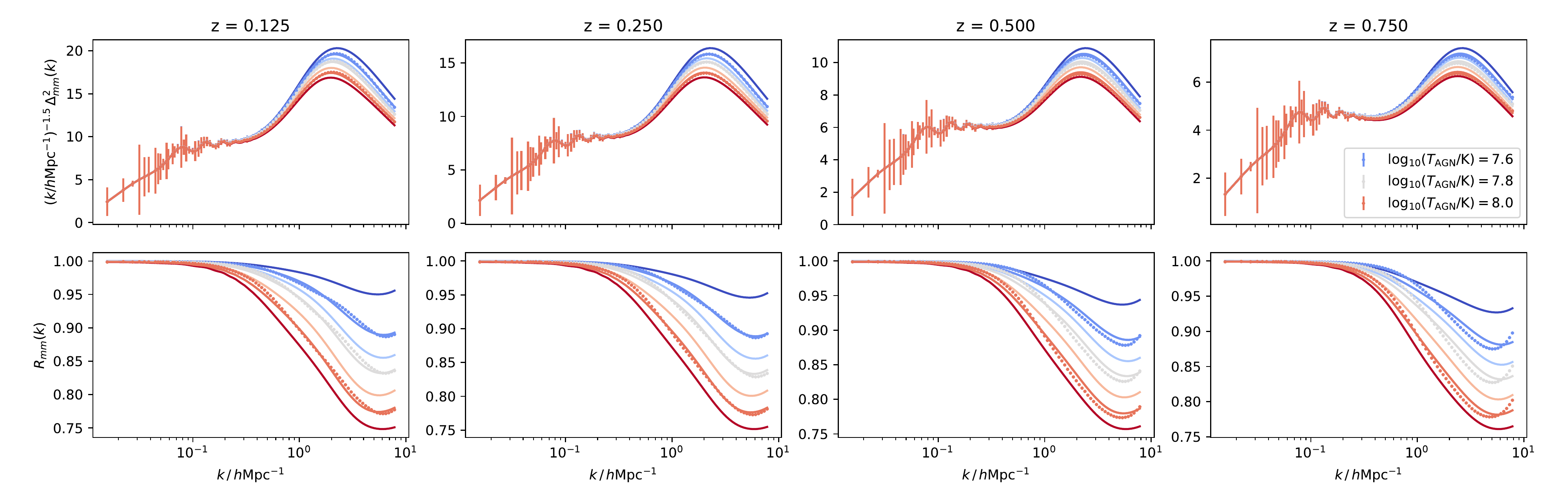}
\end{center}
\caption{Similar to Fig.~\ref{fig:model_stars}, but for the matter--matter spectrum (2). The power spectra are shown in the top row, while the response functions are shown in the bottom row. We see that the model is nearly perfect (errors at the sub-per-cent level) across a wide range of scales and for the different feedback scenarios. We see that the amount of power-spectrum suppression increases with the feedback temperature. The critical parameter that governs this in our modelling is the halo mass $M_0$, which governs the halo mass below which more than half of the cosmic baryon density is missing from the halo. In the fitting parameters we see a strong preference for higher values of $M_0$ for stronger feedback temperatures (from blue to red). How our model fares outside the range of AGN heating temperatures to which it was fitted is also shown via the lines that have no corresponding simulation points. To generate these we have interpolated and extrapolated our model parameters as a function of AGN heating temperature, which is shown in 0.1 dex intervals between $10^{7.5}$ (bluest) and $10^{8.1}\Kelvin$ (reddest), indicating that our model is robust to extrapolation.}
\label{fig:model_matter}
\end{figure*}

\subsection{Matter}

In Fig.~\ref{fig:model_matter} we present model (2), in which we fitted the parameters $\epsilon_1$, $\Gamma$ and $M_0$ to the matter--matter data only. The underlying star model, which contributes to the matter--matter spectrum, is fixed to that shown in Fig.~\ref{fig:model_stars} and discussed in the previous subsection. We are able to match the power-spectrum response seen in the simulations at the per-cent level for each of the feedback models and at all of the redshifts shown. The fitted model has a strong correlation between $M_0$ and AGN strength, which makes physical sense as stronger feedback ejects more halo gas. We also see strong trends in the fitted values of $\epsilon_1$ and $\Gamma$, which may be due to different amounts of back reaction on the dark matter and different heating of the residual halo gas. The higher AGN heating temperatures favours higher $\Gamma$, which corresponds to a less concentrated gas profile, as might be expected from more violent feedback. The upturn in the response function at small scales originates from the stellar contribution to the matter spectrum. In addition, in  Fig.~\ref{fig:model_matter} we also show how the model fares when pushed beyond the range of AGN heating temperatures to which it was fitted. To do this we linearly interpolate and extrapolate the parameters of our halo model as a function of $\log T_\mathrm{AGN}/\Kelvin$. Indeed, this is the reason to use a physically motivated halo model in the first place, as opposed to a simple fitting function or a more blind interpolation between simulation results. This demonstrates some nice physical features of our model that may not be respected by more simplistic fitting functions (such as that in \citealt{Mead2015b}). The suppression in the matter--matter power spectrum in our model originates because gas has been expelled from haloes, with the threshold mass increasing with the AGN heating temperature. Then, to a lesser extent, there are some additional effects from the non-NFW profile of the remaining gas and some back reaction on the concentration of the dark-matter halo component and then some effect from the very centrally concentrated stellar distribution which shows as an upturn at small scales $\sim 10\iMpc$. Because of this, there are some limits to how our model can behave: At one extreme we could raise the AGN heating temperature to a very high value that would result in almost all halo gas being expelled. The maximum suppression that this could have on the power would be to lower the amplitude of the one-halo term by $(\Omega_\mathrm{c}/\Omega_\mathrm{m})^2 \simeq 0.7$. At the other extreme, a low value of the AGN heating temperature would result in almost no gas expulsion and the matter--matter spectrum would be almost unaffected. These limits can be seen to be approached by the extreme temperature values shown in Fig.~\ref{fig:model_matter}. We also looked at additionally fitting the parameters  $\epsilon_2$ and $\beta$, but found that these did not produce a significant improvement in the quality of the fit.

We also investigated the predictions of our model at fixed AGN heating temperature as we vary the underlying cosmology. As shown in \cite{vanDaalen2020}, the impact of AGN feedback on the matter--matter spectrum is quite insensitive to the difference between the \planck and \wmap9 cosmologies. We find a similar insensitivity in our model, which is reassuring, but we do find a small difference in the sense that the matter--matter power suppression in the \planck cosmology is predicted to be slightly less than that in \wmap9 at fixed AGN strength, which was shown by  \cite{vanDaalen2020}. This presumably originates from the slightly different baryon fractions between the two different cosmologies, which means that slightly more halo mass is lost in the \wmap9 case. 

\begin{figure*}
\begin{center}
\includegraphics[width=18cm]{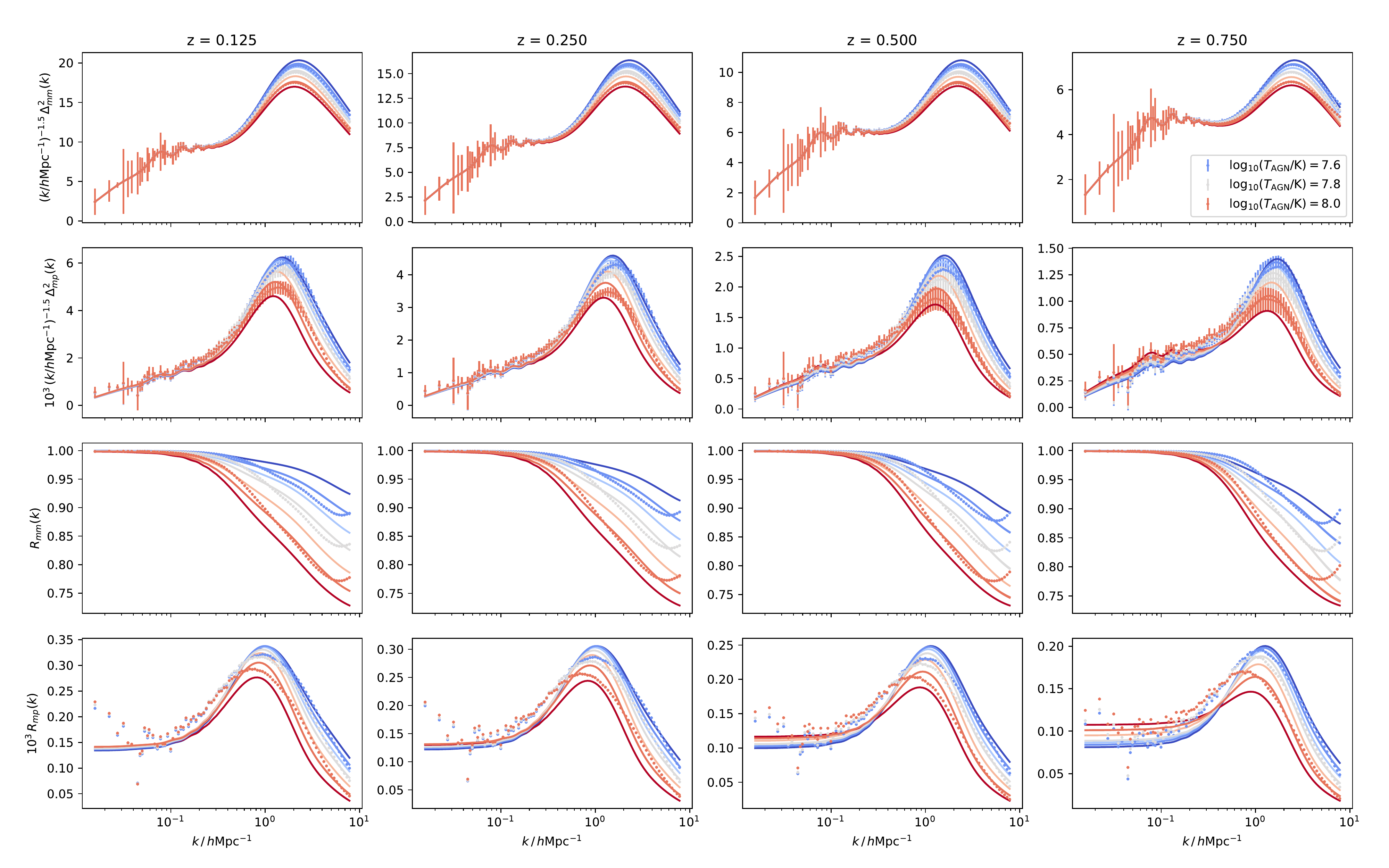}
\end{center}
\caption{As Figs.~\ref{fig:model_stars} and \ref{fig:model_matter}, but for model (3) for the matter--matter (first and third row; power and response) and matter--electron pressure spectra (second and fourth row; power and response). The fit to the matter--matter spectra is slightly degraded compared to the case when we fit the matter--matter spectra exclusively, but we still match the response at the few per cent level while simultaneously we match the matter--electron pressure spectrum well for $k>1\iMpc$. If we compare this to the error bars in the power spectra in the second row we see that our fit is good for scales where the three AGN heating temperature strengths are clearly demarcated in $k$. At larger scales we suspect the larger errors arise because the power is predominantly coming from few massive haloes. Similar to Fig.~\ref{fig:model_matter}, we show how the model fares outside the range of AGN strengths to which is was fitted. AGN heating temperature is shown in 0.1 dex intervals between $10^{7.5}\Kelvin$ and $10^{8.1}\Kelvin$, while the model was fitted between $10^{7.6}\Kelvin$ and $10^{8.0}\Kelvin$, thus demonstrating that our model is robust to extrapolation.}
\label{fig:model_matter_pressure}
\end{figure*}

\subsection{Matter and electron pressure}

In Fig.~\ref{fig:model_matter_pressure} we show the main result of this paper, model (3), which is a result of a joint fit to the matter--matter and matter--electron pressure spectra. This model could be used to predict both shear--shear and shear--$y$ tomographic correlations. We do not consider the electron pressure auto-spectrum because while testing we discovered that this auto-spectrum is particularly difficult to fit; we suspect that this is because it receives much of its contributions from only a few of the most massive haloes present in the \bahamas simulation, and therefore that the spectra measured from the simulations will have very high Poisson noise contributions. Indeed, we see large variations in this power spectrum between the three different realisations of the \bahamas, and this variation dwarfs the differences between the feedback strengths for $k<2\iMpc$. The matter--electron pressure spectrum still suffers from this problem somewhat, as can be inferred from the error bars in the second row of Fig.~\ref{fig:model_matter_pressure}, but the wavenumbers for which the variance between the $400\Mpc$ boxes is greater than the difference between the feedback models are $k<1\iMpc$. We see that we are able to simultaneously match the response functions at the few per cent level for the wavenumbers at which the feedback scenarios are clearly demarcated. We see a trend that increasing AGN heating temperature causes an increase in $\alpha$, the parameter that governs departures from simple virial temperature scaling, suggesting that gas that avoids being ejected is nevertheless heated significantly. However, the general trend is for suppressed small-scale power as AGN strength increases, which in our model arises from the fact that more gas is ejected from the more massive haloes, thus decreasing the pressure overall. In the matter--electron pressure case we note a trend that increasing AGN strength suppresses the power spectrum for scales dominated by the one-halo term; however, particularly at the higher $z$, we see that the power spectrum is relatively enhanced at larger scales. This could be because the heated gas is pushed out to larger scales by the feedback and this in turn gives an excess large-scale pressure contribution, either directly, or via shock heating the inter-galactic medium. In Fig.~\ref{fig:model_matter_pressure} we also show how our model for the matter and electron pressure fares when evaluated for AGN heating temperatures outside the range within which the model was constructed. As in Fig.~\ref{fig:model_matter} we see that the model behaves sensibly outside of the standard temperature range and thus we suggest that it can be used for cosmological and astrophysical parameter inference. We also looked at additionally fitting the parameters  $\epsilon_2$ and $\beta$, but found that these did not produce a significant improvement in the quality of the fit.

\begin{figure*}
\begin{center}
\includegraphics[width=18cm]{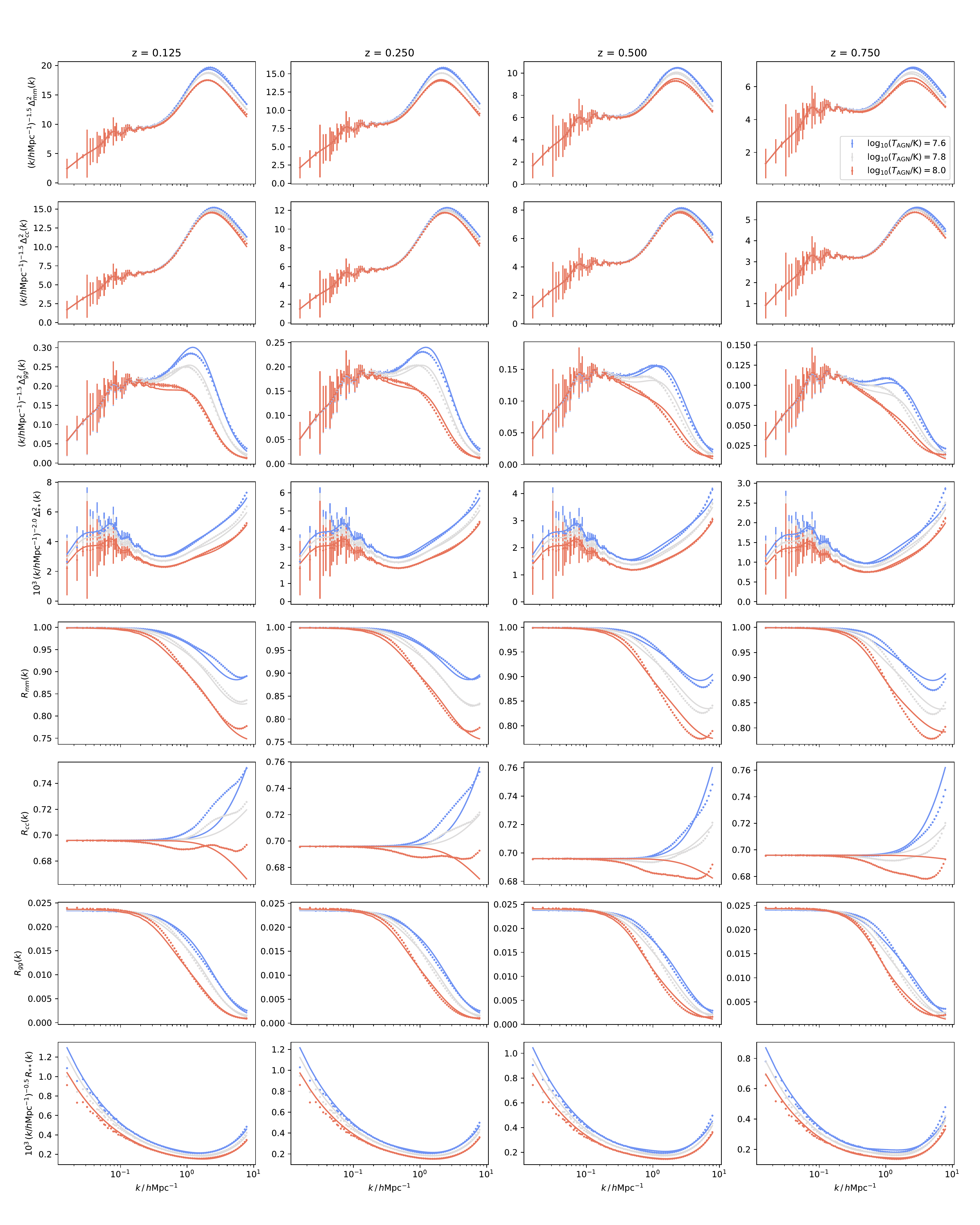}
\end{center}
\caption{As Figs.~\ref{fig:model_stars}, \ref{fig:model_matter} and \ref{fig:model_matter_pressure}, but for the best-fitting model (4) to all 10 possible power spectra of matter, CDM, gas and star fields, although we only show the 4 auto-spectra here. The top half of the figure show power spectra while the bottom half show response functions. Columns show different redshifts. Different colours are different feedback temperatures: $T=10^{7.6}$ (blue), $T=10^{7.8}$ (grey) and $T=10^{8.0}\Kelvin$ (red). Points (with errors) are from \bahamas while lines are from our halo model.}
\label{fig:model_all_matter}
\end{figure*}

\subsection{Matter, CDM, gas and stars}

\begin{figure*}
\begin{center}
\includegraphics[width=18cm]{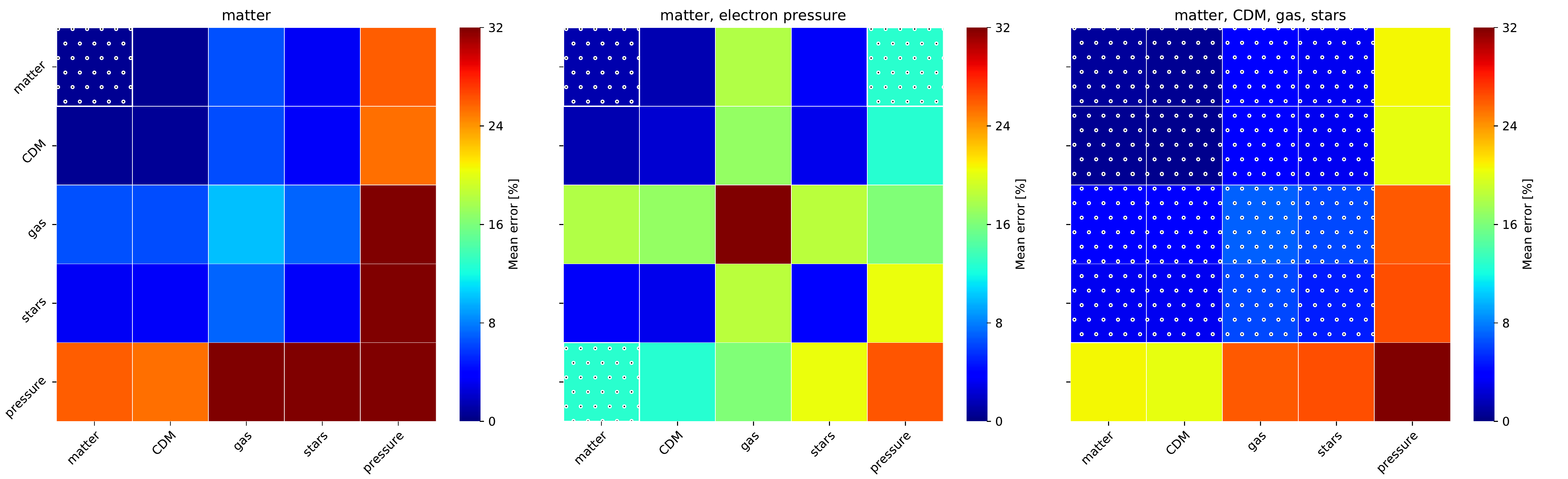}
\end{center}
\caption{The full error matrix for our fitted models calculated for each field pair for the $10^{7.8}\Kelvin$ \agn simulation. The error is averaged linearly over redshift between $z=0$ and $1$ and over $k$ logarithmically between $\simeq 0.015$ and $7\iMpc$. This matrix is symmetric because power spectra are symmetric when swapping the field labels. We show the matrix for the three models presented in this work: matter (2, left); matter, electron pressure (3, centre); matter, CDM, gas, stars (4, right). The squares that contain dots are the particular cross-spectra that are fitted in the model. In the left-hand plot, we see that in the matter model we get a reasonable fit to the power spectra all of the constituent fields. In the central plot we get a better match to to the matter--electron pressure spectrum than in the left-hand panel, but this is at the expense of any power spectra that directly involves the gas field, which shows some tension in our model between the gas and pressure modelling. In the right-hand panel, where we fit matter, CDM, gas and stars, we see that all the matter spectra are reasonable, and better than in the left-hand panel, but achieving this is at the expense of any spectra that involves the pressure field.}
\label{fig:error_matrix}
\end{figure*}

In Fig.~\ref{fig:model_all_matter} we show a model that is the result of a joint fit to all possible combinations of matter, CDM, gas and star spectra. There are 10 distinct power spectra for these 4 fields. In our halo model, the matter field is the sum of the CDM, gas and stars, and so naturally gets upweighted in our fitting when it is included alongside its constituent parts. We see that our model gets reasonable matches to the auto-spectra shown in Fig.~\ref{fig:model_all_matter} and we have also checked that the cross-spectra are well matched (see Fig.~\ref{fig:error_matrix}), with the error always being roughly (but not exactly) the mean of the errors of the auto-spectra of the two contributing fields. We see that the gas spectrum is particularly well matched, which gives us confidence in our choice of the KS profile for the gas density profile. The match to the CDM--CDM response looks less good in Fig.~\ref{fig:model_all_matter}, but note that the deviations from this being a constant are very small, of order a few per cent. Physically this implies that the CDM is not deformed very significantly when going from a gravity only to a hydrodynamic model. Indeed, the suppression in the matter--matter power spectrum is driven predominantly by the (lack of) contribution from the gas, and then the upturn at smaller scales mainly derives from the contribution of the stars (see Fig.~\ref{fig:hydro}, as well as \citealt{vanDaalen2020}). We also looked at additionally fitting the parameters $\epsilon_2$, $\beta$ and $\sigma_*$, but found that this did not produce a significant improvement in the quality of the fit.

\subsection{Model accuracy}

In Fig.~\ref{fig:error_matrix} we show the full error matrix for the \agn simulation for each of models (2), (3) and (4) but calculated for every possible power spectra of the 5 fields we have available. The matrix is $5\times5$ but only $15$ of the $25$ elements are independent due to symmetry. Each cell shows the relative difference for the model cross spectrum compared to the simulation, averaged linearly over $z$ between $0$ and $1$ and logarithmically over $k$ between $\simeq 0.015$ and $7\iMpc$. Error matrices for \lo and \hi are not shown, but are similar. We see that when we fitted only the matter--matter spectrum (2) we still have a reasonable match to the CDM, gas and stars spectra and combinations thereof. When we go from (2) to (3) and include the matter--electron pressure spectrum we improve the match to the matter--electron pressure, but at the cost of power spectra that involve the gas. This demonstrates a possible problem in our modelling in how we simultaneously model the gas density and pressure: When we go from case (2) to (4) and try to match matter, CDM, gas and stars power spectra, but ignore electron pressure we match the gas spectrum well, but this success comes at the expense of the pressure, which is ignored in this fit.

\section{Discussion}
\label{sec:summary}

\subsection{Summary}

We have presented a hydrodynamical halo model that can be used to make analytical calculations of power spectra for combinations of the matter, CDM, gas and star density fields, as well as for cross-spectra of these fields with the electron-pressure field. This model will be used in the future to provide constraints on both cosmology and feedback physics using a combination of lensing and tSZ data. The model has a number of tuneable parameters that pertain to gas physics. We provided four possible sets of values for these parameters that provide good fits to power spectra for: (1) stars, (2) matter, (3) matter and electron pressure and (4) matter, CDM, gas and stars. Parameters of these models are all fitted to power spectra measured from the \bahamas simulations for three different AGN feedback strengths, which in this case is parameterized by a single sub-grid parameter: the AGN heating temperature. By interpolating between model parameters as a function of AGN heating temperature we suggest that our model can be used in a cosmological analysis and that it is robust to extrapolation. In particular, model (2) will be useful in lensing-only studies as it is able to reproduce the deficit in matter--matter power seen in hydrodynamic simulations, compared to the \dmonly case, at the per-cent level. Model (3) provides a slightly worse match to the mater--matter power (accurate to $2$ per cent), but captures the temperature variations in the matter--electron pressure spectrum that can be directly integrated to calculate the lensing--tSZ cross correlation (accurate to $15$ per cent) as well as the lensing--lensing correlation. This accuracy is sufficient for forthcoming lensing and tSZ data sets (Tr{\"o}ster et al. in prep).

\subsection{Identified problems}

\subsubsection{Electron pressure errors}

Why is our model so much less accurate for matter--electron pressure (15 per cent) compared to matter--matter (2 per cent)? We suspect that this is partly because we report the error averaged over $\log k$ from $0.015$ to $7\iMpc$ and linearly over $z$ from $0$ to $1$. The matter--electron pressure response is noisier at large scales compared to the matter spectra, and this introduces an unavoidable source of error for our necessarily smooth halo model prediction. We may also be seeing the fact that the relatively small box size of \bahamas means that the response (or power) has not converged for $k<1\iMpc$, and it is possible that our model would provide a better fit to data from a significantly larger simulation. 

\subsubsection{Response approach}

In this paper, we fitted to the power spectrum response, which we created by dividing a measured hydrodynamical power spectrum by a gravity-only spectrum. This response cancels the Gaussian error in power spectra at large scales for the matter spectra, but it may create complicated correlations at smaller scales. In particular, this response may not be appropriate for the electron-pressure spectra, even at large scales, since the large-scale variance does not mimic that of the matter spectra. We fitted parameters of our models to the response using a least-squares approach, with no $k$ or $z$ dependent error. In the future we hope to have access to more and larger simulations that we would need in order investigate this more thoroughly and so that we may derive an appropriate error bar.

\subsubsection{Electron pressure auto spectrum}

While completing this work we tried to generate a model for the electron pressure auto-spectrum, together with matter--matter and matter--electron pressure; but this proved to be difficult and was abandoned. We discovered that the realisation-to-realisation variance in the electron pressure auto-spectrum measured from \bahamas was huge, and dwarfed the difference between feedback strengths for $k<2\iMpc$, which left us with very little signal to fit. This in turn suggests that boxes of much greater volume than the \bahamas $(400\Mpc)^3$ must be used to ascertain the feedback dependence of this spectrum on large scales. Therefore, our model should not be used to compute tSZ--tSZ power spectrum at this stage. However, with larger hydrodynamical simulations being run in the future, our approach should still be able to be used to match the power spectrum by refitting parameters. We note that it would also be useful to have more accurate calibration data for the simulations to match, as this is one of the limiting factors for \bahamas at the moment.

\subsubsection{Joint model}

We also tried to generate a joint model for matter, CDM, gas, stars and electron pressure, which has not been presented in this paper because it did not work well. We found that it was particularly difficult to simultaneously model spectra that involved the electron pressure and those that involve the gas density with the same underlying physical model. In the current modelling, these two fields are joined in two different ways: first via $M_0$, which determines the overall gas abundance as a function of halo mass, and second via $\Gamma$, the polytropic index of the KS gas profile, which jointly determines the density and pressure profiles. The KS profile is the correct model for undisturbed gas that is in hydrostatic equilibrium in an NFW halo, but this state will clearly be violated in the presence of AGN feedback. It was shown in \cite{Yan2020} that the KS model could provide a reasonable match to gas-density profiles in the \bahamas simulations, but it is likely that the polytropic link between density and pressure is not respected in the presence of feedback \citep{Battaglia2012a,Battaglia2012b}. It is probable that a better model could be generated by either changing the gas profile, for example by explicitly including heated gas, or by weakening the links between the density and pressure modelling. However, note that some link is necessary if one wants to simultaneously constrain feedback and cosmology with cross-correlation data. It is also possible that some of the inaccuracy stems from our treatment of non-thermal pressure within clusters. We have a single multiplicative parameter, $\alpha$, which increases or decreases gas temperature independent of halo mass. In future it may be useful to investigate physical models for non-thermal pressure in clusters, such as those presented in \cite{Nelson2014} or \cite{ShiX2015}.

\subsubsection{Non-linear halo bias}

The halo model provides a good description of the power spectra for two-halo dominated regimes and the one-halo dominated regimes, but it is frustrating that it fails in the transition region between these two regimes. In retrospect, the reason for this is obvious: The assumption made is that haloes are linearly tracing an underlying linear matter field, and this assumption clearly breaks down at scales that correspond to this transition: $k\sim 0.5\iMpc$ at $z=0$ for matter--matter. Linear bias is not valid at this wavenumber \citep[][]{Desjacques2018} and the two-halo term predicted by the model at these and smaller scales will be, quite simply, wrong. We suggest that a fruitful line of future inquiry would be to improve the treatment of halo bias within the standard halo-model framework \citep[\eg][]{Smith2007}. We have somewhat sidestepped this issue in this paper by fitting the \bahamas response functions with halo-model responses, rather than the power spectra directly. This is certainly of benefit because we cancel the Gaussian noise present at large-scales in the simulations. This makes the halo model response an almost perfect tool for understanding matter--matter power response for dark energy models measured in simulations without hydrodynamics \citep{Mead2017, Cataneo2019} when the underlying \emph{linear} spectrum is fixed. This would also be true in hydrodynamic simulations if we could be assured that the troubles in the transition region manifest themselves in the same way, and with the same scale dependence, for other power spectra as they do for matter--matter. Unfortunately, from Fig.~\ref{fig:power_components} we can see that this is not exactly the case, since the transition region occurs at slightly different scales for the different power spectra (particularly the scale for matter--electron pressure is quite different from matter--matter): there simply cannot be some universal $k$-dependent prescription that one could apply to solve all problems in all cases. It is therefore possible that the response is not the optimal quantity to consider when one considers the different power spectra measured in hydrodynamic simulations.

\subsubsection{Expelled gas modelling}

One further shortcoming in our modelling is in our treatment of the gas expelled from haloes. In this paper we do not consider any sort of one-halo contribution arising from expelled gas, and instead account for the effects of this gas only in the two-halo term. In reality, we know that gas expelled from a halo will still be correlated with that halo, and will be found in an extended shell outside the virial radius. Since this shell is larger than the halo virial radius, it may provide a significant contribution to the power at scales that correspond to the transition between the two- and one-halo terms in the classic halo model, and we have so far ignored this. In the early stages of this work we attempted to explicitly include a gas component outside the halo virial radius, but we found that this introduces some problems with the standard halo-model formalism: In the calculation, it is necessary to truncate the density profile of each NFW halo at the virial radius, otherwise they each have an infinite mass. If the gas halo is allowed to extend beyond the virial radius then we introduce an explicit \emph{anti-}correlation between gas and CDM because we have gas in regions where we explicitly have no CDM. This anti-correlation allows the CDM--gas cross spectrum to be negative in some regions of parameter space. While negative cross-spectra are not problematic in general, in cosmology it would be unrealistic due to the general structure of the cosmic web. Indeed, such a negative cross spectrum is not seen in simulations. It may be possible to improve this by using a different CDM halo profile with a finite mass as the radius tends to infinity, but this introduces difficulty in deciding which mass to associate with which halo when identifying haloes in simulations. The fact that we treat the ejected gas in the two-halo term as a simple multiplicative bias also generates another problem: the two-halo term for ejected gas is not suppressed at small scales relative to the shape of linear theory. For other fields, it is always suppressed by a weight that depends on the halo profile and halo bias (equation~\ref{eq:two_halo}). The details of this two-halo term suppression cannot be taken seriously because they occur at scales where the halo bias is no longer linear, however, this suppression does have the virtue of ensuring that the two-halo term is always much smaller than the one-halo term at small scales. Because the gas profiles themselves lack small-scale structure, the two-halo term for gas (and electron pressure) can be larger than the one-halo term at very small scales ($k\sim100\iMpc$). Clearly this is unphysical and would need to be addressed in further work.

\subsubsection{Important halo masses}

A further potential problem for this approach is the differing halo masses that are important for the tSZ effect compared to those important for galaxy weak lensing. As shown in Fig.~\ref{fig:power_mass_contribution}, the matter--electron pressure spectrum derives most contribution from higher-mass haloes than matter--matter. It is therefore these haloes that we learn most about from the lensing-tSZ cross correlation. However, it is lower-mass haloes that mostly affect the suppression in the power spectrum that is important for the lensing-lensing spectrum. A fruitful avenue for future research may be to learn more directly about the gas distribution from the haloes that are most relevant, \ie those of $10^{13}$ to $10^{14}\Msun$. It is difficult to do this using the tSZ-lensing cross correlation directly, and instead one may have to use some sort of clipping \citep[\eg][]{Simpson2011, Simpson2013b, Hill2013b, Giblin2018} technique to mask the influence of the highest mass haloes. We note that there is already a great deal known about the impact of baryons on the matter distribution from tSZ and X-ray studies of galaxies, groups, and clusters from targeted observational studies. Can we use this knowledge to help the weak-lensing efforts? One could imagine using the model presented in this paper to look at both the lensing and tSZ signal (or lensing--tSZ) around haloes of specific masses. This would add information about the signal as it arises from different mass bins, somehow akin to the information in Fig.~\ref{fig:power_mass_contribution}. However, this would necessitate having reliable measurements of individual halo masses and would require modelling effects such as halo mis-centring \citep[\eg][]{Yan2020}. It would also need to be ascertained whether the halo model can be trusted for calculations that are binned in halo mass and how problems with these calculations compare to problems in the standard when integrated over all halo masses.

\subsection{Conclusion}

We intend to use this model in the near future to analyse the cross correlation between the tSZ map from Planck and weak lensing from the \kids data, as well as the lensing auto correlation data (Tr{\"o}ster et al. in prep). This work has the potential to simultaneously learn about cosmology and the magnitude of AGN feedback in the Universe and therefore to inform galaxy formation modelling and future hydrodynamical simulations as well as to shrink error bars on cosmological parameters. As alluded to in this paper, the tSZ data itself also has the power to provide tight constraints on $\sigma_8$ and initial tests show that this could break degeneracies that exist in a traditional lensing analysis.

So far, we have only presented the best-fitting model parameters for each scenario and we have not investigated errors on, or degeneracies between, our model parameters, but we hope to be able to do this in future. We note that this would require us to create meaningful error bars on our response measurements.

It is clear that the fitted parameters presented here are very \bahamas specific, which may worry some readers who would prefer a model that could encompass a wider range of hydrodynamical simulations. However, we think that our model is general enough that it could be refitted to other simulations in the future. We also note \cite{vanDaalen2020} has demonstrated a strong correlation between the matter--matter response function and the gas fraction in haloes of $\sim 10^{14}\Msun$ and that the \bahamas simulations reproduce observations of gas fractions in haloes of this mass to within the observational error bar. Given that the gas content in haloes of this mass also is important for tSZ suggests that tuning our model to \bahamas was a reasonable first choice. In the future, it may be possible to rewrite our results in terms of more physical quantities, such as the halo gas fraction \cite[\eg][]{Chisari2019b, vanDaalen2020}, rather than the unphysical AGN heating temperature. It would also be desirable to confront our model with different hydrodynamical simulations, run with different underlying assumptions and methods, for example those run with the Adaptive Refinement Tree code \citep{Kravtsov2002, Nagai2007b} or those of \textsc{illustris} \citep{Vogelsberger2014}.


In the future, it may be possible to use this halo model to make predictions for higher-order statistics and these may lead to increased constraints on the model parameters. We also envisage the model being used to make predictions for other `hydrodynamical' observable quantities, such as $n$-point statistics that involve the X-ray field, which is a different function of gas temperature and density compared to the tSZ and provides complimentary information about the state of gas in the universe. The utility of the halo model that we present does not require its parameters to be fitted to simulation data, this was simply one option, and the option that we decided to pursue in this paper. In the future the model can be used to provide quick estimates for the effect of new feedback scenarios that have yet to be simulated, or to explore non-standard cosmological scenarios that have yet to be hydrodynamical simulated. It would even be possible to investigate the non-linear coupling of feedback with non-standard cosmological scenarios.

\section*{Acknowledgements}
AJM acknowledges support from a CITA National Fellowship and (together with LVW) from NSERC. AJM and TT have received funding from the Horizon 2020 research and innovation programme of the European Union under Marie Sk\l{}odowska-Curie grant agreements Nos. 702971 (AJM) and 797794 (TT). TT and CH acknowledge support from the European Research Council under grant number 647112, while IGM acknowledges the same funding under grant number 769130. CH also acknowledges support from the Max Planck Society and the Alexander von Humboldt Foundation in the framework of the Max Planck-Humboldt Research Award endowed by the Federal Ministry of Education and Research. We thank an anonymous referee for useful comments that improved the quality of this paper.

\label{lastpage}


\bibliography{meadbib}
\bibliographystyle{aa}

\appendix

\section{Evaluation of the two-halo term}
\label{app:two_halo}

There is some confusion in the literature about how exactly to numerically evaluate the integral for the halo model two-halo term:
\begin{equation}
P_{2\mathrm{H},uv}(k)=P_{\mathrm{lin}}(k)
\prod_{i=u,v}\left[\int_0^\infty b(M)W_i(M,k)n(M)\;\mathrm{d}M\right]\ .
\label{eq:two_halo_appendix}
\end{equation}
We shall define the integral appearing in equation~(\ref{eq:two_halo_appendix}) to be
\begin{equation}
I_u(M_1,M_2,k)=\int_{M_1}^{M_2} b(M)W_u(M,k)n(M)\;\mathrm{d}M\ .
\label{eq:difficult_integral}
\end{equation}
In this appendix we will specialise to the two-halo term for the matter--matter spectrum, $u=v=\mathrm{m}$, as this provides a good illustration of the source of confusion. Equation~(\ref{eq:difficult_integral}) \emph{should} be evaluated over all halo masses, $M_1\to0$ to $M_2\to\infty$. Confusion arises because most numerical implementations evaluate this integral only over a finite range of halo mass, which seems sensible given that usually only a finite range of masses are thought to be relevant. However, note that for the matter spectrum we \emph{must} have the following large-scale limit respected on physical grounds:
\begin{equation}
P_{2\mathrm{H},\mathrm{mm}}(k\to0)=P_{\mathrm{lin}}(k\to0)\ .
\label{eq:two_halo_limit}
\end{equation}
So the integral in equation~(\ref{eq:difficult_integral}) \emph{must} tend to unity when $u=\mathrm{m}$ in the $k\to0$ limit. In this case, $W_\mathrm{m}(M,k\to0)\to M/\bar\rho$ and equation~(\ref{eq:difficult_integral}) becomes
\begin{equation}
I_\mathrm{m}(M_1,M_2,k\to0)=\frac{1}{\bar\rho}\int_{M_1}^{M_2} b(M)Mn(M)\;\mathrm{d}M\ .
\label{eq:difficult_integral_lowk}
\end{equation}
If we pick sensible-seeming limits of $M_1=10^{10}\Msun$ and $M_2=10^{16}\Msun$ then we will find, for a standard cosmological model and a standard mass function at $z=0$, that $I_\mathrm{m}(M_1=10^{10}\Msun,M_2=10^{16}\Msun,k\to0) \simeq 0.67 $. The problem is that, if taken literally, most popular mass functions have a large amount of the total matter contained in very low mass haloes, and so the convergence of equation~(\ref{eq:difficult_integral}) is slow as $M_1$ is lowered. The upper limit of $M_2$ is not usually a problem as long as it is reasonably high, say $10^{16}\Msun$ at $z=0$, which is effectively infinite here.

How to proceed? One thing that should \emph{absolutely not} be done is to `renormalise' the mass function by multiplying it by a constant to enforce that $I_\mathrm{m}(M_1,\infty,k\to0)=1$ for a particular choice of $M_1$. Physically, this amounts to removing low-mass haloes and putting all their mass into higher-mass haloes. In an example where $I_\mathrm{m}(M_1,\infty,k\to0)=0.67$ this would amount to increasing by $\sim 50$ per cent the numbers of haloes between the integration limits! Using a mass function `renormalised' in this way will drastically change the scale dependence in the two- and one-halo terms because it increases the number of haloes above $M_1$.

There are two choices for how to proceed: The first \citep[\eg][]{Yoo2006, Cacciato2012} is to assume that $W_\mathrm{m}(M<M_1,k)=M/\bar\rho$, which is equivalent to taking low-mass haloes to have delta-function profiles. One can then numerically evaluate equation~(\ref{eq:difficult_integral}) above the limit of $M_1$ and define a new function
\begin{equation}
A(M_1)=1-I_\mathrm{m}(M_1,\infty,k\to0)\ ,
\end{equation}
which is the missing part of the integral from below $M_1$ (this only needs to be computed once). One can then \emph{add} $A(M_1)$ every time equation~(\ref{eq:difficult_integral}) is evaluated:
\begin{equation}
\begin{split}
I_\mathrm{m}(M_1,\infty,k)\to &
\int_{M_1}^{\infty} b(M)W_\mathrm{m}(M,k)n(M)\;\mathrm{d}M \\
&+A(M_1)\ .
\end{split}
\label{eq:solution_one}
\end{equation}
Note that this correction is additive, not multiplicative. This approximation is sufficient as long as the physical \emph{shapes} of haloes at and below $M_1$ do not influence the calculation; however, it is only the appropriate correction for equation~(\ref{eq:difficult_integral}) for matter. The second choice is to alter the mass function so that the missing signal from haloes below $M_1$ is considered to be in haloes of mass exactly $M_1$ \citep{Schmidt2016}. This amounts to making the substitution
\begin{equation}
n(M)\to n'(M)=n(M)+\frac{A(M_1)\Dirac{M-M_1}}{b(M_1)M_1/\bar\rho}\ ,
\end{equation}
in equation~(\ref{eq:difficult_integral}). This reduces to a different additive correction to equation~(\ref{eq:difficult_integral}) of
\begin{equation}
\begin{split}
I_u(M_1,\infty,k)\to &
\int_{M_1}^{\infty} b(M)W_u(M,k)n(M)\;\mathrm{d}M \\
&+A(M_1)\frac{W_u(M_1,k)}{M_1/\bar\rho}\ .
\end{split}
\label{eq:solution_two}
\end{equation}
This approximation is sufficient provided the physical shapes of haloes below $M_1$ do not influence the calculation, but it maintains scale-dependence in the two-halo term compared to equation~(\ref{eq:solution_one}). We prefer the second option because it applies evaluations of equation~(\ref{eq:difficult_integral}) for any field $u(\mathbf{r})$, not just for matter (note the `general field' subscript $u$ in equation~\ref{eq:solution_two}, compared to the `matter' subscript $\mathrm{m}$ in equation~\ref{eq:solution_one}). We apply this mass-function alteration only to the two-halo term in all of our calculations. 

Note that the problem of slow convergence with respect to the lower integration limit does not usually affect the one-halo term for matter spectra because low-mass haloes contribute very little to the one-halo term, whose amplitude and shape depends on $\langle M \rangle$ (notation from Section~\ref{sec:k_to_zero}) which is $\sim10^{13.5}\Msun$ at $z=0$ for a standard cosmological model, and therefore is usually included in a sensible integration range for equation~(\ref{eq:difficult_integral}).

In cases where the power spectrum is unaffected by low-mass haloes then the discussion in this appendix is not relevant. For example, for the galaxy--galaxy number density power you only need to compute the integrals down to the halo mass below which the haloes contain no galaxies (obviously). In the case of the electron-pressure field, so little pressure comes from low mass haloes that this correction can be ignored. Considering the discussion in this appendix will, however, be relevant for any spectrum that involves the matter field because the low-mass haloes contribute significantly to the overall mass.

\section{Power spectrum measurements from hydrodynamical simulations}
\label{app:power_measurement}

In this paper we measure the power spectrum of overdensity fluctuations and electron pressure using particle data from hydrodynamic simulations. We do this using a Fast-Fourier Transform (FFT) algorithm applied to Cartesian fields, defined on a regular mesh, that are created from particle data. Generally, we consider a mesh-defined field $u(\mathbf{r})$ where $u_i$ is the contribution to $u(\mathbf{r})$ per particle located at $\mathbf{r}_i$
\begin{equation}
u(\mathbf{r})=\sum_{i=1}^N u_i \Dirac{\mathbf{r}-\mathbf{r}_i}\ ,
\label{eq:field_of_particles}
\end{equation}
with $N$ being the total number of particles contributing to $u(\mathbf{r})$.

To create density-contrast fields from particles in hydrodynamical simulations it is necessary to assign a different density per particle to the mesh because hydrodynamical particles may have different masses $m_i$. We therefore write
\begin{equation}
\frac{\rho(\mathbf{r})}{\bar\rho}={\sum_{i=1}^N \frac{m_i}{\bar{m}}} \Dirac{\mathbf{r}-\mathbf{r}_i}\ ,
\label{eq:density_field}
\end{equation}
where $\bar{m}$ is the expected mean mass in a cell. In our definitions of overdensity, this division to create a density contrast is always by the mean mass of all matter, not only of the specific species being considered. The situation is more complicated when creating the electron pressure fields from the particles. Hydrodynamic gas particles (labelled with $i$) come tagged with an internal energy ($k_\mathrm{B}T_i$) and a local density ($\rho_i$). To convert the particle gas temperature to a contribution to the total electron pressure per particle, $P_{\mathrm{e},i}$, a quantity that can be summed to create an electron-pressure field on mesh via
\begin{equation}
P_\mathrm{e}(\mathbf{r})=\sum_{i=1}^N P_{\mathrm{e},i} \Dirac{\mathbf{r}-\mathbf{r}_i}\ ,
\label{eq:electron_pressure_field}
\end{equation}
we use the ideal gas law:
\begin{equation}
P_{\mathrm{e},i}V=N_{\mathrm{e},i}k_\mathrm{B}T_i
\label{eq:ideal_gas_law_electrons}
\end{equation}
where $N_{\mathrm{e},i}$ is the total number of free electrons in the gas particle in the mesh-cell volume $V$:
\begin{equation}
N_{\mathrm{e},i}=\frac{m_i}{m_\mathrm{p}}\frac{1}{\mu_\mathrm{e}}\ ,
\label{eq:electron_number}
\end{equation}
where $m_i$ is the gas particle mass, $m_\mathrm{p}$ is the proton mass and $\mu_\mathrm{e}$ is the number of free electrons per proton mass. In creating the electron-pressure field from the particles we also calculate the particle hydrogen number density 
\begin{equation}
n_{\mathrm{H},i}=\frac{f_\mathrm{H} \rho_i}{m_\mathrm{p}}\ ,
\end{equation}
where $f_\mathrm{H}\simeq0.752$ is the hydrogen gas mass fraction. We exclude particles with $n_\mathrm{H}>0.1\,\mathrm{cm}^{-3}$ as these particles are presumably cold, neutral and should eventually form stars. We checked that imposing this exclusion does not greatly affect our results.

$P_{\mathrm{e},i}$ from equation~(\ref{eq:ideal_gas_law_electrons}) is directly related to the quantity $\Upsilon_i$ used in \cite{Roncarelli2006,Roncarelli2007} and \cite{McCarthy2018}:
\begin{equation}
\Upsilon_i=\sigma_\mathrm{T}\frac{k_\mathrm{B}T_i}{m_\mathrm{e}c^2}\frac{m_i}{\mu_\mathrm{e} m_\mathrm{p}}\ ,
\label{eq:Upsilon}
\end{equation}
where $m_\mathrm{e}$ is the electron mass and $\sigma_\mathrm{T}$ is the Thompson scattering cross section. $\Upsilon_i$ differs from $P_{\mathrm{e},i}$ only by a factor of volume and some constants and has dimensions of area rather than pressure.

In \nbody simulations it is well known that the raw measured power spectrum receive an unphysical additive contribution from the finite number of particles -- so-called discreteness or `shot' noise. This contribution arises due to the automatic correlation of particles with themselves at zero separation. In the correlation function this contribution is confined to zero separation, but in $P(k)$ the contribution is spread evenly over all wavenumbers by the Fourier transform (the Fourier transform of a delta function is a constant). In a standard \nbody simulation with equal-mass particles the shot noise contribution to the matter--matter $P(k)$ spectrum is:
\begin{equation}
S=\frac{L^3}{N}\ ,
\label{eq:standard_shot_noise}
\end{equation}
where $N$ is the total number of particles in the simulation and $L$ is the box size. As expected, the shot noise contribution is larger when there are fewer particles. It is common practice to subtract this contribution from power spectra measured from \nbody simulations since in the limit of an infinite number of particles it would vanish.

From hydrodynamical simulations we are interested in calculating auto- and cross-spectra between between different fields generated from the particles; for example, density, pressure or temperature fields. In the case that the fields to be correlated are generated from distinct sets of particles then the shot noise contribution is zero (\eg gas--star density). However, in other cases the fields might be generated using (a subset of) same particles (\eg matter--star density) or from different properties of the same particles (\eg gas density--electron pressure, which both come from the same hydrodynamical gas particles). There is also the confounding issue that each particle does not provide the same contribution to the eventual field (\eg individual stellar and gas particles have different masses, gas particles contribute different pressures).

The additive shot-noise contribution for the cross spectrum, $P_{uv}(k)$, between two fields $u(\mathbf{r})$ and $v(\mathbf{r})$, comprised of particles, where a subset $\tilde{N}$ of those particles contribute to \emph{both} fields $u(\mathbf{r})$ and $v(\mathbf{r})$, can be shown to be
\begin{equation}
S_\mathrm{uv}=\frac{L^3}{M^2}\sum_{i=1}^{\tilde{N}} u_i v_i\ ,
\label{eq:hydro_shot_noise}
\end{equation}
where $L$ is the simulation box size and $M$ is the total number of mesh cells. Note that the shot-noise contribution is not really a function of $M$, as this will always cancel with factors of $M$ that come from $u_i$ and $v_i$. Equation~(\ref{eq:hydro_shot_noise}) can be derived by taking the Fourier transform of equation~(\ref{eq:field_of_particles}) and then considering the part of the power spectrum that would still arise even if the particle positions $\mathbf{r}_i$ are completely uncorrelated, so that the only contribution is from the self-correlation of each particle with itself.

Let us demonstrate that equation~(\ref{eq:hydro_shot_noise}) reduces to more standard formulae: If we are interested in shot noise in the matter power spectrum we start from the particle masses from which we create a density field. The density-contrast field generated from the particles is given by equation~(\ref{eq:density_field}). We can write $\bar{m}$ as the total mass in the simulation divided by the total number of mesh cells:
\begin{equation}
\bar{m}=\frac{1}{M}\sum_{i=1}^{N} m_i\ .
\end{equation}
Using equation~(\ref{eq:hydro_shot_noise}) and considering shot-noise in the matter--matter power spectrum $\mathrm{m}$ we get:
\begin{equation}
S_\mathrm{mm}=L^3\frac{\sum_{i=1}^N m_i^2}{\left[\sum_{i=1}^N m_i \right]^2}\ .
\label{eq:shot_noise_nonequal_mass}
\end{equation}
If all the particle have equal masses then equation~(\ref{eq:shot_noise_nonequal_mass}) reduces to equation~(\ref{eq:standard_shot_noise}) as required.

\begin{figure}
\begin{center}
\includegraphics[height=8.5cm, angle=270]{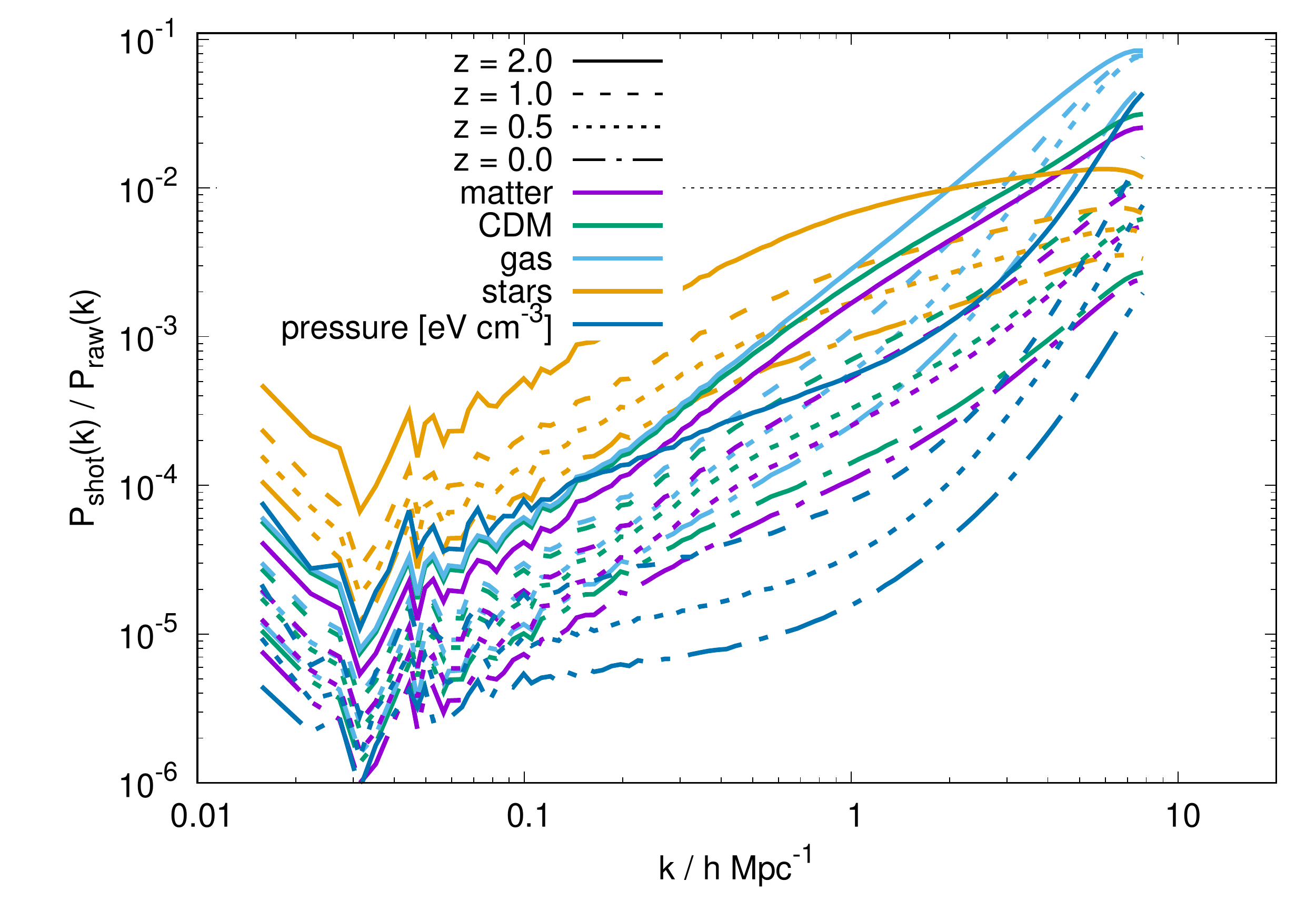}
\end{center}
\caption{The shot-noise contributions to auto-spectra of the various fields measured from the \agn \bahamas simulation. We show the ratio of the shot-noise to the auto-spectra as a function of scale for matter (purple), CDM (green), gas (blue), stars (orange) and electron pressure (yellow). The dashed line shows the cutoff for a one per cent contribution. We see that shot noise is most important for the gas power spectrum at higher redshifts, where it can contribute as much as $8$ per cent of the raw measured power for the smallest scales shown. Generally shot noise becomes less important as the simulation evolves and structure develops.}
\label{fig:shot_noise}
\end{figure}

In Fig.~\ref{fig:shot_noise} we show the contribution of shot noise to the raw auto-spectra measured for the different fields from the \agn \bahamas simulations. Subtracting shot noise is most important for the gas spectrum, which makes sense because gas is the most diffuse component and therefore the one for which discreteness effects are most obvious. At $z=2$ at $k=7\iMpc$ shot noise can makes up 8 per cent of the raw measured gas--gas spectrum. Shot noise is also important at a similar level in some of the cross-spectra that we measure. For all of the results presented in this paper shot noise has been subtracted from the power spectra.

\section{Variations in halo-model parameters}
\label{app:variations}

\begin{figure*}
   \begin{center}
   \includegraphics[height=18cm, angle=270]{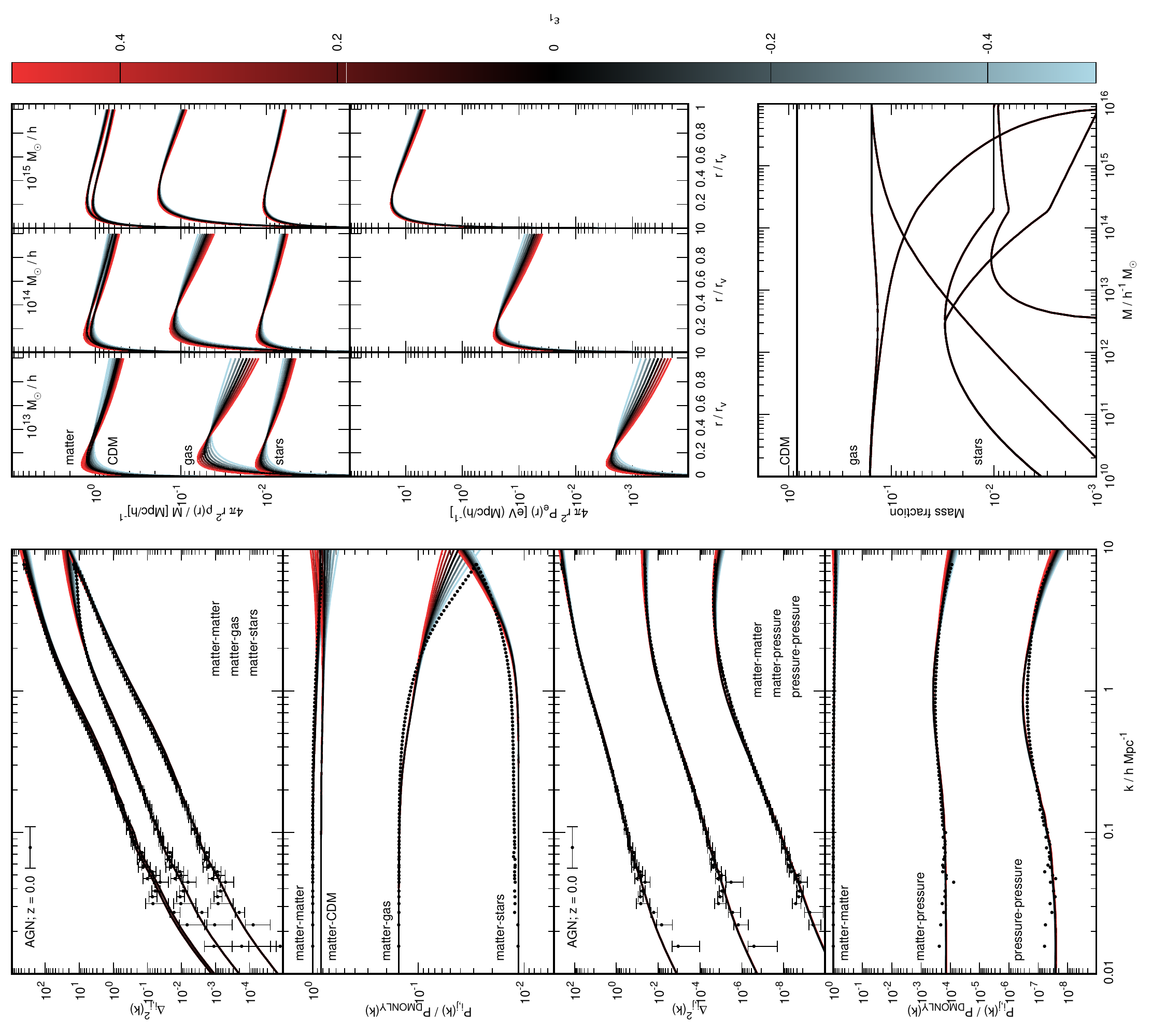}
   \end{center}
   \caption{The effect of varying $\epsilon_1$, which governs deviations in halo concentration for low-mass haloes that have ejected their gas, on halo profiles (top right two panels), halo mass fractions (bottom right) and power spectra and responses (left column; top left two panels for matter spectra, bottom left two panels for pressure). Increasing $\epsilon_1$ increases halo concentration and this generally boosts power. From the top right we see this has the largest impact for lower mass haloes because these have ejected most of their gas. From the left-hand panels, this has comparatively more of an impact on power spectra involving gas than on the other spectra. We can also see similar effects on spectra that involve the electron pressure.}
\label{fig:epsilon_variation}
\end{figure*}

\begin{figure*}
   \begin{center}
   \includegraphics[height=18cm, angle=270]{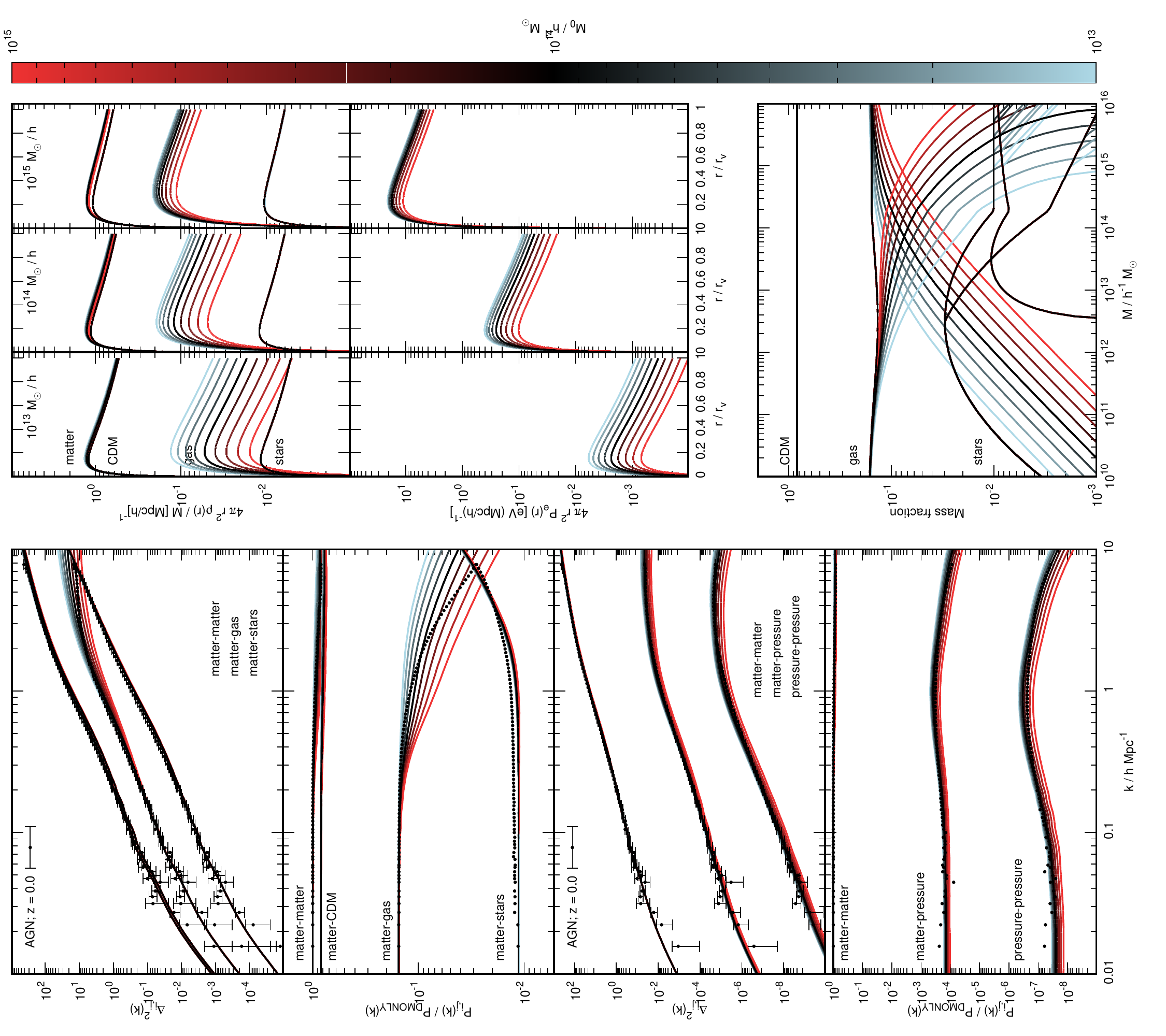}
   \end{center}
   \caption{Same as Fig.~\ref{fig:epsilon_variation} but for $M_0$, which governs the halo mass below which haloes have lost most of their gas. As can be seen in the bottom-right panel, changing $M_0$ changes the split of gas between the bound component, which dominates high-mass haloes and the unbound component, which is the case for low-mass haloes that have jettisoned most of their gas. The effects of this on the bound halo profiles can be seen in the top-right panel. Decreasing $M_0$ adds to the amplitude of the gas density profiles and this boosts power in power spectra involving gas. We see a similar, but more limited effect on spectra involving the electron pressure. This limitation arises because low-mass haloes are lower temperature, so contribute much less to spectra involving the electron pressure than they do to those involving the gas density. There is also an overall amplitude affect on the pressure spectra because the unbound gas is taken to have a fixed $T_\mathrm{w} = 10^{6}\Kelvin$ that contributes power on large scales.}
   \label{fig:M0_variation}
\end{figure*}

\begin{figure*}
   \begin{center}
   \includegraphics[height=18cm, angle=270]{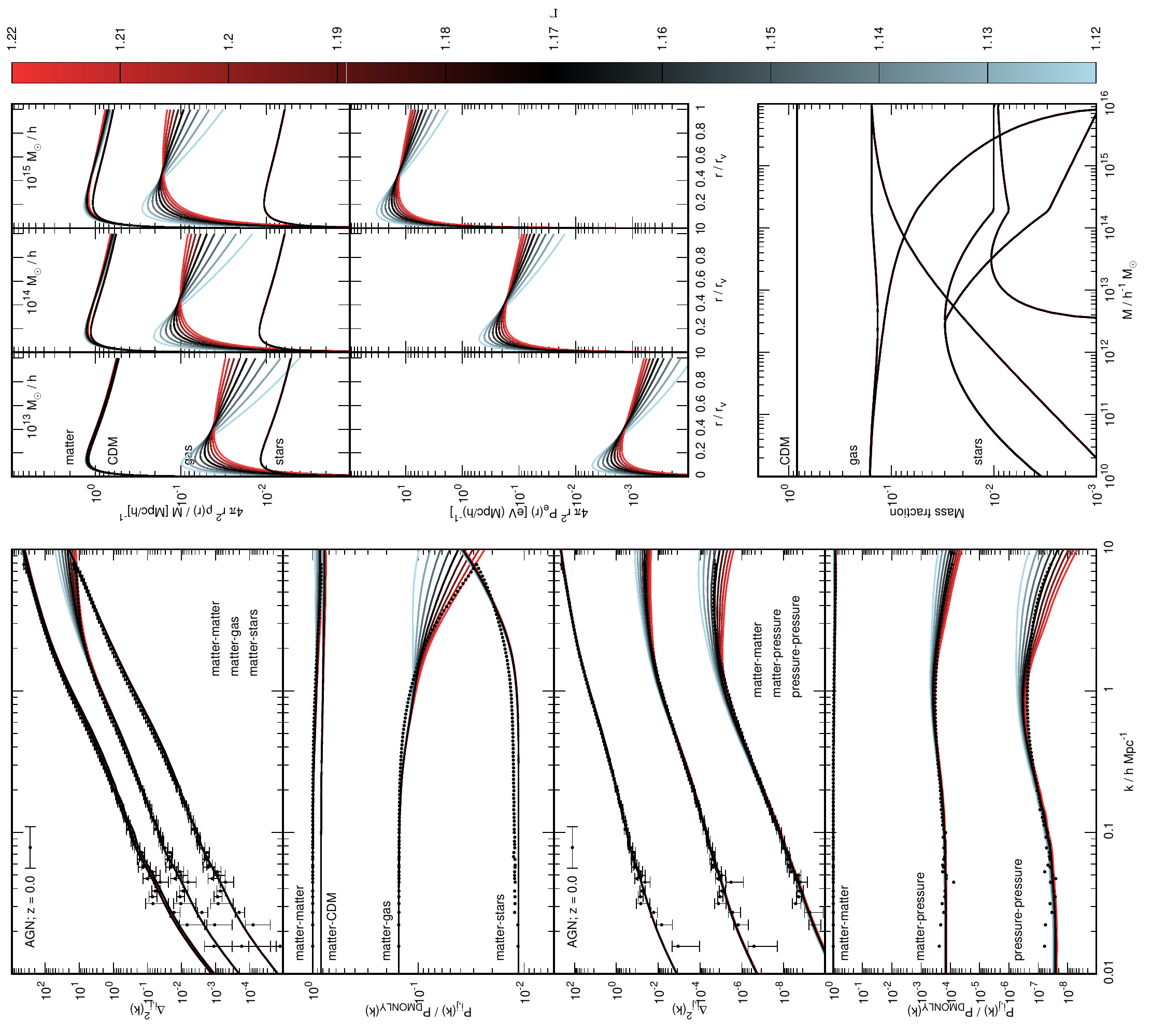}
   \end{center}
   \caption{Same as Fig.~\ref{fig:epsilon_variation} but for $\Gamma$, the polytropic index for the gas. Increasing $\Gamma$ makes the gas profile less concentrated, as can be seen in the halo-profile plots (top right). If $\Gamma$ is lowered then gas is more concentrated and the power spectrum of gas is boosted at small scales. Effects on power spectra that do not involve gas are much smaller. The same trends are seen in the electron-pressure profile and spectra involving the electron pressure, which are also sensitive to this parameter as the gas pressure is partly determined by the gas density.}
   \label{fig:Gamma_variation}
\end{figure*}

\begin{figure*}
   \begin{center}
   \includegraphics[height=18cm, angle=270]{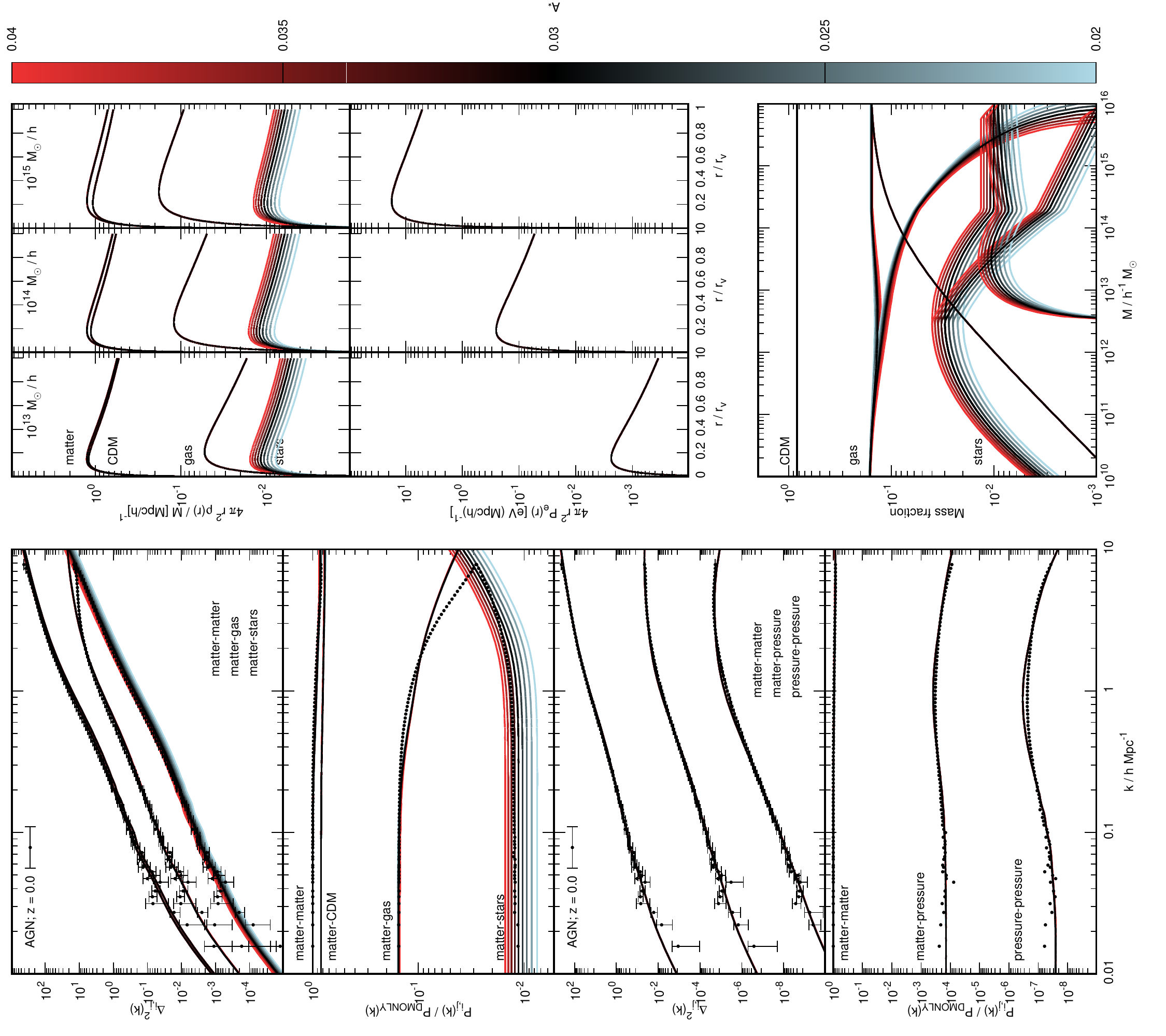}
   \end{center}
   \caption{Same as Fig.~\ref{fig:epsilon_variation}, but for $A_*$, which governs the peak-star-formation efficiency in haloes. The lower-right panel shows that $A_*$ effects the overall abundance of stellar mass, but not the ratio in central and satellite stellar mass. There is some back reaction on the gas abundance that occurs in order to keep the baryon fraction universal, so that halo gas is slightly depleted as a result of this boost, so there is a small effect on gas profiles. The effect on the halo profiles can be seen in the top right panel. Increasing $A_*$ boosts star-formation efficiency, which has a fairly uniform effect on the amplitude of any power spectra involving the star density as can be seen in the top two panels in the left-hand column. There is an effect on the matter--matter spectrum at small scales as stars start to dominate the power. The effect on electron pressure is minimal, but not non-existent due to the gas abundance being affected.}
   \label{fig:Astar_variation}
\end{figure*}

\begin{figure*}
   \begin{center}
   \includegraphics[height=18cm, angle=270]{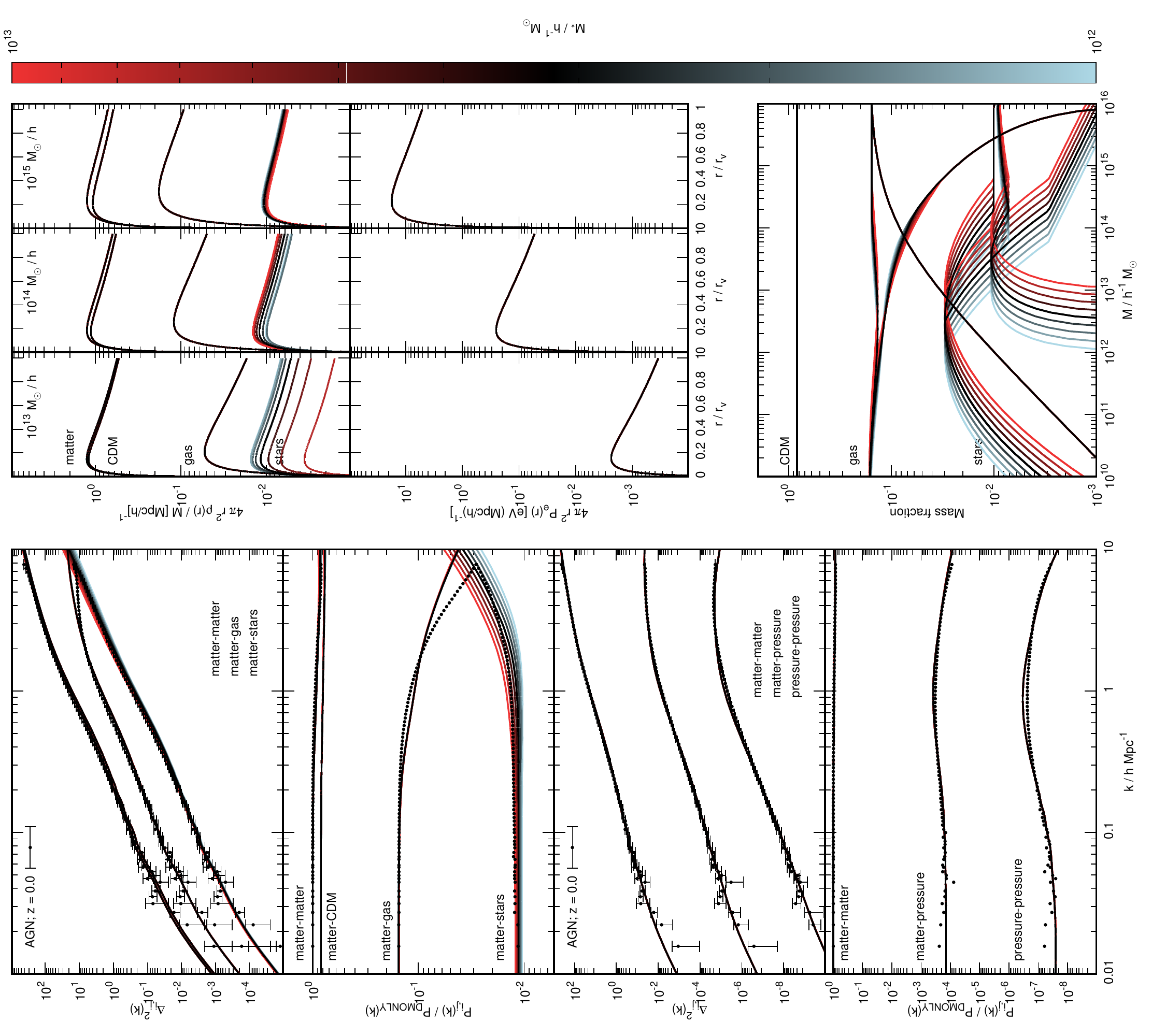}
   \end{center}
   \caption{Same as Fig.~\ref{fig:epsilon_variation} but for $M_*$, the halo mass at which star formation efficiency peaks. Changing this parameter shifts the halo-stellar-mass fraction to lower and higher masses, as can be seen in the bottom-right panel, there is a small back-reaction effect on the gas abundance. Changing $M_*$ indirectly affects the halo profiles at different masses, as can be seen in  the top right panel, but note that the delta-function component of the stellar halo profile is not shown in this plot. The power spectra involving stars are affects in a more scale-dependent manner as can be seen in the top-left two panels.} 
   \label{fig:Mstar_variation}
\end{figure*}

\begin{figure*}
   \begin{center}
   \includegraphics[height=18cm, angle=270]{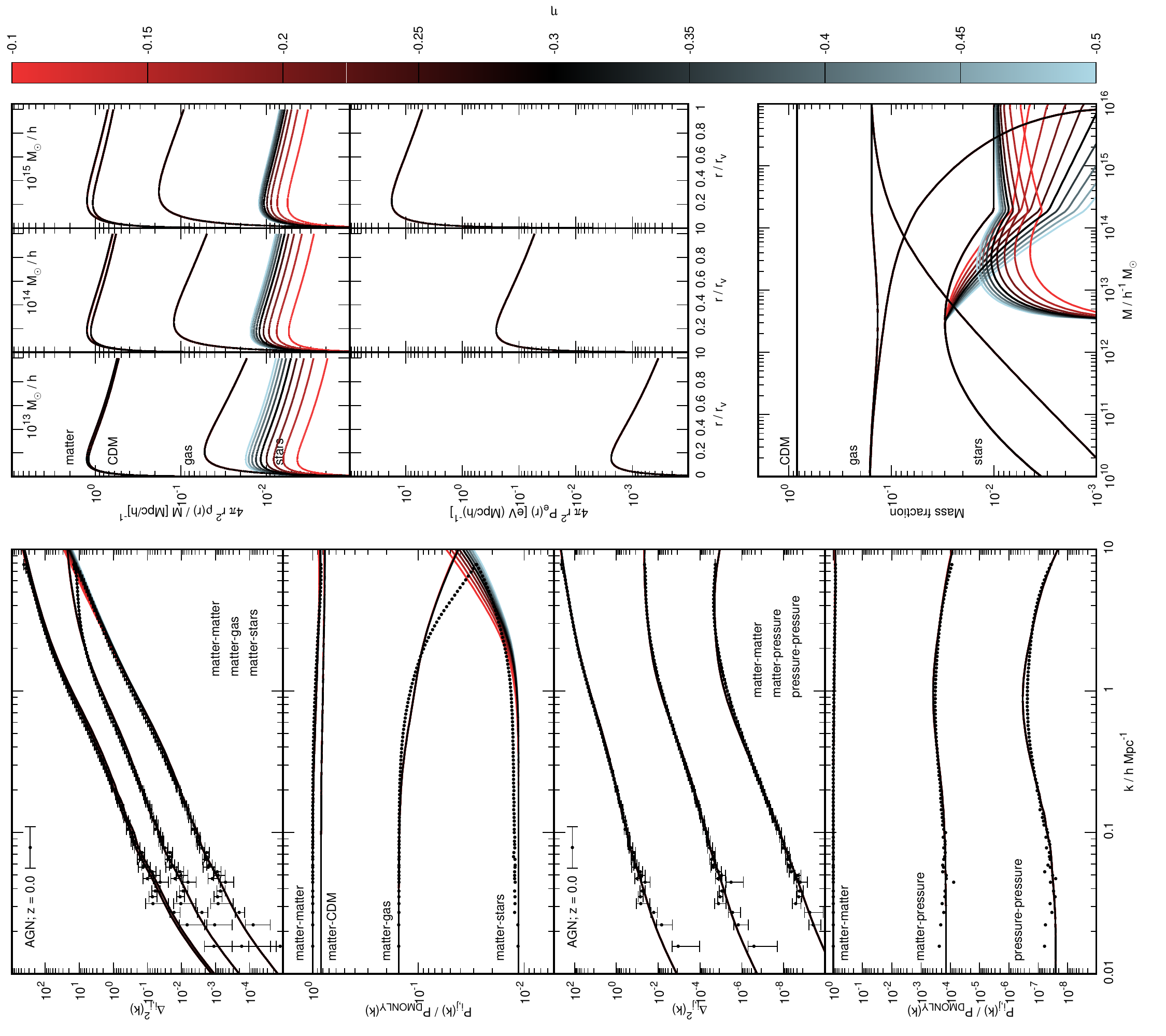}
   \end{center}
   \caption{Same as Fig.~\ref{fig:epsilon_variation} but for $\eta$, which determines the split of stellar halo mass between central and satellite galaxies for $M>M_*$, where $M_*=10^{12.5}\Msun$ in this figure. In the lower-right panel we see that increasing $\eta$ from the fiducial $-0.3$ means that more mass is put into central galaxies, whereas decreasing it places more mass in satellites. As a result of this, the amplitude of the stellar halo is changed, as can be seen in the top-right panel, with decreasing $\eta$ putting more mass into the NFW part of the profile and thus boosting the halo amplitude. This has a scale-dependent effect on the matter--stars power spectrum, as can be seen in the top-left two panels.}
   \label{fig:eta_variation}
\end{figure*}

\begin{figure*}
   \begin{center}
   \includegraphics[height=18cm, angle=270]{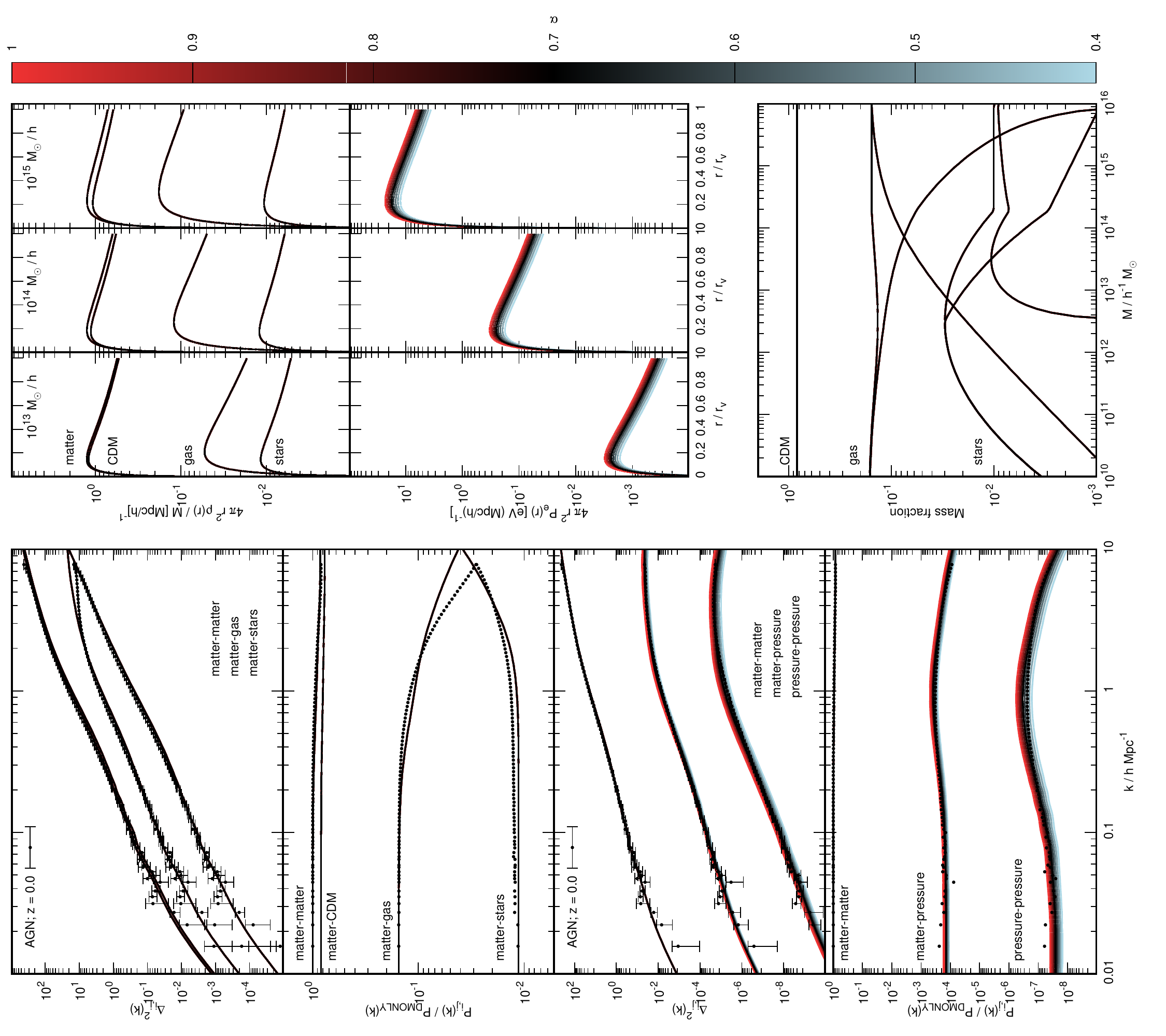}
   \end{center}
   \caption{Same as Fig.~\ref{fig:epsilon_variation} but for $\alpha$. $\alpha$ governs deviations of the halo temperature from the virial relation. We see that this affects the pressure profiles (middle right) by uniformly scaling the amplitude, because the pressure is directly proportional to temperature; increasing the temperature increases the pressure. We can see the effect of this on the pressure power spectra in the lower left two panels where the effect is not uniform due to the complicated dependence of the power spectra on the underlying halo populations in the two- and one-halo terms, which have different scalings with temperature.}
   \label{fig:alpha_variation}
\end{figure*}

\begin{figure*}
   \begin{center}
   \includegraphics[height=18cm, angle=270]{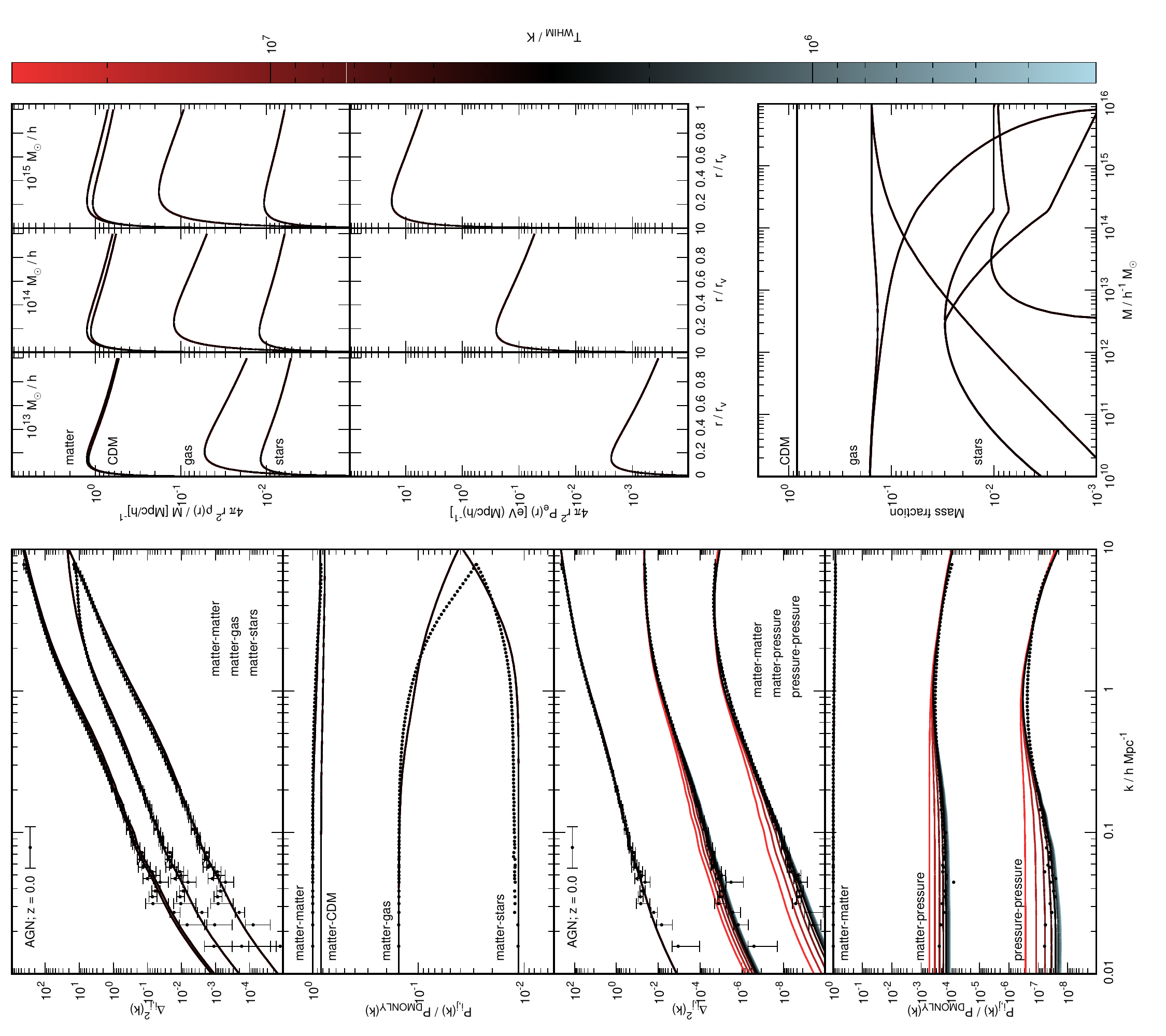}
   \end{center}
   \caption{Same as Fig.~\ref{fig:epsilon_variation} but for $T_\mathrm{w}$. Varying $T_\mathrm{w}$ only affects the temperature of the unbound gas that contributes to the two-halo term only. We see that as we increase the gas temperature we boost the two-halo term. Note that there is a floor to the effect that lowering $T_\mathrm{w}$ can have, which is because the one-halo term contributes at large scales in the pressure spectra too, and lowering the temperature of this gas beyond a certain point means that only the two-halo term becomes dwarfed by the one-halo term at large scales.}
   \label{fig:Tw_variation}
\end{figure*}

In this appendix we present figures that demonstrate the effect of varying our fitted parameters on the halo profiles and on the eventual power spectra that we are interested in. We show the effect of varying $A_*$ (Fig.~\ref{fig:Astar_variation}), $M_*$ (Fig.~\ref{fig:Mstar_variation}), $\eta$ (Fig.~\ref{fig:eta_variation}), $\epsilon_1$ (Fig.~\ref{fig:epsilon_variation}), $\Gamma$ (Fig.~\ref{fig:Gamma_variation}), $M_0$ (Fig.~\ref{fig:M0_variation}), $\alpha$ (Fig.~\ref{fig:alpha_variation}) and $T_\mathrm{whim}$ (Fig.~\ref{fig:Tw_variation}) on the power spectra, response, halo density profiles and halo mass fractions. These plots are extremely useful if one wants to gain physical insight on the effect of changing the hydrodynamical parameters on the eventual Fourier Space statistics, which can otherwise be quite difficult to understand.

We see that increasing $A_*$ increases the amplitude of the stellar content of haloes independently of their mass, and this has an almost scale-independent effect on any power spectra that includes the star field. There is a small back reaction on the gas because stars can only be added at the expense of gas, but this is very small because stars are a small fraction of the available halo baryons, even in the most extreme cases. Changing $M_*$ changes the halo mass at which star formation peaks, which has a more scale-dependent effect on the eventual power spectra as stellar mass is shifted into haloes with different structural properties. Altering $\eta$ changes the distribution of stellar mass between the extended satellite galaxy part and the central galaxy. This has no effect at the largest scales, but effects the one-halo term. Shifting more stellar mass into central galaxies makes this term more and more shot-noise like.

$\epsilon_1$ governs the concentration modification of lower-mass galaxies that have ejected most of their gas. This has no effect on the power spectra at large scales, but changes them at small scales where there is sensitivity to the halo concentration. $\Gamma$ governs the gas density and pressure profiles and has large effects on the halo profiles. Decreasing $\Gamma$ makes the gas and pressure profiles more centrally concentrated and therefore boosts the associated power spectra at small scales. Altering $M_0$ changes the mass corresponding to haloes that have lost half of their gas (with lower-mass haloes having lost more than half). This has a comparatively large effect on spectra, with a strong scale dependent effect on the matter--gas (no change at large scales) and a more scale-independent effect on spectra involving the pressure.

Increasing $\alpha$ increases the temperature of gas that is bound to the host halo. This boosts the pressure signal in a perfectly scale-independent way. There is no effect whatsoever on any matter spectra since these are completely insensitive to gas temperature. Conversely, increasing $T_\mathrm{w}$ boosts the temperature of the ejected gas, which is correlated only on large scales and thus only affects the large-scale, two-halo portions of power spectra that involve the pressure.


\begin{figure*}
   \begin{center}
   \includegraphics[height=18cm, angle=270]{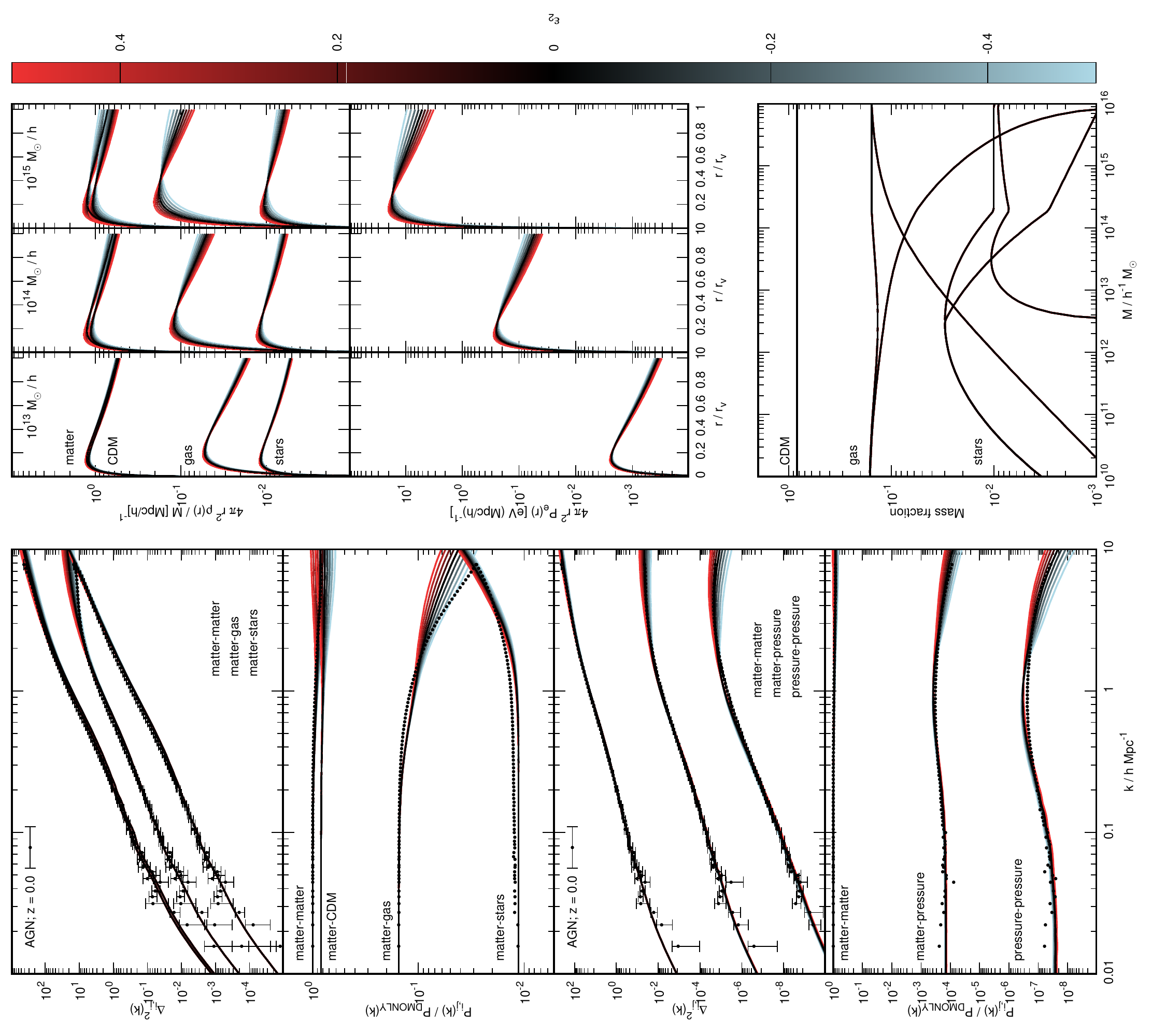}
   \end{center}
   \caption{Same as Fig.~\ref{fig:epsilon_variation} but for $\epsilon_2$, which governs deviations in halo concentration for high-mass haloes that have retained their gas. Increasing $\epsilon_2$ increases halo concentration and this generally boosts power. From the top right we see this has the largest impact for higher mass haloes because these have retained most of their gas. From the left-hand panels, this has comparatively more of an impact on power spectra involving gas than on the other spectra. We can also see some effects on spectra that involve the electron pressure.}
\label{fig:epsilon2_variation}
\end{figure*}

\begin{figure*}
   \begin{center}
   \includegraphics[height=18cm, angle=270]{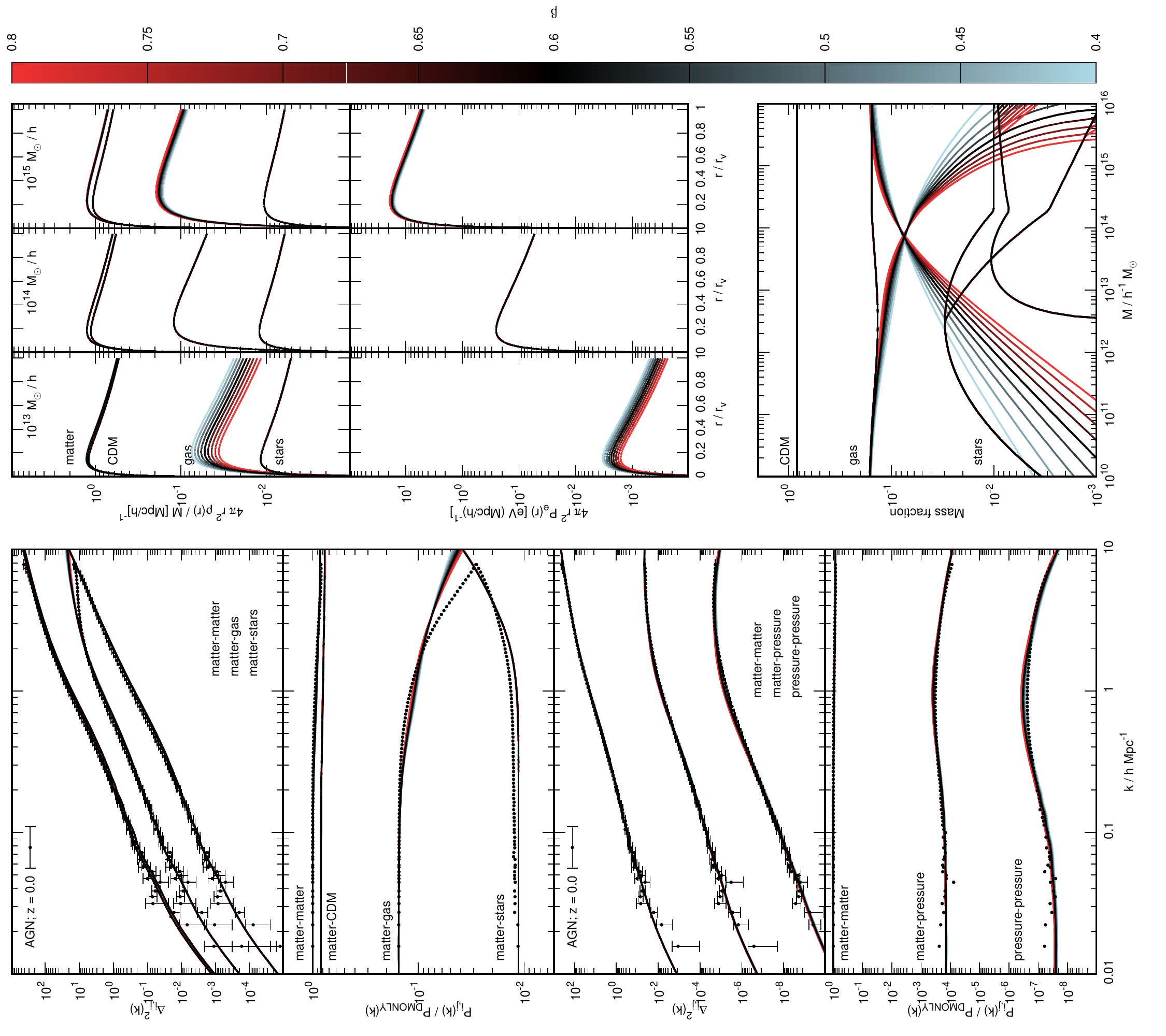}
   \end{center}
   \caption{Same as Fig.~\ref{fig:epsilon_variation} but for $\beta$, the power-law index governing the decline in halo gas fraction for low-mass haloes. As can be seen from the bottom-right panel, increasing $\beta$ makes the decline in gas fraction sharper, and this has a comparatively large effect on low-mass haloes as can be seen in the other panels on the right-hand side of the figure. High mass haloes are affected in the opposite sense because the gas fraction is forced to be $0.5$ at $10^{14}\Msun$. The effects on the power spectra can be seen in the left-hand panels, where the matter--gas spectrum can be seen to be most affected.}
   \label{fig:beta_variation}
\end{figure*}

\begin{figure*}
   \begin{center}
   \includegraphics[height=18cm, angle=270]{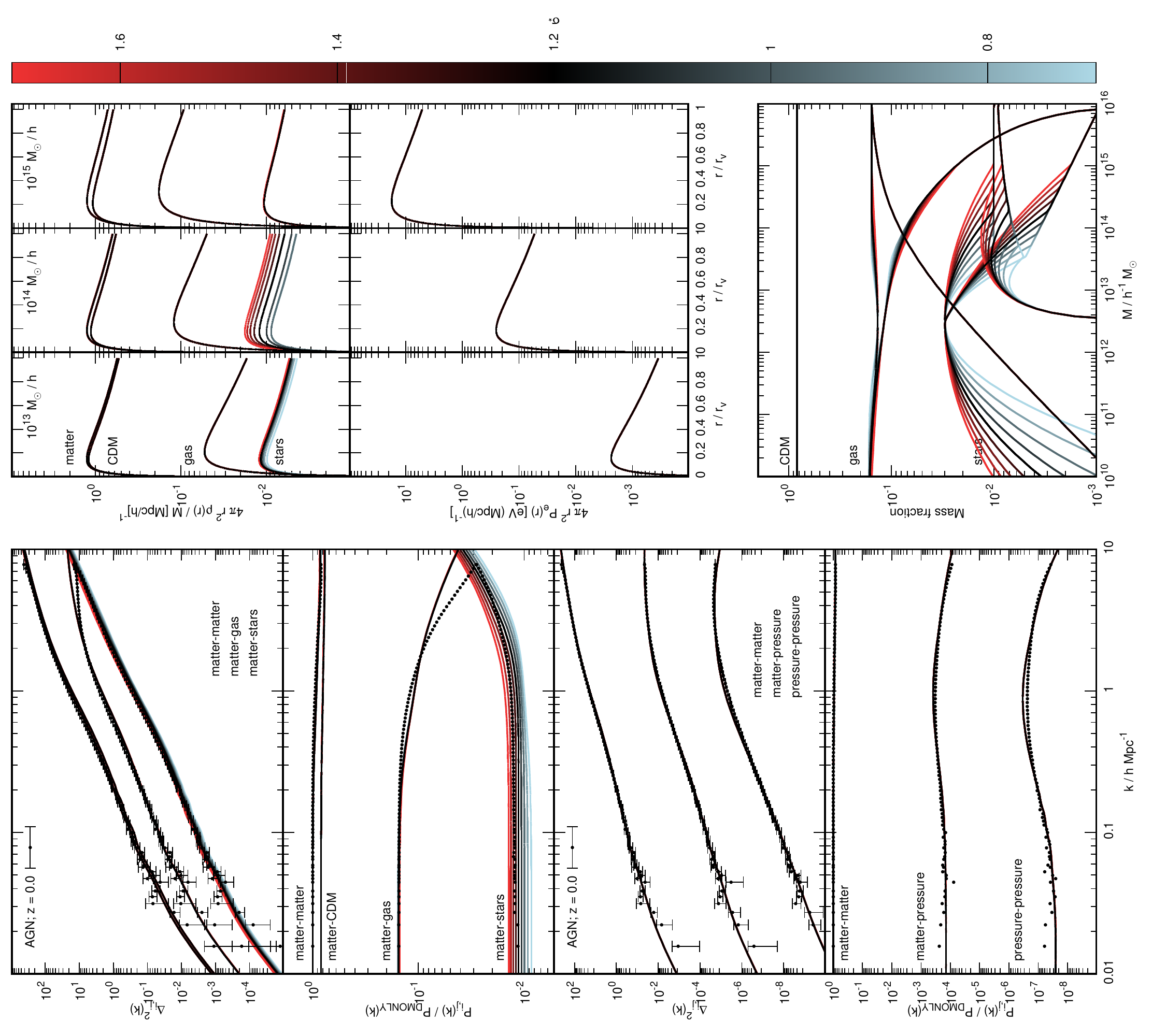}
   \end{center}
   \caption{Same as Fig.~\ref{fig:epsilon_variation} but for $\sigma_*$, which governs the width of the stellar-mass distribution around the peak (which is at $M_* = 10^{12.5}\Msun$ here). As can be seen in the bottom-left panel, decreasing $\sigma_*$ shrinks the stellar-mass distribution, resulting in fewer stars in both higher and lower mass haloes than $M_*$. The top-right panel shows that this has the biggest impact on $10^{14}\Msun$ haloes, not higher-mass haloes due to the saturation we impose at high masses. In the left-hand panels we see that the effects of this are constrained to be on power spectra involving the stars. There is only a tiny effect on gas spectra which is due to baryonic mass conservation.}
   \label{fig:sstar_variation}
\end{figure*}

\section{Projected fields}
\label{app:projections}

In this paper we are primarily interested in the cross correlation between tSZ and various lensing quantities. We use the metric convention of \cite{Bartelmann2001}:
\begin{equation}
\mathrm{d}s^2=\mathrm{d}t^2-a^2[\mathrm{d}r^2+f_k^2(r)(\mathrm{d}\theta^2+\sin^2\theta\mathrm{d}\phi^2)]\ ,
\label{eq:metric}
\end{equation}
where $a$ is the scale factor, $r$ is the comoving distance, and $f_k$ is the comoving, angular-diameter distance.

The weak-lensing signal is produced by the bending of light by the large-scale structure of the Universe. With some approximations, gravitational lensing is conveniently summarised by the convergence field $\kappa$, which can be written as
\begin{equation}
\kappa(\mathbf{\theta})=\frac{3}{2}\Om\left(\frac{H_0}{c}\right)^2\int_0^{r_\mathrm{H}}\frac{f_\mathrm{k}(r)}{a(r)}q(r)\delta_\mathrm{m}(r,\mathbf{\theta})\;\mathrm{d}r\ .
\label{eq:kappa}
\end{equation}
Here $q(r)$ is the lensing-efficiency kernel, which weights redshifts along the line-of-sight according to their contribution to the total lensing
\begin{equation}
q(r)=\int_{z(r)}^\infty p(z')\frac{f_k[r'(z')-r]}{f_k[r'(z')]}\;\mathrm{d}z'\ ,
\label{eq:lensing_efficiency}
\end{equation}
where $p(z)$ is the normalised redshift distribution of lensed galaxies. If one is interested in CMB lensing then $p(z)=\Dirac{z-z_*}$ where $z_*$ is the redshift of the last-scattering surface. In this case:
\begin{equation}
q(r,r_*)=\frac{f_k(r_*-r)}{f_k(r_*)}\ ,
\end{equation}
 where $r_*$ is the comoving distance to $z_*$.
 
 The tSZ signal in the CMB is produced when high energy, free electrons increase the energy of passing, cool CMB photons via inverse Compton scattering. This results in a boost in the effective temperature of the CMB at the location of the electrons that is in proportion to the electron `pressure', usually expressed in terms of the dimensionless Compton `$y$' parameter:
\begin{equation}
y(\mathbf{\theta})=\frac{\sigma_\mathrm{T}}{m_\mathrm{e}c^2}\int_0^{r_\mathrm{H}} \frac{P_\mathrm{e}(r,\mathbf{\theta})}{a^3(r)}\;{a(r)\mathrm{d}r}\ ,
\label{eq:y}
\end{equation}
where $\sigma_\mathrm{T}$ is the Thompson scattering cross section, $m_\mathrm{e}$ is the electron mass, $c$ is the speed of light. The integral is taken over comoving distance between us and the particle horizon, $r_\mathrm{H}$. The electron pressure can be written as $P_\mathrm{e}=n_\mathrm{e}\kb T_\mathrm{e}$ where $n_\mathrm{e}$ is the comoving electron number density and $T_\mathrm{e}$ is the physical electron temperature. The factors of $a$ convert our comoving $r$ and electron pressure to a physical quantities.

Given a general 2D field $U(\mathbf{\theta})$ that is the projected version of a 3D field $u(r,\mathbf{\theta})$ via some projection kernel $X_U(r)$:
\begin{equation} 
U(\mathbf{\theta})=\int_0^{r_\mathrm{H}} X_U(r) u(r,\mathbf{\theta})\;\mathrm{d}r
\label{eq:general_projection}
\end{equation}
 we can write the angular power spectrum between two such projected fields, $U$ and $V$, in terms of the 3D power spectrum of $u$ and $v$ using the Limber approximation (\citealt{Kaiser1992}) as
\begin{equation}
C_{UV}(\ell)=\int_0^{r_\mathrm{H}}\frac{X_U(r)X_V(r)}{f^2_k(r)} P_{uv}\left(k(r),z(r)\right)\;\mathrm{d}r\ ,
\label{eq:limber}
\end{equation}
where $k(r)=(\ell+1/2)/f_k(r)$ and $z(r)$ corresponds to the redshift at comoving distance $r$. We have increased the low-$\ell$ accuracy of the approximation to $\order{\ell^{-2}}$ by including the lowest-order correction $\ell\to\ell+1/2$ \citep{LoVerde2008}. For the lensing projection kernel we have
\begin{equation}
X_\kappa(r)=\frac{3}{2}\Omega_m\left(\frac{H_0}{c}\right)^2\frac{f_k(r)q(r)}{a(r)}\ ,
\label{eq:projection_lensing}
\end{equation}
and we must project the 3D matter--matter power spectrum, $P_\mathrm{mm}(k)$. For $y$ projection we have
\begin{equation}
X_y(r)=\frac{\sigma_\mathrm{T}}{m_\mathrm{e}c^2}\frac{1}{a^2(r)}\ ,
\label{eq:projection_y}
\end{equation}
and we must project the 3D matter--electron pressure power spectrum, $P_\mathrm{mP}(k)$, to obtain the lensing-$tSZ$ cross correlation.

There has been some recent discussion on the validity of the Limber approximation for the interpretation of cosmological observations. We note that comparisons between this approximation and a full calculation show that the Limber approximation is valid given the scales we are interested in, and that we will be interested in in the near future. We point the interested reader towards \cite{Hill2013b}, \cite{Kilbinger2017} and \cite*{Lemos2017}. Note that the accuracy of the Limber approximation is generally \emph{better} for $C_{\kappa y}(\ell)$ compared to for $C_{\kappa\kappa}(\ell)$ due to the broader projection kernel for $y$ compared to that of $\kappa$.

\begin{figure*}
\begin{center}
\includegraphics[height=18cm, angle=270]{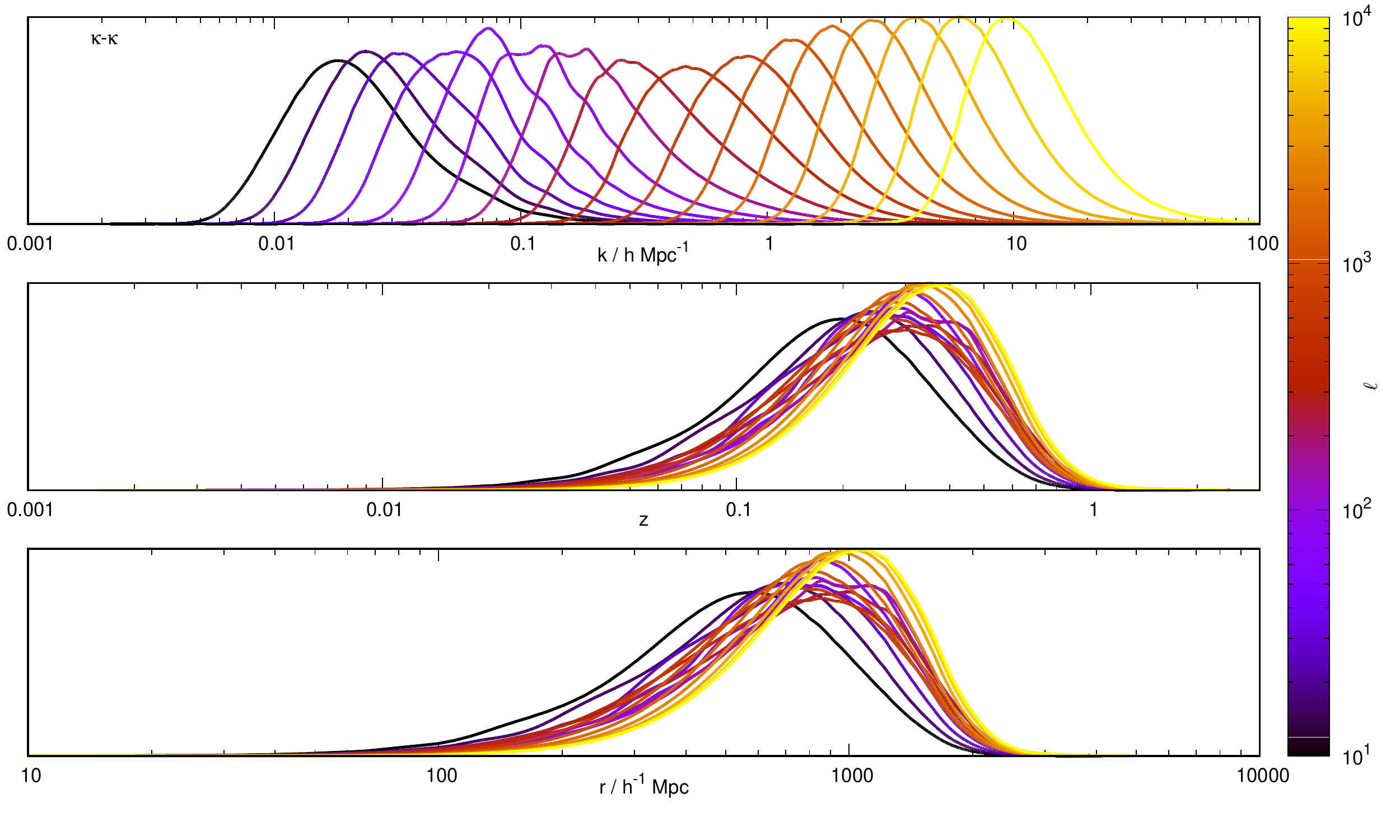}
\includegraphics[height=18cm, angle=270]{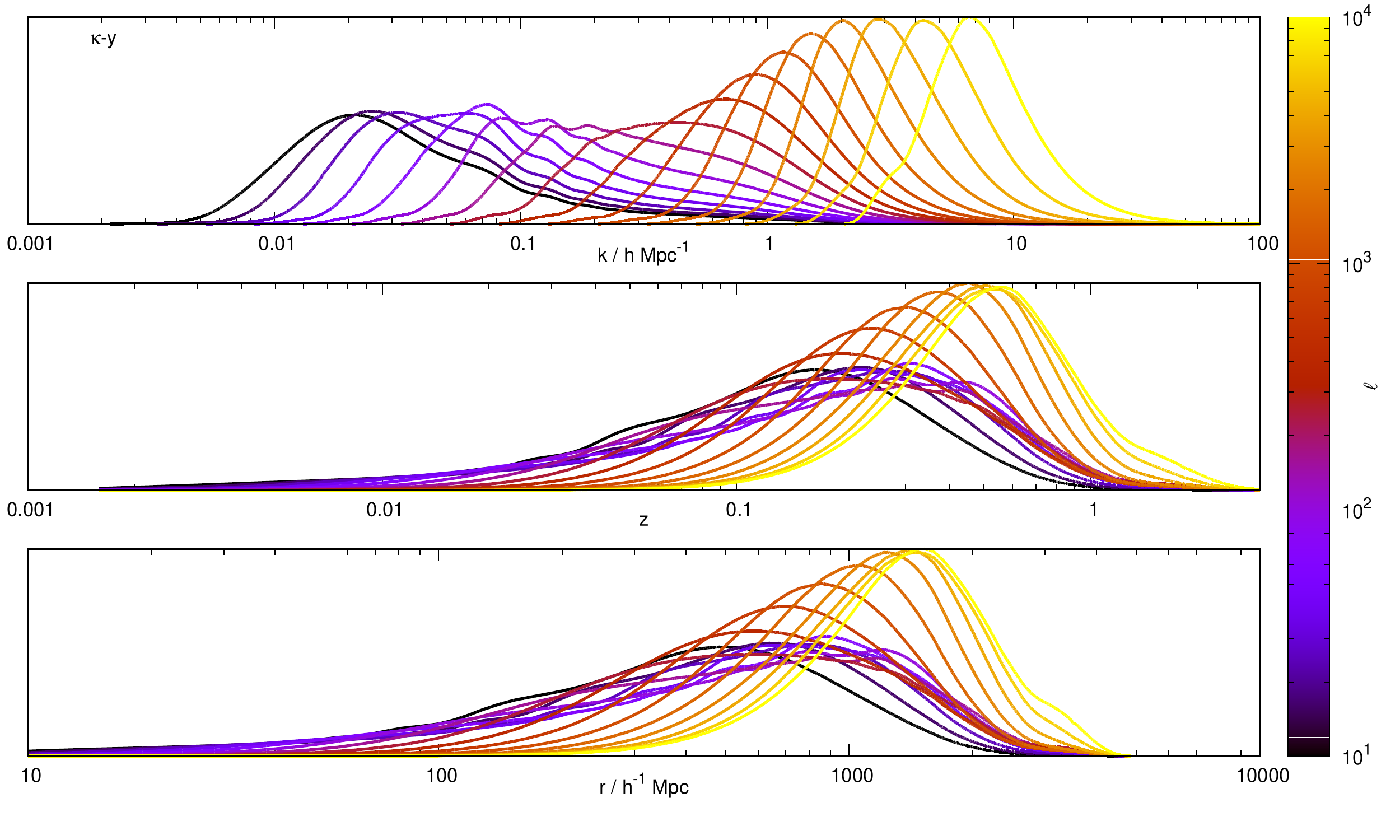}
\end{center}
\caption{The normalised contribution to each $\ell$ of  $C(\ell)$ for $\kappa$--$\kappa$ (top) and $\kappa$--$y$ (bottom) as a function of scale $k$, redshift $z$ and (comoving) distance $r$. The lensing is taken from  \kids galaxies for redshifts $z=0.1\to0.9$. Each of these curves is the normalised integrand in equation~(\ref{eq:limber}) re-expressed as a function of either $k$, $z$ or $r$ (so $\mathrm{d}C(\ell)/\mathrm{d}k$, $\mathrm{d}C(\ell)/\mathrm{d}z$, $\mathrm{d}C(\ell)/\mathrm{d}r$). In practice, we carry out our integration in $r$ between the observer and $z(r)=9$ and we checked that our integrations are robust to this upper limit. The power spectra being integrated are set to zero for $k<10^{-4}\iMpc$ and $k>10^2\iMpc$ and again our results are robust to this. Note the generally broader and lower--$z$ contributions for $\kappa$--$y$ compared to those for $\kappa$--$\kappa$.}
\label{fig:ell_contributions}
\end{figure*}

The contributions to $C_{\kappa\kappa}(\ell)$ and $C_{\kappa y}(\ell)$ as a function of $k$, $z$ and $r$ are shown in Fig.~\ref{fig:ell_contributions} for different $\ell$. Note that, for a given $\ell$, $C_{\kappa y}(\ell)$ receives its contributions from lower $z$ and $r$ compared to $C_{\kappa\kappa}(\ell)$, and also from a much broader range of $k$, $z$ and $r$. To generate this figure we used a $p(z)$ distribution (equation~\ref{eq:lensing_efficiency}) taken from \kids galaxies for redshifts $z=0.1\to0.9$ \citep{Joudaki2017b}.

\end{document}